\documentclass[%
 reprint,
nofootinbib,
 amsmath,amssymb,
 aps,
 prd,
floatfix,
]{revtex4-2}

\usepackage{amsmath}
\usepackage{amssymb}
\usepackage{graphicx} 
\usepackage{leftindex}
\usepackage{musicography}
\usepackage{soul}
\usepackage{color}

\usepackage{dcolumn}

\newcommand{\bs}[1]{\boldsymbol{#1}}
\newcommand{\bvec}[1]{\vec{\boldsymbol{#1}}}
\newcommand{\bdel}{\vec{\boldsymbol{\nabla}}}
\newcommand{\avg}[1]{\langle #1 \rangle}

\definecolor{amber}{rgb}{1.0, 0.49, 0.0}

\newcommand{\tnorm}{|\vec{\tau}|}
\newcommand{\znorm}{|\vec{\zeta}|}

\newcommand{\kform}[2]{\underset{\substack{\sim\\[-2pt]\scriptscriptstyle #1}}{#2}}
\newcommand{\twoform}[1]{\underset{\substack{\sim\\[-2pt]\scriptscriptstyle 2}}{#1}}
\newcommand{\oneform}[1]{\underset{\substack{\sim\\[-2pt]\scriptscriptstyle}}{#1}}

\begin{document}

\title{General Relativistic and Non-Inertial Effects on Magnetic Reconnection Rates}
\author{Jake Willard}
\affiliation{Dartmouth College}
\author{Yi-Hsin Liu}
\affiliation{Dartmouth College}

\begin{abstract}
    In this paper, we offer a theoretical contribution to the study of magnetic reconnection beyond the assumptions made within the standard ``special relativistic" framework, which assumes an inertial system moving within a flat spacetime. In contrast to previous works in this direction, we introduce for the first time a fully gauge-independent foundation for the theory of reconnection: our definition and description of the process, as well as our definition of the ``rate of reconnection," does not depend on a choice of reference frame or coordinate system. Because of this, the real physical differences that result from gravity or the system having net linear acceleration can be cleanly separated from the mere differences in how physics appears within different types of coordinate systems. We explore these real differences by extending the ``mesoscale" theory of reconnection introduced by \citet{Liu2017}, in which the rate is determined not only by physics within the diffusion region but also by ideal and force-free physics within an upstream mesoscale region. We find that our theory accounts for three separate ``non-inertial and curved spacetime" (NICS) effects, including a ``gravitationally induced magnetic tension" effect, caused by the gravitational distortion of the plane of reconnection that forces the magnetic field-lines to be more or less curved than they would be in flat spacetime. We then apply our generalized mesoscale theory to a simplified description of an accretion plasma near an astrophysical compact object, and arrive at two simple conclusions: (1) whether the rate is increased or decreased by gravity is determined purely by the orientation of the system relative to the gravitational source, with the scenario in which the outflow jet is aimed towards or away from the gravitational source producing the greatest possible increase in the rate; (2) the magnitude of the gravitational effect on the rate scales with the dimensionless parameter $\bar{\kappa} \equiv r_s\Delta y^2/r^3$, where $r$ is the radial position of the system relative to the Schwarzschild radial coordinate, $r_s$ is the Schwarzschild radius of the gravitational source, and $\Delta y$ is the scale length that defines the ``mesoscale" in our theory.
\end{abstract}

\maketitle

\section{Introduction} \label{Section::Intro}

Magnetic reconnection is a ubiquitous process seen in magnetized plasmas across a wide diversity of scales and situations. Conceptually, it is a process whereby anti-parallel magnetic fields are dragged together such that the field lines ``appear" to break in half at the magnetic null point and then reconnect, forming field lines with higher field-line curvature. The tension in the magnetic field downstream will accelerate the plasma to potentially very high speeds, creating jets and a large release in energy \citep{Priest2000}. Reconnection is the cause of solar flares, substorms in planetary magnetospheres, is believed to drive high energy emissions in black hole magnetospheres, and more \citep{Hesse2020}. 

Astrophysical contexts involving plasmas in the vicinity of compact objects suggests the need for a theory of magnetic reconnection that is compatible with general relativity. Progress in this direction has been limited to the attempt to generalize the Sweet-Parker model beyond the assumptions of an inertial system in flat spacetime \citep{Asenjo2017, Comisso2018}. One major limitation of these initial works is their reliance on a representation of the system containing non-physical degrees of freedom which are difficult to disentangle from the real physics. The interpretation of the results has therefore become a subject for debate, with \citet{Shen2024} arguing that the entirety of what \citet{Comisso2018} declares as a gravitational effect on the rate of reconnection is in fact a non-physical artifact of their choice of coordinate system. They further argue that the relevant spatial scales involved in magnetic reconnection are simply far too small for a general relativistic approach to substantially differ from a special relativistic approach. 

These critiques highlight two flaws in the way the problem has so far been considered. Firstly, there are the non-physical degrees of freedom that exist within a standard description of magnetic reconnection. These degrees of freedom derive from the gauge freedom of general relativity, which determine in part the behavior of electric and magnetic fields. Since a magnetic field will behave differently in a general relativistic context than in a special relativistic context for reasons which are not entirely physical, a reliance on such a representation of the process will likely always result in some confusion over the physical interpretation of one's results. The second flaw is the expectation implicit within the Sweet-Parker model itself that the rate of reconnection will be determined only by physics at the scale of the diffusion region itself. It is important to note that the skepticism of Ref.\@ \citep{Shen2024} towards to necessity of a general relativistic theory of reconnection is only reasonable if this assumption is correct. Going beyond the Sweet-Parker, it was shown by \citet{Liu2017} that the inclusion of a broader mesoscale region in addition to the diffusion region is necessary to self consistently explain fast reconnection rates.

This paper aims to make four new contributions. Firstly, in Section \ref{Section::Theory} we provide a fully gauge-invariant foundation for a theory of magnetic reconnection in which the system is itself defined independently from any reference frame or choice of coordinate system. In this description, the rate of reconnection will then not be a relative quantity which depends in any way on how the system is observed, but a scalar parameter quantifying the invariant characteristics of the Faraday 2-form $\twoform{F}$. We will then present a natural coordinate representation of steady-state reconnection defined locally around its central world-sheet. 

In Section \ref{Section::Overview}, we will explore the new electromagnetic behavior that is included in the fully general relativistic framework. This new behavior derives from both the real physical content of general relativity that is technically excluded by special relativity (gravity) as well as the behavior specific to a system which reaches a steady-state within a non-inertial frame of reference, which is \textit{de facto} excluded by the standard formalism of special relativity. The combination of effects seen in a non-inertial system within a curved spacetime (NICS) we will refer to as ``NICS effects." The three primary NICS effects on the rate of reconnection include a dilation/contraction of time scales associated with light propagation through the system (and by extension, the propagation of magnetic flux), a gravitational distortion of the system`s 3D shape which changes the equilibrium value of the plasma flow speed, and a ``gravitationally induced magnetic tension" caused by the gravitational distortion of the plane of reconnection itself which effectively contributes to the curvature of magnetic field-lines. 

In Section \ref{Section::GeneralizedFormulae}, we then apply this foundation to derive a leading-order correction to the formula for the reconnection rate derived by \citet{Liu2017} accounting for the local NICS effects. This theory is a ``two-scale" theory, meaning that it finds the rate to be determined not solely by physics within the diffusion region, but also by mesoscale physics upstream. We will only explore the way that NICS effects enter in at the mesoscale, accepting as valid the argument made by Ref.\@ \citep{Shen2024} that dissipation-scale physics is likely unaffected by these considerations. Finally, in Section \ref{Section::CentralBodyProblem} we use our formula to explore the NICS effects which arise for magnetic reconnection in the vicinity of massive compact objects. We attempt to do so in the simplest way possible, by assuming a static and spherically symmetric gravitational source and an accretion plasma which has zero angular momentum relative to this source. We ultimately derive a very simple expression for the average rate of reconnection as a function of the system's orientation within the gravitational field, and its size relative to the spacetime curvature scale. 

Throughout the paper, we will represent the electrodynamics using Lorentz-Heaviside units, all speeds will be normalized to the speed of light such that $c=1$, and we will adopt the convention that $\eta_{\mu \nu} \doteq \text{diag}(-1,1,1,1)$. Any 4-vector with components $v^{\mu}$ will be expressed in coordinate free notation simply as $\vec{v}$, and any higher rank tensor with components $T_{\mu \nu \cdots}^{\alpha \beta \cdots}$ will be denoted as $\mathbb{T}$. The exterior derivative will be denoted as $\oneform{d}$, and $k$-forms will be denoted as $\kform{k}{T}$. In Section \ref{Subsection::GaugeConstraints}, special emphasis will be given to vectors and vector components which are tangent to a specific spacelike 2D surface, and these will be expressed in boldface (e.g, $\bvec{B}_r$ and $\bvec{H}_r$). 

\section{Theoretical Foundations} \label{Section::Theory}

It is well known that the representation of electromagnetism given in terms of vector fields $\vec{E}$ and $\vec{B}$ contains non-physical degrees of freedom. Within the standard framework of special relativity, these degrees of freedom are merely the freedom to choose any inertial frame of reference. The choice of inertial frame will greatly influence the values of $\vec{E}$ and $\vec{B}$, but will not influence at all the form of Maxwell's Equations which govern their behavior, since these equations are preserved by Lorentz transformations. The choice of frame therefore does not affect at all the physical meaning of the fields, which is independent of the non-physical choices made in how to represent the physics. By contrast, the gauge freedom of general relativity encompasses a far greater flexibility in coordinate representations, and the form of Maxwell's Equations will no longer be independent of that choice. Here, one choice of gauge conditions will lead to $\vec{E}$ and $\vec{B}$ fields which behave in entirely different ways from an alternative choice, which makes the correct physical interpretation of this behavior difficult. We provide more detail on this issue in Appendix \ref{Appendix::MaxwellsEquations}.

Magnetic reconnection is conventionally defined as a type of behavior exhibited by the magnetic field. This is problematic within a fully general relativistic framework, since a part of this behavior is something the physicist chooses. In this section, we will therefore show how the essential characteristics of the process can be defined independently of any coordinate representation. We will then establish a natural set of gauge constraints which are appropriate for the case of steady-state reconnection, and will then clarify the meaning of our definitions in terms of this representation. 

\subsection{Gauge-Invariant View of Reconnection} \label{Subsection::InvariantReconnectionDefinition}

Our aim in this section is to establish a definition of magnetic reconnection and its rate using only the invariant characteristics of the Faraday 2-form $\twoform{F}$. In other words, we will provide an alternative conception of the process which is not strictly reliant on its decomposition within local reference frames into separate electric and magnetic vector fields. For the sake of intuition, it will be helpful to make reference to the components of $\twoform{F}$ within a ``natural" choice of local frame. At each spacetime event, let the vector field $\vec{e}_t$ be the 4-velocity of such a frame, where we assume that this roughly describes the motion through spacetime of the system. Then let $\{\vec{e}_t,\vec{e}_x,\vec{e}_y,\vec{e}_z \}$ be an orthonormal basis at each spacetime event, i.e.\@, $g^{\mu \nu} = \eta^{\hat{\alpha} \hat{\beta}} (\vec{e}_{\hat{\alpha}})^\mu(\vec{e}_{\hat{\beta}})^\nu$, where $\hat{\alpha},\hat{\beta} \in \{t,x,y,z \}$; ``hatted'' indices are reserved for orthonormal basis through out the paper. We will also define a ``co-basis" $\{ \oneform{e}^t, \oneform{e}^x, \oneform{e}^y, \oneform{e}^z \}$ such that $\oneform{e}^{\hat{\alpha}}(\vec{e}_{\hat{\beta}}) = \delta^{\hat{\alpha}}_{\hat{\beta}}$, or equivalently, $g_{\mu \nu} = \eta_{\hat{\alpha} \hat{\beta}} (\oneform{e}^{\hat{\alpha}})_\mu (\oneform{e}^{\hat{\beta}})_\nu$. If $\vec{E}$ and $\vec{B}$ are 4-vectors representing the electric and magnetic fields within this local frame, $\twoform{F}$ has the decomposition,
\begin{equation} \label{Statement::FOrthonormal}
\begin{split}
    \twoform{F} =& B^x \oneform{e}^y \wedge \oneform{e}^z + B^y \oneform{e}^z \wedge \oneform{e}^x + B^z \oneform{e}^x \wedge \oneform{e}^y \\
    +& E^x \oneform{e}^x \wedge \oneform{e}^t + E^y \oneform{e}^y \wedge \oneform{e}^t + E^z \oneform{e}^z \wedge \oneform{e}^t.
\end{split}
\end{equation}

\begin{figure*}
    \centering
    \includegraphics[width=0.9\linewidth]{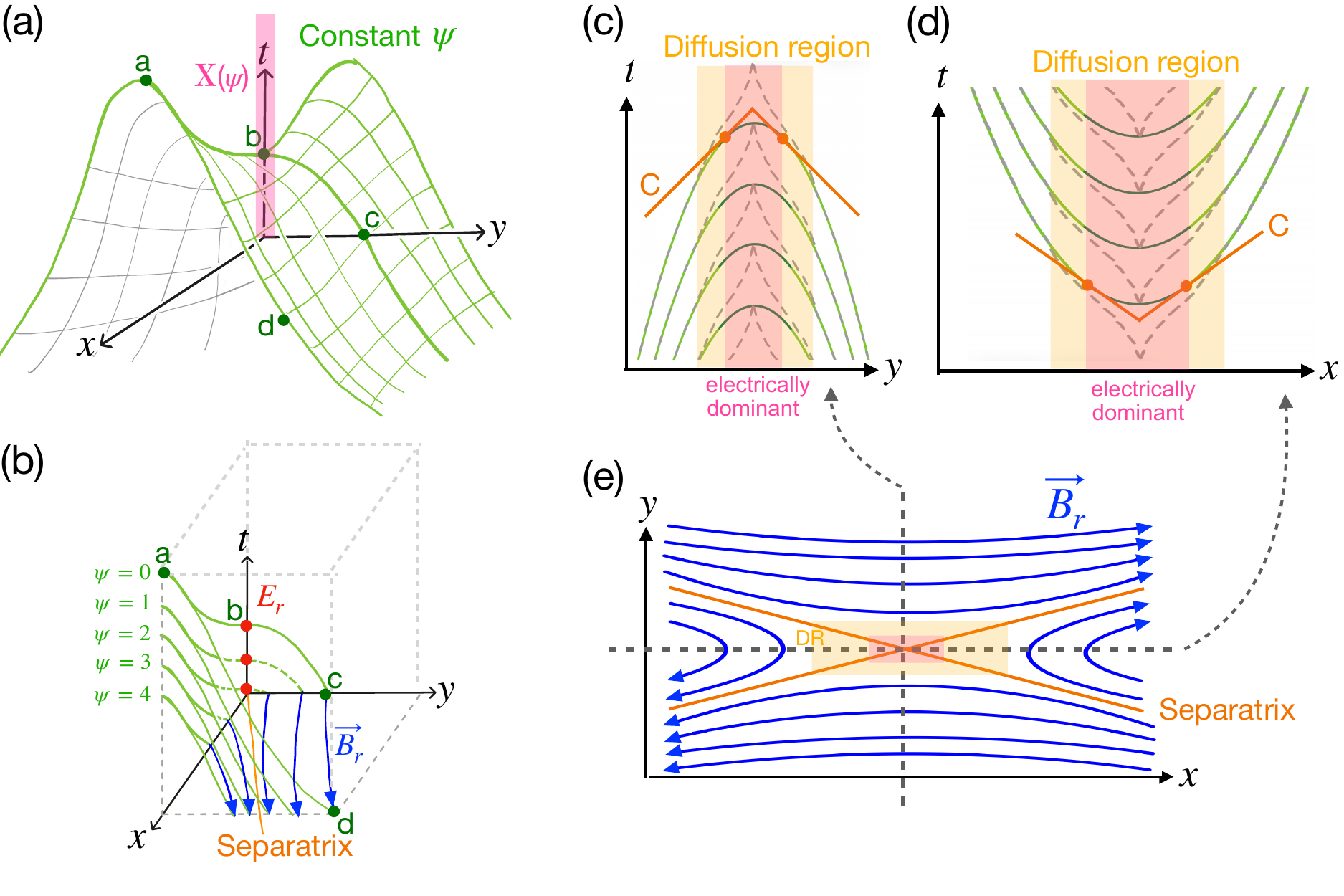}
    \caption{Illustration of the field-sheet structure of $\twoform{F_r}$ which characterizes the reconnection process. (a) Depiction of the field-sheets themselves within a constant $z$ surface. The pink shaded region is the electrically dominant region, approximated as a curve $\mathtt{X}(\psi)$ in Section \ref{Subsection::RateDefinition} (b) Illustration of the relation between field sheets and reconnecting magnetic field $\bvec{B}_r$ and out-of-plane electric field $E_r$ in the first  octant of panel (a). (c) $x=0$ cross section of the field-sheet structure. Solid green curves are field-sheets, and dashed gray curves are the world-lines of the plasma inflow. The yellow shaded region is the diffusion region, and the the pink shaded region is the electrically dominant region. This is defined as the region where $|\vec{\nabla}\psi|^2<0$, and is therefore bounded between the two points where the light curves (i.e., $\pm 45^\circ$ orange lines) are tangent with the field-sheets. (d) $y=0$ cross section of the field-sheet structure and the world-lines of plasma outflow. (e) a constant $t$ cross section of the field-sheets, which gives the expected magnetic field-line geometry for magnetic reconnection. Here the orange diagonal lines depict the magnetic separatrix.}
    \label{fig::ReconnectionInSpacetime}
\end{figure*}

First, consider a simplified case where $\vec{e}_z$ represents the ``out-of-plane" direction, so that the ``reconnecting components" of $\vec{B}$ are $B^x=F_{y z}$ and $B^y = F_{zx}$, and the component of $\vec{E}$ induced by their reconnection is $E^z=F_{zt}$. Note that they are the $z$ components of $\twoform{F}$ and these components are entirely contained within the 1-form $\twoform{F} \cdot \vec{e}_{z} = B^x\oneform{e}^y - B^y \oneform{e}^x - E^z \oneform{e}^t$. Using this insight, now we will assume that the system is ``2D" in the sense that there exists a spacelike vector field $\vec{\zeta}$ such that all of the physics is invariant under the flow of $\vec{\zeta}$. Formally, it requires the Lie derivative along $\vec{\zeta}$ of any physical tensor $\mathbb{T}$ to vanish, i.e., $\mathcal{L}_{\vec{\zeta}} \mathbb{T} = 0$. The ``out-of-plane" direction is now given by the symmetry of the system, and the ``reconnecting part" of $\twoform{F}$ is contained within the 1-form $\oneform{d}\psi \equiv \twoform{F} \cdot \vec{\zeta}$, where $\psi$ will be called the ``flux function." This identity assumes implicitly that $\twoform{F} \cdot \vec{\zeta}$ is a closed form, but using Cartan's ``magic formula" \citep{Hehl2003}, we can show that this fact is a consequence of this translational symmetry,
\begin{equation}
    \mathcal{L}_{\vec{\zeta}} \twoform{F} = \vec{\zeta} \cdot (\oneform{d}\twoform{F}) + \oneform{d}( \vec{\zeta} \cdot \twoform{F})=  -\oneform{d}(\twoform{F} \cdot \vec{\zeta})= 0,
\end{equation}
where $\oneform{d}\twoform{F}=0$ because of the Gauss-Faraday Law. We will show that all of the essential characteristics of the process of reconnection and its rate can be purely defined in relation to the level surfaces of constant $\psi$. 

Within a spacelike submanifold of spacetime (i.e.\@, a ``time slice"), the constant-$\psi$ surfaces correspond to the ``flux surfaces" commonly referenced in ``2D" plasma physics theories. That is, they have the meaning of ``in-plane" magnetic field lines that are extended along $\vec{\zeta}$ to form 2D surfaces, whose time evolution then describes a 3D surface in spacetime. We will illustrate this meaning by first introducing a decomposition of $\twoform{F}$ that follows from a choice of spatial coordinate $z$ ``adapted" to $\vec{\zeta}$, such that $\oneform{d}z(\vec{\zeta})=1$ and $|\vec{\nabla}z|^2>0$. Within any coordinate system containing $z$, $\vec{\partial}_z =\vec{\zeta}$, and therefore $(\oneform{d}\psi)_\mu=\partial_{\mu}\psi = F_{\mu \nu} (\vec{\partial}_z)^{\nu} = F_{\mu z}$, which picks up all the $z$-components of $\twoform{F}$, as expected. We define $\twoform{F_g}$ as the part of $\twoform{F}$ having ``no $z$-component," that is, $(\twoform{F_g})_{\mu z}=0$ for all $\mu$. $\twoform{F}$ then has the decomposition,
\begin{equation}\label{F_decomposition}
\begin{split}
   \twoform{F} &= \oneform{d}\psi \wedge \oneform{d}z + \twoform{F_g} \\
    &= \twoform{F_r} + \twoform{F_g},     
\end{split}
\end{equation}
where $\twoform{F_r}=\oneform{d}\psi \wedge \oneform{d}z$ captures the essential electromagnetic field components critical to the definition of reconnection rates.
If we suppress the $z$ dimension (i.e., on a constant-$z$ surface) in our visualization, we will see the 2-form $\twoform{F_r}$ as layers of constant-$\psi$ structure associated with the 1-form $\oneform{d}\psi$. One of these continuous 2D surfaces (with no boundary) is illustrated in Fig.\@ \ref{fig::ReconnectionInSpacetime}(a). 

By comparing Eq.\@ (\ref{F_decomposition}) directly with Eq.\@ (\ref{Statement::FOrthonormal}), the meaning of the ``spatial" components of $\vec{\nabla}\psi$ must have the meaning of the in-plane magnetic field, whereas the time variation in $\psi$ is related to the out-of-plane electric field. To make this association more obvious, assume an inertial system within a flat spacetime, where $\{t ,x,y, z \}$ is a ``Minkowskian" coordinate system corresponding to the system's ``rest frame." Then, $\oneform{d}x^{\hat{a}} = \oneform{e}^{\hat{a}}$, and $\oneform{d}\psi$ can be decomposed as $\oneform{d}\psi=\nabla_{\hat{a}}\psi\oneform{d}x^{\hat{a}}$. 
Therefore,
\begin{equation} 
\begin{split}
    \twoform{F} =& (\partial_x\psi \oneform{d} x + \partial_y \psi \oneform{d}y + \partial_t \psi \oneform{d}t) \wedge \oneform{d}z+\twoform{F_g} \\
    =& \partial_y \psi \oneform{e}^y \wedge \oneform{e}^z - \partial_x \psi \oneform{e}^z \wedge \oneform{e}^x - \partial_t \psi \oneform{e}^z \wedge \oneform{e}^t+\twoform{F_g} \\
\end{split}
\end{equation}
Comparing with Eq.\@ (\ref{Statement::FOrthonormal}), we get
\begin{equation} \label{Result::BandEMinkowski}
    \hookrightarrow B^x = \partial_y \psi, \quad B^y =-\partial_x \psi, \quad E^z = -\partial_t \psi.
\end{equation}
Note that $\twoform{F_g}$ contains none of the information of these field components.
We will call the 2D surfaces having constant $\psi$ and $z$ ``field-sheets," which is a term we borrow from \citet{Gralla2014}, and draw further connection to in Appendix \ref{Appendix::FieldSheets}. The relation between the field-sheet structure and magnetic reconnection geometry follows intuitively from Eq.\@ (\ref{Result::BandEMinkowski}), and is shown in Fig.\@ \ref{fig::ReconnectionInSpacetime}(b) and \ref{fig::ReconnectionInSpacetime}(e); the intersection of field sheets with a spatial 2D surface ($(x,y)$ here) traces out the in-plane magnetic field ($\bvec{B}_r$) lines and the spacing between these blue lines is inversely proportional to this in-plane magnetic field strength (i.e., Eq.\@ (\ref{Result::BandEMinkowski})). The spacing between the red dots, where the $t$-axis pierces the field sheets is inversely proportional to the strength of the out-of-plane electric field, $E_r$, at the reconnection X-line (i.e., Eq.\@ (\ref{Result::BandEMinkowski})).

Each field-sheet describes the ``motion" of a pair of field-lines flowing towards the diffusion region (as illustrated in Fig.\@ \ref{fig::ReconnectionInSpacetime}(c)) and a new pair of field-lines flowing away from the diffusion region (as illustrated in Fig.\@ \ref{fig::ReconnectionInSpacetime}(d)). i.e., imaging the motion of blue magnetic field lines in Fig.\@ \ref{fig::ReconnectionInSpacetime}(b) when one moves the $t$-slice upward.  
The ideal Ohm's Law, $\vec{U} \cdot \twoform{F}=0$, requires that $\psi$ is a constant of the bulk plasma motion, since $\vec{U} \cdot \twoform{F} \cdot \vec{\zeta}=\oneform{d}\psi(\vec{U}) = 0$. Therefore, the diffusion region is represented in Fig.\@ \ref{fig::ReconnectionInSpacetime}(c) and (d) as the region where the world-lines of the plasma (the gray dashed curves) begin to deviate from the field-sheets. Also shown in Fig.\@ \ref{fig::ReconnectionInSpacetime}(c)-(d) is the region where $\twoform{F_r}$ become electrically dominant, i.e.\@, $|\twoform{F_r}|^2=(B^x)^2 + (B^y)^2-(E^z)^2 < 0$. This region is necessitated if there is to exist an in-plane magnetic null point featuring a non-zero (and non-ideal) electric field. Note that this electrically dominant region is also defined simply by $|\vec{\nabla}\psi|^2<0$, or equivalently, as the region where the field-sheets are spacelike. Outside of this region, the field-sheets are timelike and can therefore be conceptualized as ``world-sheets" of the field-line motion. This conceptualization breaks down when the field-sheets become spacelike, where ``field-line motion" ceases to have any invariant meaning.

Fig.\@ \ref{fig::ReconnectionInSpacetime} depicts only the cross section of the (3D) constant-$\psi$ surfaces within a single constant-$z$ surface, and the illustrations were given the intuitive justification under the assumption of an inertial system in flat spacetime, which led to the relations in Eq.\@ (\ref{Result::BandEMinkowski}). However, the essential geometric characteristics which these illustrations represent can be stated generally. We can assert that, in general, magnetic reconnection features a region where $|\vec{\nabla}\psi|^2<0$, and that the constant-$\psi$ surfaces ``pass through" this region so as to result in ``saddle-shaped" field-sheets, as in Fig.\@ \ref{fig::ReconnectionInSpacetime}(a). In other words, they enter the region tangent to ``in-flowing" light surfaces (i.e., orange lines of $\pm 45^\circ$ in Fig.\@ \ref{fig::ReconnectionInSpacetime}(c)), and exit the region tangent to ``out-flowing" light surface (i.e., orange lines of $\pm 45^\circ$ in Fig.\@ \ref{fig::ReconnectionInSpacetime}((d)).

\subsection{Covariant Definition of the Rate} \label{Subsection::RateDefinition}

\begin{figure}
    \centering
    \includegraphics[width=1.0\linewidth]{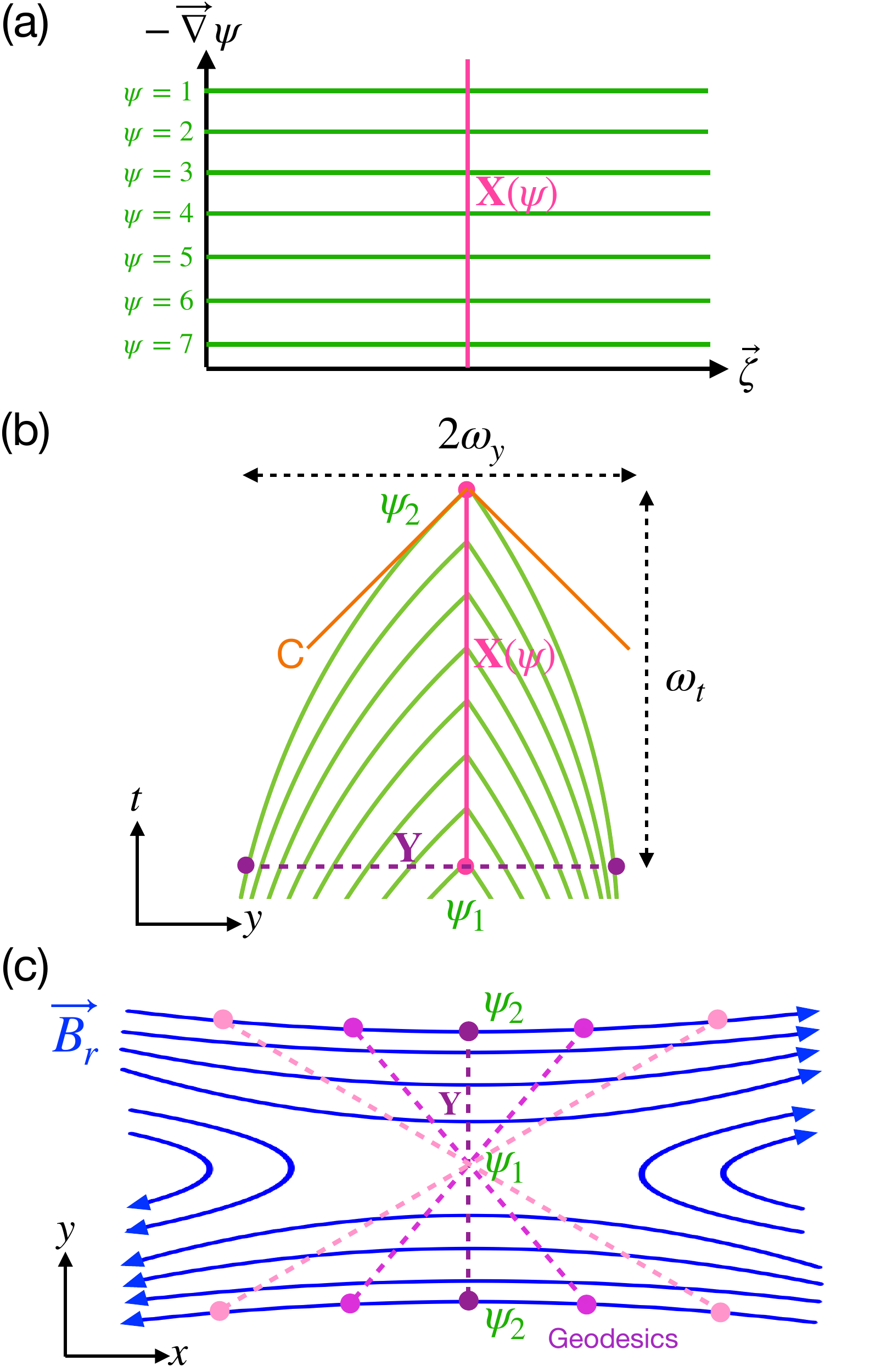}
    \caption{Illustration of the geometric constructions which define the reconnection rate. In (a), a range of constant-$\psi$ surfaces is shown within the X-sheet, spanned by $\vec{\nabla}\psi$ and $\vec{\zeta}$. The world-line $\mathtt{X}(\psi)$ is also shown to flow orthogonal to $\vec{\zeta}$ and parallel to $\vec{\nabla}\psi$. In (b), the green curves represent the range of constant-$\psi$ curves given by $\psi\in[\psi_1,\psi_2]$ within a spacetime diagram, which also contains the curve $\mathtt{Y}$ and the segment of $\mathtt{X}(\psi)$ connecting $\psi=\psi_1$ to $\psi=\psi_2$. In (c), the blue curves (i.e., in-plane magnetic field lines) represent the same range of constant-$\psi$ curves $\psi\in[\psi_1,\psi_2]$ within a ``time slice", and the dashed curves represent geodesics passing through $\mathtt{X}(\psi_1)$ (the middle point) orthogonally to the X-sheet and end on the surfaces $\psi=\psi_2$ on both sides. $\mathtt{Y}$ is defined as the shortest of these dashed curves.}
    \label{fig::XYIllustration}
\end{figure}

The electrically dominant region where $|\vec{\nabla}\psi|^2<0$ is generally ``slab-shaped," that is, it is ``thin" in the $x$ and $y$ dimensions and ``long" in the $z$ and $t$ dimensions. For our definition of the rate, we will assume that this region is sufficiently ``thin" that it can effectively be treated as a 2D sheet spanned by $\vec{\nabla}\psi$ and $\vec{\zeta}$. In Fig.\@ \ref{fig::ReconnectionInSpacetime}, this corresponds to the assumption that the pink shaded region in panels (a) and (c)-(d) can be approximated as a world-line, and the region in panel (e) as a single point. We will call this 2D sheet the ``X-sheet," since it is essentially the ``world-sheet" of the in-plane magnetic ``X-line." In Fig.\@ \ref{fig::XYIllustration}(a), the constant-$\psi$ surfaces are depicted within the X-sheet. Since $\oneform{d} \psi(\vec{\zeta})=\vec{\zeta}\cdot \twoform{F} \cdot \vec{\zeta}=0$, these are also the flow-lines of $\vec{\zeta}$. If we ``slice" the X-sheet orthogonally to $\vec{\zeta}$, the result is a world-line we will denote as $\mathtt{X}(\psi)$, which is a flow-line of $\vec{\nabla}\psi$ as shown in Fig.\@ \ref{fig::XYIllustration}(a). Since these events are precisely the ``moments of reconnection," the rate of reconnection ought to be defined such that it quantifies the proper time intervals along $\mathtt{X}(\psi)$. 

It will be helpful to briefly review what quantity the term ``rate of reconnection" conventionally refers to in reconnection studies. By Faraday's Law, the rate of change of the total upstream magnetic flux must be equal to the induced ``electromotive force" EMF along the X-line $d\Phi/dt = E^{z} w_z$, where $w_z$ is the length of the X-line. If $w_t$ is the time interval over which an amount of magnetic flux $\Phi$ reconnects, then $\Phi = \avg{E^z} w_t w_z$, where the average is taken over a patch of the X-sheet with dimensions $(w_t,w_z)$. At any given moment in time, the total upstream magnetic flux is $\Phi = \avg{B^x} w_y w_z $, where this average is taken over a patch of the $(y,z)$ plane upstream having dimension $(w_y,w_z)$. What is conventionally called the reconnection rate is actually a quantity of speed obtained by normalizing the EMF $\Phi/w_t$ against $\Phi/w_y$, yielding $w_y/w_t = \avg{E^z}/\avg{B^x}=\avg{\partial_t \psi}/\avg{\partial_y \psi}$. That is, the rate is just the distance interval $w_y$ over which a certain range in $\psi$ is distributed spatially, divided by the time interval $w_t$ over which that same range in $\psi$ is distributed temporally along $\mathtt{X}(\psi)$. 

Making this notion gauge-invariant is simple. We first define $w_t$ as exactly the proper time interval along $\mathtt{X}(\psi)$, as depicted in Fig.\@ \ref{fig::XYIllustration}(b), over which a specific range of $\psi$ undergoes reconnection. This range will be $\psi \in [\psi_1,\psi_2]$, where we assume that $\mathtt{X}(\psi_2)$ lies in the future light cone of $\mathtt{X}(\psi_1)$ (i.e.\@, the field-sheets of $\psi=\psi_1$ reconnect before $\psi=\psi_2$). We define $w_t\big|_{\psi_1}^{\psi_2}$ to be the proper time interval along $\mathtt{X}(\psi)$ between the events $\mathtt{X}(\psi_1)$ and $\mathtt{X}(\psi_2)$. Let $\mathtt{Y}\big|_{\psi_1}^{\psi_2}$ be the shortest geodesic curve which intersects the event $\mathtt{X}(\psi_1)$ orthogonally to the X-sheet (that is, orthogonally to $\vec{\nabla}\psi$ and $\vec{\zeta}$) and meets the surface $\psi=\psi_2$ at both endpoints, as shown in Fig.\@ \ref{fig::XYIllustration}(c). We then define the length of $\mathtt{Y}\big|_{\psi_1}^{\psi_2}$ as $2w_y\big|_{\psi_1}^{\psi_2}$, as in Fig.\@ \ref{fig::XYIllustration}(b). Note that $w_y$ is effectively the average length scale corresponding to the range $\psi \in [\psi_1,\psi_2]$ between both ``inflow streams." Finally, we define the rate of reconnection,
\begin{equation}
    \mathcal{R}\Big|_{\psi_1}^{\psi_2} \equiv \left( \frac{w_y}{w_t} \right) \Big|_{\psi_1}^{\psi_2}.
\end{equation}
Importantly, this definition only references to the covariant structure of field sheets, $\oneform{d}\psi=\twoform{F_r}\cdot\vec{\zeta}$, independent of the choice of coordinates and/or observers. Therefore, this definition of reconnection rate is covariant.

\subsection{Natural Gauge Constraints for ``2D" and Stationary Reconnection} \label{Subsection::GaugeConstraints}

In addition to the assumption of ``2D" reconnection, our theory will also assume that the process reaches a steady state. This means that there are now two vector fields which represent translational symmetries of the system in spacetime: a spacelike vector $\vec{\zeta}$ defined such that $\mathcal{L}_{\vec{\zeta}} \mathbb{T} = 0$ for any physical tensor $\mathbb{T}$, and now a timelike vector $\vec{\tau}$ defined such that $\mathcal{L}_{\vec{\tau}} \mathbb{T} = 0$. In order for our theory to be consistent, we must be able to apply this constraint to the spacetime metric $g_{\mu \nu}$. Therefore, $\vec{\zeta}$ and $\vec{\tau}$ must be Killing vectors, satisfying the equations \citep{Kriele2011},
\begin{equation} \label{Statement::KillingVectors}
\begin{split}
    &\mathcal{L}_{\vec{\zeta}} g_{\mu \nu} = \nabla_{\mu} \zeta_{\nu} + \nabla_{\nu} \zeta_{\mu} = 0, \\
    &\mathcal{L}_{\vec{\tau}} g_{\mu \nu} = \nabla_{\mu} \tau_{\nu} + \nabla_{\nu} \tau_{\mu} = 0.
\end{split}
\end{equation}
Our gauge constraint will be that our coordinate system includes a time coordinate $t$ and spatial coordinate $z$ such that $\vec{\partial_t} \equiv \vec{\tau}$ and $\vec{\partial_z} \equiv \vec{\zeta}$. Eq.\@ (\ref{Statement::KillingVectors}) then implies $\partial_t g_{\mu \nu} = 0$ and $\partial_z g_{\mu \nu} = 0$.  We will assume that $\vec{\tau}$ and $\vec{\zeta}$ are orthogonal to each other, and also that they are each ``hypersurface orthogonal," which simply means that there is a unique way to satisfy the constraint such that $\vec{\nabla} t \propto \vec{\tau}$ and $\vec{\nabla} z \propto \vec{\zeta}$, and the spacetime line element then takes the form,
\begin{equation} \label{Statement::GaugeConstraint}
    ds^2 = -\tnorm^2 dt^2 + \bs{g}_{ij} d\bs{x}^i d\bs{x}^j + \znorm^2 dz^2,
\end{equation}
where $\tnorm = \sqrt{|\tau_{\mu} \tau^{\mu}|}$, $\znorm = \sqrt{|\zeta_{\mu} \zeta^{\mu}|}$, and $\bs{g}_{ij}$ is the induced metric in the 2D surfaces of constant $z$ and $t$, and $\bs{x}^i$ are coordinates parameterizing these surfaces. We will now adopt the convention that vectors which lie within the plane of reconnection will be denoted using boldface. One should note that hypersurface orthogonality is not necessarily a realistic assumption, and we intend to further generalize our theory beyond this limitation in a later work.

\begin{figure}
    \centering
    \includegraphics[width=0.8\linewidth]{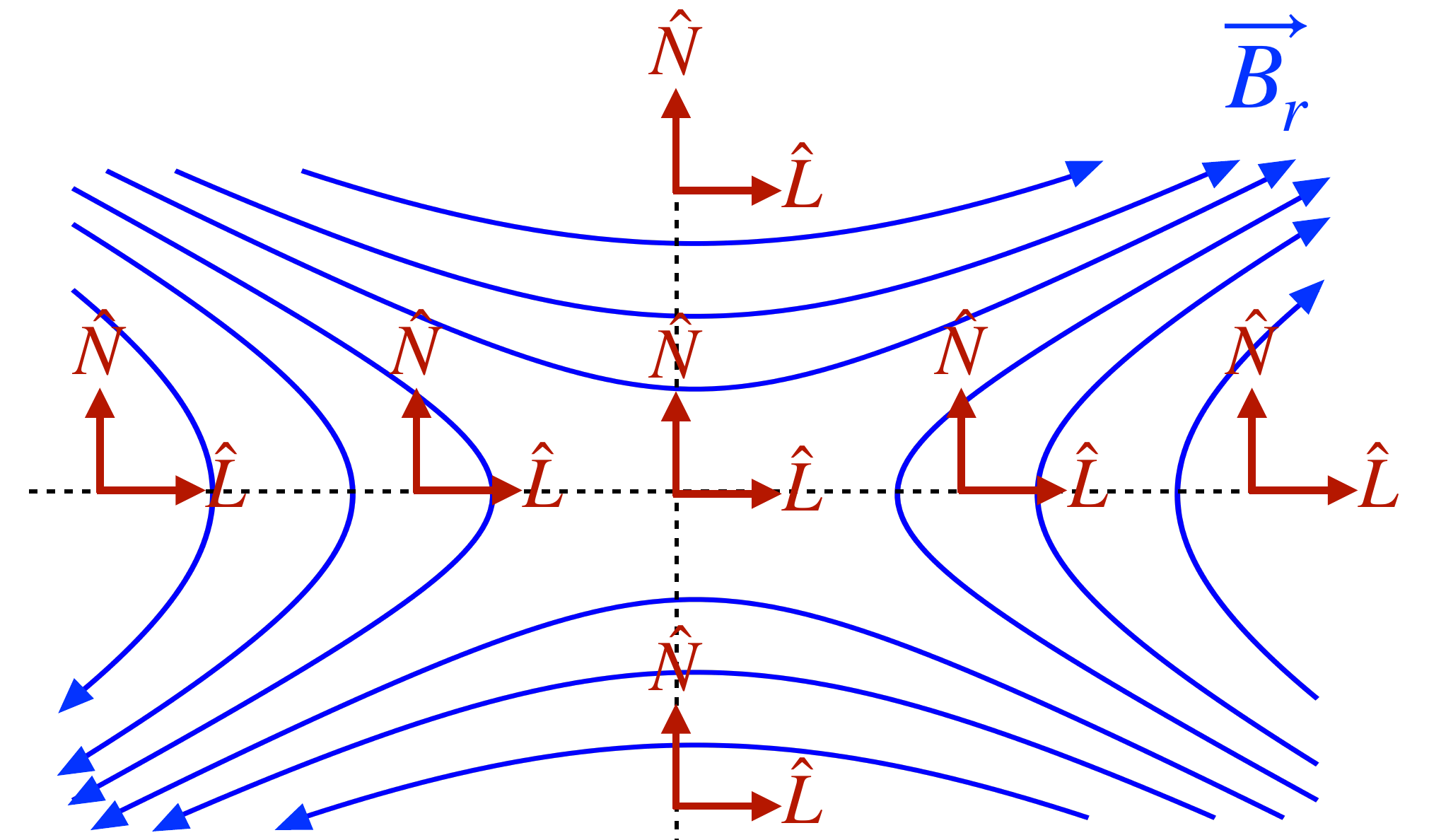}
    \caption{Illustration of the ``inflow line" (vertical dashed line) and ``outflow line" (horizontal dashed line) in relation to the orthonormal basis vectors $\hat{\bs{L}}$ and $\hat{\bs{N}}$. Although in Eq.\@ (\ref{Statement::FCoordinateDecomposition}) it is assumed that these vectors are defined everywhere within the plane of reconnection, we do not specify their exact orientation except along the intersecting curves shown. Note that the ``inflow line" will become the $y$-axis, and the ``outflow line" the x-axis, in Section \ref{Subsection::Coordinates}.}
    \label{fig::LNVisual}
\end{figure}

This choice of gauge determines a representation of the reconnection process within a 2D plane, corresponding to any of the submanifolds orthogonal to $\vec{\tau}$ and $\vec{\zeta}$, which we will call the ``plane of reconnection." In relation to this plane, we will define an orthonormal set of basis vectors familiar in reconnection studies \citep{Denton2016, Denton2018}. In the plane itself, we define a vector $\hat{\bs{L}}$ aligned with the plasma outflow, and a vector $\hat{\bs{N}}$ aligned with the plasma inflow (normal to the current sheet), which are ``orthonormal" such that $\bs{g}_{ij} = \bs{L}_i\bs{L}_j + \bs{N}_i\bs{N}_j$. Out of the plane, we define a unit vector $\hat{M} \equiv \vec{\zeta}/\znorm$. If we also define the vector $\hat{T} \equiv \vec{\tau}/\tnorm$, then $\{\hat{T}, \hat{\bs{L}}, \hat{\bs{N}}, \hat{M} \}$ is a complete orthonormal basis. $\vec{E}$ and $\vec{B}$, which are the electric and magnetic fields within the local frame of $\hat{T}$, can then be decomposed into ``in-plane" and ``out-of-plane" components,
\begin{equation} \label{Statement::EBDecomposition}
\begin{split}
    &\vec{B} \equiv \bvec{B}_r + B_g \hat{M}, \\
    &\vec{E} \equiv \bvec{E}_g + E_r \hat{M}.
\end{split}
\end{equation}
$\bvec{B}_r$ are the reconnecting components of $\vec{B}$, and $B_g$ is the ``guide field." Conversely, $E_r$ is the electric field component induced by the reconnection process, and $\bvec{E}_g$ are the ``in-plane" electric components induced by an evolving guide field. This decomposition corresponds precisely to the decomposition of the Faraday 2-form $\twoform{F} = \twoform{F_r} + \twoform{F_g}$ introduced in Section \ref{Subsection::InvariantReconnectionDefinition}. If we adopt the convention where, given a vector $\vec{v}$, $\oneform{v}$ denotes the ``dual" 1-form satisfying $\oneform{v}(\vec{u})\equiv \vec{v} \cdot \vec{u}$, then the ``co-basis" corresponding to $\{\hat{T}, \hat{\bs{L}}, \hat{\bs{N}}, \hat{M} \}$ is simply $\{\oneform{T}, \oneform{\bs{L}}, \oneform{\bs{N}}, \oneform{M} \}$, and $\twoform{F_r}$ and $\twoform{F_g}$ have the decomposition,
\begin{equation} \label{Statement::FCoordinateDecomposition}
\begin{split}
    \twoform{F_r} =& \text{Rot}(\bvec{B}_r) \wedge \oneform{M} + E_r \oneform{M} \wedge \oneform{T} \\
    \twoform{F_g} =& B_g \oneform{\bs{L}} \wedge \oneform{\bs{N}} + \oneform{\bs{E}_g} \wedge \oneform{T}
\end{split}
\end{equation}
where $\text{Rot}(\bvec{v})\equiv \bvec{v} \cdot (\oneform{L} \wedge \oneform{N}) = \bs{v}^L\oneform{\bs{N}} - \bs{v}^N \oneform{\bs{L}}$ is the 1-form representing a $\pi/2$ rotation of $\bvec{v}$ within the plane of reconnection. To relate the above expression for $\twoform{F_r}$ to its definition in Section \ref{Subsection::InvariantReconnectionDefinition} $\twoform{F_r} \equiv \oneform{d} \psi \wedge \oneform{d}z$, note that $\oneform{T}=g_{\mu \nu}T^{\mu} \oneform{d}x^{\nu}=-\tnorm \oneform{d}t$ and $\oneform{M} = g_{\mu \nu} M^{\mu} \oneform{d}x^{\nu}=\znorm \oneform{d}z$. Therefore,
\begin{equation} \label{Result::CoordinatedPsi}
\begin{split}
    \oneform{d}\psi = \twoform{F} \cdot \vec{\zeta} =& \znorm \left(\text{Rot}(\bvec{B}_r) - E_r \oneform{T} \right) \\
    =& \znorm \text{Rot}(\bvec{B}_r) + \znorm \tnorm E_r \oneform{d}t
\end{split}
\end{equation}
Note that, while $\partial_t \psi$ is a uniform function in a steady state (since $\vec{\nabla}\partial_t \psi = \partial_t \vec{\nabla}\psi=0$), $E_r = \partial_t \psi/(\znorm \tnorm)$ will be non-uniform in general.

Within the plane of reconnection, the X-sheet corresponds to the point where $\bvec{B}_r=0$, which we will denote as $\mathtt{x}$ and call the ``X-point." The field-line structure of $\bvec{B}_r$ surrounding this point is shown in Fig.\@ \ref{fig::LNVisual}, which also illustrates the orientation of the vectors $\hat{\bs{L}}$ and $\hat{\bs{N}}$ along the ``inflow" and ``outflow" lines, which are defined as the two curves within the plane of reconnection which intersect at $\mathtt{x}$ and are everywhere orthogonal to $\bvec{B}_r$. Since the X-point is ``stationary" within the plane of reconnection with respect to both $\vec{\tau}$ and $\vec{\zeta}$, the world-line $\mathtt{X}(\psi)$ is a flow-line of $\hat{T}$, and the ``X-line" is a flow-line of $\hat{M}$. Therefore, the 4-acceleration of the X-point is $\bvec{a} \equiv \nabla_{\hat{T}} \hat{T} \big|_{\mathtt{x}}$, and the curvature of the X-line is $\bvec{h} \equiv \nabla_{\hat{M}} \hat{M} \big|_{\mathtt{x}}$. Note that these are effectively the two extrinsic ``principal curvatures" of the X-sheet (i.e., Fig. \@\ref{fig::XYIllustration}(a) with $\vec{\nabla}\psi\ \|\ \hat{T}$ and $\vec{\zeta}\ \|\ \hat{M}$). In general, the curvatures of the flow-lines of $\hat{T}$ and $\hat{M}$ are encoded in the first derivatives of the functions $\tnorm$ and $\znorm$,
\begin{equation}
\begin{split}
    \nabla_{\hat{T}} T^{\nu}  &=  \left(T^{\mu} \partial_{\mu} T^{\nu} + \Gamma^{\nu}_{\alpha \beta} T^{\alpha} T^{\beta} \right) \\
    &= \left( \frac{1}{|\vec{\tau}|} \partial_t (\delta_t^{\nu} /|\vec{\tau}|) + \frac{1}{|\vec{\tau}|^2}\Gamma_{tt}^{\nu} \right) \\
    &= \frac{g^{\nu \lambda}}{2|\vec{\tau}|^2} \left(\partial_t g_{\lambda t} + \partial_t g_{\lambda t} - \partial_{\lambda} g_{tt}  \right) \\
    &= \nabla^{\nu} \ln(|\vec{\tau}|), \\ \\
    \nabla_{\hat{M}} M^{\nu} &= \left(M^{\mu} \partial_{\mu} M^{\nu} + \Gamma^{\nu}_{\alpha \beta} M^{\alpha} M^{\beta} \right) \\
    &= \left( \frac{1}{\znorm} \partial_t (\delta_z^{\nu} /\znorm) + \frac{1}{\znorm^2}\Gamma_{zz}^{\nu} \right) \\
    &= \frac{g^{\nu \lambda}}{2\znorm^2} \left(\partial_z g_{\lambda z} + \partial_z g_{\lambda z} - \partial_{\lambda} g_{zz}  \right) \\
    &= -\nabla^{\nu} \ln(|\vec{\zeta}|).
\end{split}
\end{equation}
We will want to treat the X-sheet as the central world-sheet which represents the overall motion of the system. $\bvec{a}$ should therefore be thought of as the net linear acceleration of the system itself, and $\bvec{h}$ as encapsulating its 3D shape. A theory of reconnection should treat these as independent variables one derives from the physical situation itself, and so their relationship to $\tnorm$ and $\znorm$ must be imposed as additional constraints on the coordinate system,
\begin{equation} \label{Statement::SecondaryGaugeConstraint}
\begin{split}
    &\vec{\nabla} \ln(|\vec{\tau}|) \Big|_{\mathtt{x}} = \bvec{a}, \\
    &\vec{\nabla} \ln(|\vec{\zeta}|) \Big|_{\mathtt{x}} = -\bvec{h}.
\end{split}
\end{equation}

\subsection{The Stationary Rate of Reconnection} \label{Subsection::StationaryRate}

We will assume that the ``inflow line" is a geodesic curve. Comparing Fig.\@ \ref{fig::XYIllustration}(c) with Fig.\@ \ref{fig::LNVisual}, it is intuitive that the curve $\mathtt{Y}\big|_{\psi_1}^{\psi_2}$ with $\psi\big|_{\mathtt{x}}=\psi_1$ is a segment of the inflow line with endpoints on the field-line $\psi=\psi_2$. Since $\partial_t \psi$ has a uniform value within the plane of reconnection, $\Delta \psi\equiv |\psi_2-\psi_1|$ is stationary. For a given $\Delta \psi$, $\mathtt{Y}\big|_{\psi_1}^{\psi_2}$ is then a stationary curve within the plane of reconnection, which will be denoted simply as $\mathtt{Y}[\Delta \psi]$. We will parameterize the points along $\mathtt{Y}[\Delta \psi]$ by the parameter $s$, defined such that $|s|$ measures the arc length along the curve from the point $\mathtt{x}$ and $s \rightarrow -s$ is a reflection about the point $\mathtt{x}$. We will then make a distinction between the two sides of the inflow line lying on opposite sides of the current sheet, by labeling $\mathtt{Y}^+$ as the points corresponding to $s>0$ and $\mathtt{Y}^-$ corresponding to $s<0$. Recall that $w_y$ is defined as half the total length of $\mathtt{Y}[\Delta \psi]$. We will now define $w_y^{\pm}$ as the total length of $\mathtt{Y}^{\pm}$, such that $w_y \equiv (w_y^+ + w_y^-)/2$. We can then express the rate as the average of ``one-sided" rates,
\begin{equation}
    \mathcal{R} = \frac{w_y}{w_t} = \frac{1}{2} \left(\frac{w_y^+}{w_t} + \frac{w_y^-}{w_t} \right).
\end{equation}
We will find a relationship between these one-sided rates and the in-plane ``drift speed" $\bs{V}_d \equiv |E_r|/|\bs{B}_r|$ along the inflow line which, under the frozen-in condition, is identical with the speed of the bulk plasma inflow. First, we express the drift speed in terms of the flux function $\psi$,
\begin{equation} \label{Statement::DriftSpeed}
    \bs{V}_d \equiv \frac{|E_r|} {|\bs{B}_r|} = \Big|\frac{\partial_t \psi / (\tnorm \znorm)}{\partial_s \psi / \znorm}\Big| = \Big| \frac{\partial_t \psi}{\partial_{\bar{s}} \psi} \Big|,
\end{equation}
where $\bar{s}$ is defined such that $d\bar{s} \equiv \tnorm^{-1}ds$, the meaning of which will be clarified shortly. Again using the fact that $\partial_t\psi$ must be a uniform, we then define $\bs{\bar{V}}_d^{\pm}$ as the ``harmonic average" of $\bs{V}_d$ over $\bar{s}$ along $\mathtt{Y}^{\pm}$,
\begin{equation}
\begin{split}
    \bs{\bar{V}}_d^{\pm} &\equiv \left(\frac{1}{\pm \Delta \bar{s}^{\pm}}\int_{0}^{\pm \Delta \bar{s}^{\pm}} \frac{1}{\bs{V}_d(\bar{s})}d\bar{s}\right)^{-1} \\
    &= \Big|\frac{\partial_t \psi}{\langle \partial_{\bar{s}} \psi\rangle^{\pm}}\Big|,
\end{split}
\end{equation}
where $\langle\rangle^{\pm}$ denotes an average taken over $\bar{s}$ along $\mathtt{Y}^{\pm}$. We will now explicitly compute the parameter $\Delta \psi$ by integrating $\oneform{d}\psi$ along $\mathtt{X}(\psi)$ (the world-line of $\mathtt{x}$) and along $\mathtt{Y}^{\pm}$,
\begin{equation}
\begin{split}
    \Delta \psi = \partial_t \psi \frac{w_t}{\tnorm_{\mathtt{x}}} = \int_{\mathtt{Y}^{\pm}} \oneform{d}\psi &= \Delta \bar{s}^{\pm} \langle \partial_{\bar{s}} \psi \rangle^{\pm} = \Delta \bar{s}^{\pm} \partial_t \psi/ \bs{\bar{V}}_d^{\pm}, \\
    \hookrightarrow w_t &= \tnorm_\mathtt{x} \Delta \bar{s}^{\pm} / \bs{\bar{V}}_d^{\pm}, \\
    \hookrightarrow \frac{w_y^{\pm}}{w_t} &= \frac{w_y^{\pm}}{\tnorm_{\mathtt{x}}\Delta \bar{s}^{\pm}} \bar{\bs{V}}_d^{\pm},
\end{split}
\end{equation}
where $\Delta \bar{s}^{\pm} \equiv \pm \bar{s}(\pm w_y^{\pm})$. We will then define $\mathcal{C}^{\pm} \equiv w_y^{\pm}/(\tnorm_{\mathtt{x}} \Delta \bar{s}^{\pm})$, and the rate of reconnection is given by,
\begin{equation} \label{Result::StationaryRate}
    \mathcal{R} = \frac{1}{2}\left(\bs{\bar{V}}_d^+ \mathcal{C}^+ + \bs{\bar{V}}_d^- \mathcal{C}^- \right).
\end{equation}
We will call $\mathcal{C}^{\pm}$ the ``effective speed of light." Light always propagates at the same speed relative to any \textit{local} frame of reference. However, although light always travels the same ``infinitesimal" distance over any ``infinitesimal" proper time interval, light does not necessarily travel over \textit{finite} distances in the expected amount of \textit{finite} proper time. It is well know that light signals are delayed by gravitational fields. It is also well known that accelerating frames experience a similar time delay if they are accelerating away from the signals that are sent towards them. Both of these effects are examples of NICS effects which will amount to a difference in the effective speed of light. In Section \ref{Section::Overview}, we will further clarify how $\mathcal{C}^{\pm}$ actually depends on local gravitational effects and the system's net acceleration. 

Equation (\ref{Result::StationaryRate}) clarifies the relationship between our scalar definition of the reconnection rate and more conventional definitions, which take the rate to be identical with the speed of the plasma inflow relative to observers that are co-moving with the system. We will find that in the absence of NICS effects, $\mathcal{C}^{\pm} \rightarrow 1$, and the rate approaches the harmonic average over inflow speeds relative to the rest frame of the X-point. This is qualitatively consistent with conventional theories, in which the rate is often \textit{defined} as the speed of the inflow plasma evaluated at a point upstream. While our theory remains \textit{quantitatively} inconsistent with past theories in this limit, it is only because our theory approaches an average over inflow speeds rather than its exact value at some chosen point.

\subsection{Fermi Coordinate Representation} \label{Subsection::Coordinates}

 We will now fully define a coordinate system for the spacetime locally around the X-sheet, where the \textit{extrinsic} curvatures of the X-sheet (vectors $\bvec{a}$ and $\bvec{h}$) and the \textit{intrinsic} curvature of spacetime (the Riemann curvature tensor $\mathbb{R}$) both appear explicitly as independent variables determining the metric tensor. First, we parameterize the plane of reconnection itself using Riemann coordinates $x$ and $y$ such that the components $\bs{g}_{ij}$ are given by the well-known formula \citep{Sternberg2013},
\begin{equation} \label{Statement::RiemannMetricFormula}
    \bs{g}_{ij} = \delta_{ij} - \frac{1}{3}\left(R_{iajb} \right)\Big|_{\mathtt{x}} \bs{x}^a \bs{x}^b + \mathcal{O}(|\bvec{x}|^3),
\end{equation}
where $(x,y)=(0,0)$ is the X-point $\mathtt{x}$ and $\bvec{x} \equiv x\vec{\partial_x} + y \vec{\partial_y}$. Throughout the rest of the paper, the Latin indices $\{i,j,k,\dots \}$ index only the coordinates $x$ or $y$, whereas the Greek indices will index all spacetime coordinates. Note that $\bs{g}_{xx}=1$ ($|\vec{\partial_x}|=1$) along $y=0$ and $\bs{g}_{yy}=1$ ($|\vec{\partial_y}|=1$) along $x=0$. Therefore, if we choose to align these coordinates such that the $x$-axis is the ``outflow line" and the $y$-axis is the ``inflow line" $\mathtt{Y}$, then,
\begin{equation} 
\begin{split}
    &\hat{\bs{L}}\Big|_{y=0}=\vec{\partial_x}, \quad \oneform{\bs{L}} \Big|_{y=0}=\oneform{d}x,\\
    &\hat{\bs{N}}\Big|_{x=0}=\vec{\partial_y}, \quad \oneform{\bs{N}} \Big|_{x=0}=\oneform{d}y.
\end{split}
\end{equation}
Using Eq.\@ (\ref{Statement::FCoordinateDecomposition}) along the inflow line, 
\begin{equation} \label{Statement::FInflowDecomposition}
    \twoform{F_r} \Big|_{\mathtt{Y}} = \bs{B}^L \oneform{d}y \wedge \oneform{M} + E_r \oneform{M} \wedge \oneform{T}.
\end{equation}

Next, we will fully constrain the $t$ and $z$ coordinates. Since $\tnorm$ and $\znorm$ vary only within the plane of reconnection, the remaining (non-physical) degree of freedom can be fixed simply by specifying $\tnorm_{\mathtt{x}}=1$ and $\znorm_{\mathtt{x}}=1$. Combined with the Riemann coordinates $\vec{\boldsymbol{x}}$, the metric components are now the same as the Minkowski metric $\eta_{\mu \nu}$ everywhere along the X-sheet. This is known as Fermi coordinates \citep{Lee2021}, and in a neighborhood around the X-sheet, the components of $g_{\mu \nu}$ are,
\begin{equation} \label{Result::FermiMetricFormula}
\begin{split}
    &g_{tt} = -(1 + \vec{\boldsymbol{x}} \cdot \vec{\boldsymbol{a}})^2 + S_{ij} \boldsymbol{x}^i \boldsymbol{x}^j + \mathcal{O}(|\vec{\boldsymbol{x}}|^3), \\
    &g_{xx} = 1 - \frac{1}{3} \mathcal{K} y^2 + \mathcal{O}(|\vec{\boldsymbol{x}}|^3),\\ 
    &g_{yy} = 1 - \frac{1}{3} \mathcal{K} x^2 + \mathcal{O}(|\vec{\boldsymbol{x}}|^3), \\ 
    &g_{xy} = \frac{1}{3}\mathcal{K} xy + \mathcal{O}(|\vec{\boldsymbol{x}}|^3), \\
    &g_{zz} = (1 - \vec{\boldsymbol{x}} \cdot \vec{\boldsymbol{h}})^2 + Q_{ij} \boldsymbol{x}^i \boldsymbol{x}^j + \mathcal{O}(|\vec{\boldsymbol{x}}|^3),
\end{split}
\end{equation}
where $g_{xx}$, $g_{yy}$, and $g_{xy}$ follow from Eq.\@ (\ref{Statement::RiemannMetricFormula}), and $g_{tt}$ and $g_{zz}$ are derived in Appendix \ref{Appendix::FermiGttGzz}. $\mathbb{S}$, $\mathbb{Q}$, and $\mathcal{K}$ are derived from $\mathbb{R}$, 
\begin{equation} \label{Statement::DefinitionOfSQK}
\begin{split}
    &S_{\mu \nu} \equiv -R_{\alpha \mu \beta \nu} T^{\alpha} T^{\beta} \Big|_{\mathtt{x}},\\
    &Q_{\mu \nu} \equiv -R_{\alpha \mu \beta \nu} M^{\alpha} M^{\beta} \Big|_{\mathtt{x}}, \\
    &\mathcal{K} \equiv R_{\mu \alpha \nu \beta} N^{\mu} L^{\alpha} N^{\nu} L^{\beta} \Big|_{\mathtt{x}}.
\end{split}
\end{equation}
$\mathbb{S}$ is the tidal stress tensor \citep{Hobson2011}, which produces an acceleration on the system whose sign depends on its orientation with respect to the gravitational source. $\mathbb{Q}$ can be thought of as a spatial analogue of the tidal stress tensor: it measures the deviation between two nearby geodesics that are initially parallel and in the direction of $\hat{M}$. $\mathcal{K}$ is the intrinsic Gaussian curvature at $\mathtt{x}$ of the plane of reconnection itself\footnote{Technically, $\mathcal{K}$ as defined in Eq. (\ref{Statement::DefinitionOfSQK}) is called the ``sectional curvature" at the point $\mathtt{x}$ associated with the tangent subspace spanned by $\{\hat{\bs{L}}, \hat{\bs{N}} \}$ \citep{Sternberg2013} In other words, $\mathcal{K}$ is the Gaussian curvature at $\mathtt{x}$ of the 2D geodesic surface which is tangent to $\hat{\bs{L}}$ and $\hat{\bs{N}}$ at the point $\mathtt{x}$. As discussed in Appendix \ref{Appendix::FermiGttGzz}, the plane of reconnection itself is precisely this surface, since geodesics that are tangent to the surface at one point are tangent at all other points.}. 

Using $\tnorm = \sqrt{-g_{tt}}$ and $\znorm = \sqrt{g_{zz}}$, it is useful to derive the following formulae,
\begin{equation} \label{Result::TAndZFormulae}
\begin{split}
    \tnorm &= 1 + \bs{a}_i \bs{x}^i + \frac{1}{2}\left(\bs{a}_i \bs{a}_j - S_{ij} \right) \bs{x}^i \bs{x}^j + \mathcal{O}(|\bvec{x}|^3),\\
    \znorm &= 1 - \bs{h}_i \bs{x}^i + \frac{1}{2}\left( \bs{h}_i \bs{h}_j + Q_{ij} \right) \bs{x}^i \bs{x}^j + \mathcal{O}(|\bvec{x}|^3),\\
    \ln(\tnorm) &= \bs{a}_i \bs{x}^i - \frac{1}{2}S_{ij} \bs{x}^i \bs{x}^j + \mathcal{O}(|\bvec{x}|^3),\\
    \ln(\znorm) &= -\bs{h}_i \bs{x}^i + \frac{1}{2} Q_{ij} \bs{x}^i \bs{x}^j + \mathcal{O}(|\bvec{x}|^3).\\
\end{split}
\end{equation}
We also provide the non-zero Christoffel symbols truncated to order $\mathcal{O}(|\bvec{x}|)$, which we derive in Appendix \ref{Appendix::FermiChristoffel},
\begin{equation}\label{Result::ChristoffelSymbols}
\begin{split}
    \Gamma^t_{ti} &= \left(\bvec{a} - \mathbb{S} \cdot \bvec{x} \right)_i, \\
    \Gamma^z_{zi} &= \left(-\bvec{h} + \mathbb{Q} \cdot \bvec{x} \right)_i, \\
    \Gamma^i_{tt} &= \bs{a}^i + \left(\bs{a}^i\bs{a}_j -S^i_j  \right) \bs{x}^j,\\
    \Gamma^i_{zz} &= \bs{h}^i - \left(Q^i_j + \bs{h}^i\bs{h}_j \right) \bs{x}^j,\\
    \Gamma^x_{yi} &= \frac{1}{3} \mathcal{K} \left(2x \delta_{yi} - y\delta_{xi} \right),\\
    \quad \Gamma^y_{xi} &= \frac{1}{3}\mathcal{K} \left(2y\delta_{xi} - x\delta_{yi} \right).
\end{split}
\end{equation}

Because $g_{yy}=1$ everywhere along the inflow line $x=0$, $s=y + \mathcal{O}(y^3)$, where $s$ is the inflow line arc length parameter defined in Section \ref{Subsection::StationaryRate}. We will define the coordinate $\bar{y}$ via the relationship $d\bar{y}=\tnorm^{-1} dy$, such that $\bar{s} = \bar{y} + \mathcal{O}(y^3)$. Using Eq.\@ (\ref{Result::TAndZFormulae}), this coordinate has the explicit formula which will be needed in the following section,
\begin{equation} \label{Result::YToYBar}
\begin{split}
    &\frac{d\bar{y}}{dy} = \tnorm^{-1} = 1 - \bs{a}_yy + \frac{1}{2}\left(S_{yy} - \bs{a}_y^2 \right)y^2 + \mathcal{O}(y^3), \\ \\
    &\hookrightarrow \bar{y} = y - \frac{1}{2}\bs{a}_y y^2 + \frac{1}{6}\left(S_{yy} - \bs{a}_y^2 \right) y^3 + \mathcal{O}(y^4).
\end{split}
\end{equation}

\section{Overview of Electromagnetic NICS Effects} \label{Section::Overview}

In the previous section, we provided a foundation for analyzing reconnection which can account for NICS effects. In this section, we will break down the essential ways that NICS effects change electromagnetic behavior. First, we will show how the acceleration of the system $\bvec{a}$ and the tidal tensor $\mathbb{S}$ influence the effective speed of light $\mathcal{C}^{\pm}$. Along the way, we will show how the consideration of light propagation naturally leads to a more generalized notion of ``symmetric reconnection." Following this, we will explore how the gravitational distortion of the system's 3D shape, represented by the tensor $\mathbb{Q}$, affects the spatial distribution of $\bs{V}_d$ in a steady-state. Finally, we will derive a form of Ampere's Law describing the in-plane magnetic field, and explain the inclusion of the intrinsic Gaussian curvature $\mathcal{K}$ in this equation as producing a ``gravitationally induced magnetic tension."

\subsection{The Effective Speed of Light and ``Symmetric Reconnection"} \label{Subsection::LightSpeed}

\begin{figure}
    \centering
    \includegraphics[width=1.0\linewidth]{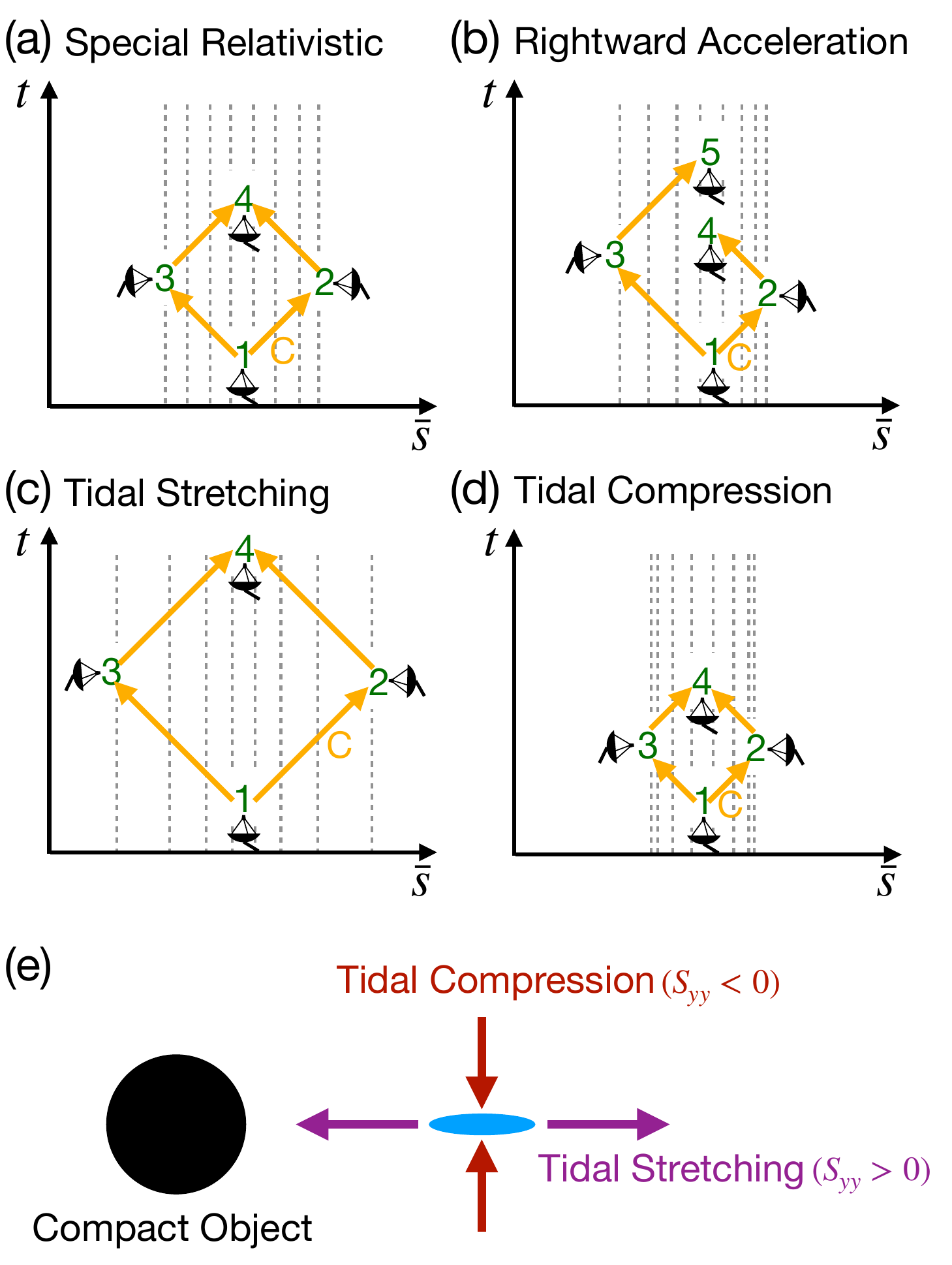}
    \caption{Illustration of the dilation and contraction of time intervals which determine the effective speed of light. In panels (a)-(d), the yellow curves represent the light signals sent and received by the ``messengers" described in Section \ref{Subsection::LightSpeed}, which always have the same slope in coordinates $(t,\bar{s})$. The vertical dashed lines are ``equidistant" points, that is, they are evenly spaced in $s$. The events labeled ``2" and ``3" are therefore the same distance apart across all four panels. In panel (b), the distances \textit{appear} compressed in front of the accelerating system and stretched out behind it, which corresponds to the contraction and dilation of the light travel times (i.e., the time interval $t_4-t_1$ and $t_5-t_1$, respectively). In panels (c)-(d), the tidal dilation and contraction of the time intervals is shown, which for the same tidal tensor $\mathbb{S}$ will depend only on the orientation of the system relative to a gravitational source, shown in panel (e). If the $y$-axis is aligned with the ``radial direction" (purple arrows), $S_{yy}>0$ and the situation is that of panel (c). Conversely, if the $y$-axis is ``angular," the situation is that of panel (d).}
    \label{fig::LSVisual}
\end{figure}

Relative to the coordinate time $t$, light does not travel at the expected speed of unity, but rather has the coordinate speed $ds/dt = \pm \tnorm$. Now the significance of $\bar{s}$ becomes clear: since $d\bar{s} \equiv \tnorm^{-1}ds$, light propagation satisfies $d\bar{s}/dt=\pm1$. Lets now suppose that in our system there are three ``messengers" located at $s=-w_y^{-}$, $s=0$, and $s=w_y^+$, similar to the setup illustrated by Fig.\@ \ref{fig::LSVisual}(a)-(d). At one moment, the messenger at $\mathtt{x}$ sends out one signal towards each of the other messengers. The change in coordinate time $t$ along the trajectory of the signals must be the same as the change in $\bar{s}$, which we have defined as $\Delta \bar{s}^{\pm}$. After this amount of time $t$ elapses after the signal is first sent, the other two messengers receive the message and instantly reply by sending signals of their own towards $\mathtt{x}$. The proper time which elapses for the messenger at $\mathtt{x}$ between the sending of the first messages and receiving the replies is $2\tnorm_{\mathtt{x}}\Delta \bar{s}^{\pm}$. Recall that the rate of reconnection is defined using the proper time interval $w_t$ over which the upstream flux is transported along a curve of length $w_y^+$ on one side and $w_y^-$ on the other. If we measure in the same way the ``rate" at which light propagates along the inflow line, dividing the distance scale $2w_y^{\pm}$ by the proper time interval $2\tnorm_{\mathtt{x}} \Delta \bar{s}^{\pm}$, the result is precisely the parameter $\mathcal{C}^{\pm}$. Using Eq.\@ (\ref{Result::YToYBar}),  it is given explicitly by the formula,
\begin{equation}
\begin{split}
    \mathcal{C}^{\pm} \equiv& \frac{w_y^{\pm}}{\tnorm_{\mathtt{x}} \Delta \bar{s}^{\pm}}  = \left(\frac{\pm\bar{y}(\pm w_y^{\pm})}{w_y^{\pm}} \right)^{-1} \\
    =& 1 \pm \frac{1}{2}\bs{a}_y w_y^{\pm} - \frac{1}{6}\left(S_{yy} - \bs{a}_y^2 \right)(w_y^{\pm})^2 \\
    &+ \mathcal{O}((w_y^{\pm})^3).
\end{split}
\end{equation}
If $\bs{a}_y>0$, $\mathcal{C}^+$ is the effective speed of light ``in front" of the accelerating system and $\mathcal{C}^-$ is the effective speed ``behind" the system. In front of the system, the light crossing time is contracted by the acceleration which causes $\mathcal{C}^+$ to increase. Behind the system, the light crossing time is dilated, causing $\mathcal{C}^-<1$ to decrease. Conversely, the gravitational ``tidal effect" is symmetric between the two sides, but its sign depends on the orientation of the system relative to a gravitational source. If the inflow line is aligned towards/away from a gravitational source, then we expect $S_{yy}>0$ corresponding to ``tidal stretching," which dilates the light crossing times and decreasing $\mathcal{C}^{\pm}$. However, if the inflow line is rotated from this orientation by $\pi/2$, the effect is instead ``tidal compression" with $S_{yy}<0$, which instead contracts the light crossing times and increases $\mathcal{C}^{\pm}$.

One common simplifying assumption in special relativistic studies is to assert that the system is in some way symmetric under reflection about the $x$-axis. Usually, it is assumed that the magnetic field $\bvec{B}_r$ has this symmetry. If we were to express this same principle in a covariant way, we would instead say that an observer at $\mathtt{x}$ should be unable to distinguish between the field-sheet structure one side of the current sheet from the other if their knowledge of the system is limited by the sending and receiving of light signals. Instead of the system of three messengers described above, imagine now that there is a continuum of messengers at all points along the inflow line. The messenger at $\mathtt{x}$ decides to ``ping" a specific field-sheet, and so they send out a signal in both directions along the inflow line with a specific value of $\psi$ encoded. All other messengers evaluate $\psi$ at their location the moment they receive the message, and if the values are equal, they instantly reply with a message sent back towards $\mathtt{x}$. In the general case, the signals sent in reply to the original message will be received by the first messenger at different events, and from this time delay the messenger at $\mathtt{x}$ will be able to distinguish between the field-sheet structure on opposite sides of the current sheet. The meaning of ``symmetric reconnection" that we propose is then that, no matter which value of $\psi$ the first messenger chooses to ``ping," the replies to this ping must always be received at exactly the same event. In a steady state, the assumption of symmetric reconnection is encapsulated by the simple statement $\psi(\bar{s}) = \psi(-\bar{s})$. It directly follows that $\Delta \bar{s}^+=\Delta \bar{s}^-$ and $\bs{\bar{V}}_d^+ = \bs{\bar{V}}_d^-$, and so we can define $\bar{w}_y \equiv \Delta \bar{s}^{\pm}$ and $\bar{\bs{V}}_d \equiv \bar{\bs{V}}_d^{\pm}$.

Note that $\Delta \bar{s}^{\pm}$ only differs from $w_y^{\pm}$ as a result of the NICS effects $\bs{a}_y$ and $S_{yy}$, which scale with the system size $w_y$. Symmetric reconnection therefore approaches the statement $w_y^+=w_y^-=w_y$ in the limit that the system size $w_y$ goes to zero. We will expand $w_y^{\pm}$ around this limit,
\begin{equation}
    w_y^{\pm}=w_y \pm C_1w_y^2 \pm C_2w_y^3 + \mathcal{O}(w_y^4).
\end{equation}
Note that this ansatz is chosen to obey $(w_y^+ + w_y^-)/2 \equiv w_y$. We will then use Eq.\@ (\ref{Result::YToYBar}) to determine the coefficients $C_1$ and $C_2$,
\begin{equation} \label{Result::CoefficientDetermination}
\begin{split}
    \bar{w}_y = \bar{y}(w_y^+) =& w_y + C_1w_y^2 \\
    &+ C_2w_y^3 - \frac{1}{2}\bs{a}_y \left(w_y^2 + 2C_1 w_y^3\right) \\
    &+ \frac{1}{6}\left(S_{yy} - \bs{a}_y^2 \right)w_y^3 + \mathcal{O}(w_y^4),\\
    \bar{w}_y =  -\bar{y}(-w_y^-) =& w_y - C_1w_y^2 \\
    &- C_2w_y^3 + \frac{1}{2}\bs{a}_y \left(w_y^2 - 2C_1w_y^3 \right) \\
    &+ \frac{1}{6}\left(S_{yy} - \bs{a}_y^2 \right)w_y^3 + \mathcal{O}(w_y^4),\\
    \hookrightarrow 0 =& 2C_1 w_y^2 + 2C_2w_y^3\\
    &- \bs{a}_yw_y^2 + \mathcal{O}(w_y^4).
\end{split}
\end{equation}
The coefficients are therefore $C_1=\bs{a}_y/2$ and $C_2=0$, which means that $w_y^{\pm} = w_y\left(1 \pm (\bs{a}_y w_y^2)/2 \right) + \mathcal{O}(w_y^4)$. Substituting back into Eq.\@ (\ref{Result::CoefficientDetermination}), we obtain a formula for the symmetric rate of reconnection,
\begin{equation} \label{Result::SymmetricRate}
\begin{split}
    \mathcal{R} &= \bs{\bar{V}}_d \frac{\mathcal{C}^+ + \mathcal{C}^-}{2} = \bs{\bar{V}}_d \frac{w^+_y + w_y^-}{2\bar{w}_y} = \bar{\bs{V}}_d \left(\frac{\bar{w}_y}{w_y}\right)^{-1} \\
    &= \bar{\bs{V}}_d \left(1 - \frac{1}{6}S_{yy}w_y^2 + \frac{2}{3}\bs{a}^2_y w_y^2 \right) + \mathcal{O}(w_y^3).
\end{split}
\end{equation}

\subsection{The Gravitational ``3D" Effect on the Drift Speed} \label{Subsection::3DEffect}

For a 2D system in a steady-state, there is a certain consequence of Faraday's Law which will be necessary to understanding the effect of the system's 3D shape. We will first illustrate this effect in a simple non-relativistic context.

\begin{figure}
    \centering
    \includegraphics[width=0.6\linewidth]{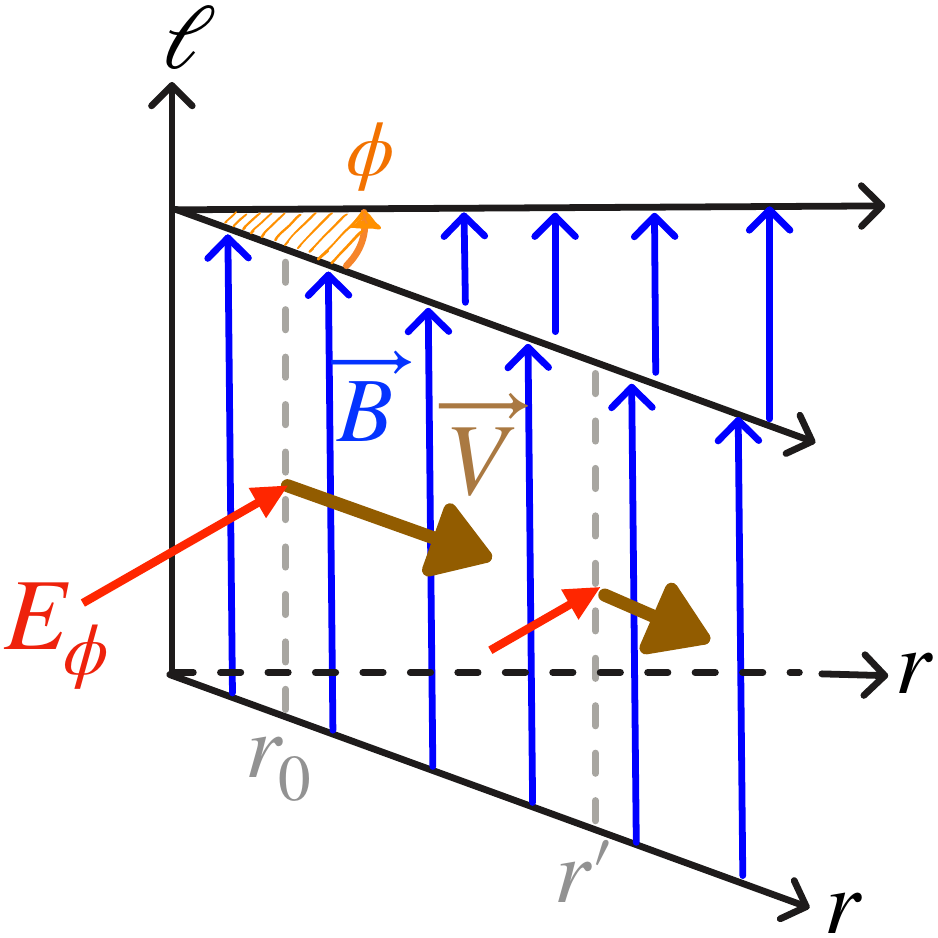}
    \caption{Illustration of the cylindrically-symmetric system described in Section \ref{Subsection::3DEffect}. As the distance between neighboring planes of constant $\phi$ increases with $r$, the equilibrium value of the induced electric field decreases corresponding with the constant EMF $E_{\phi}2\pi r$, thereby decreasing the drift speed in the case of a uniform magnetic flux density along $\vec{l}$.}
    \label{fig::3DEffect}
\end{figure}

As illustrated by Fig.\@ \ref{fig::3DEffect}, consider a cylindrically-symmetric system, such that $\vec{\zeta}=\vec{\partial_{\phi}}$ where $\phi$ belongs to the cylindrical coordinate system along with the radical coordinate $r$ and the vertical coordinate $l$. Within the region $r < r_0$, we will say that there is initially a large amount of magnetic flux oriented entirely along $l$. As the system evolves, we then suppose that this flux is transported by an ideal plasma out of the cylinder $r=r_0$ it was initially confined within at a constant rate. By Faraday's Law, the rate of change of magnetic flux through the disk $r < r_0$ must be a constant value equal to the ``electromotive force" (EMF) $E_{\phi} 2\pi r_0$, where $E_{\phi}$ is the electric field induced by the ideal flow. We now assume that for $r>r_0$, the plasma flow as a whole eventually reaches a steady state. For this to be the case, the magnetic flux through \textit{any} disk defined by $r<r^{\prime}$ where $r^{\prime}>r_0$ must change at exactly the same rate, and so $E_{\phi}2\pi r_0=E_{\phi} 2\pi r^{\prime}$ for any $r^{\prime}>r_0$. In the steady-state region, then, we have $\partial_r(E_{\phi} r)=0$. The drift speed is then $\bs{V}_d = E_{\phi}/\bs{B}_{l} = (E_{\phi}r)/(r\bs{B}_l)=\text{constant}/(r\bs{B}_l)$. There is then a natural tendency for the plasma to flow quicker at smaller $r$ and slower at larger $r$.

\begin{figure}
    \centering
    \includegraphics[width=1.0\linewidth]{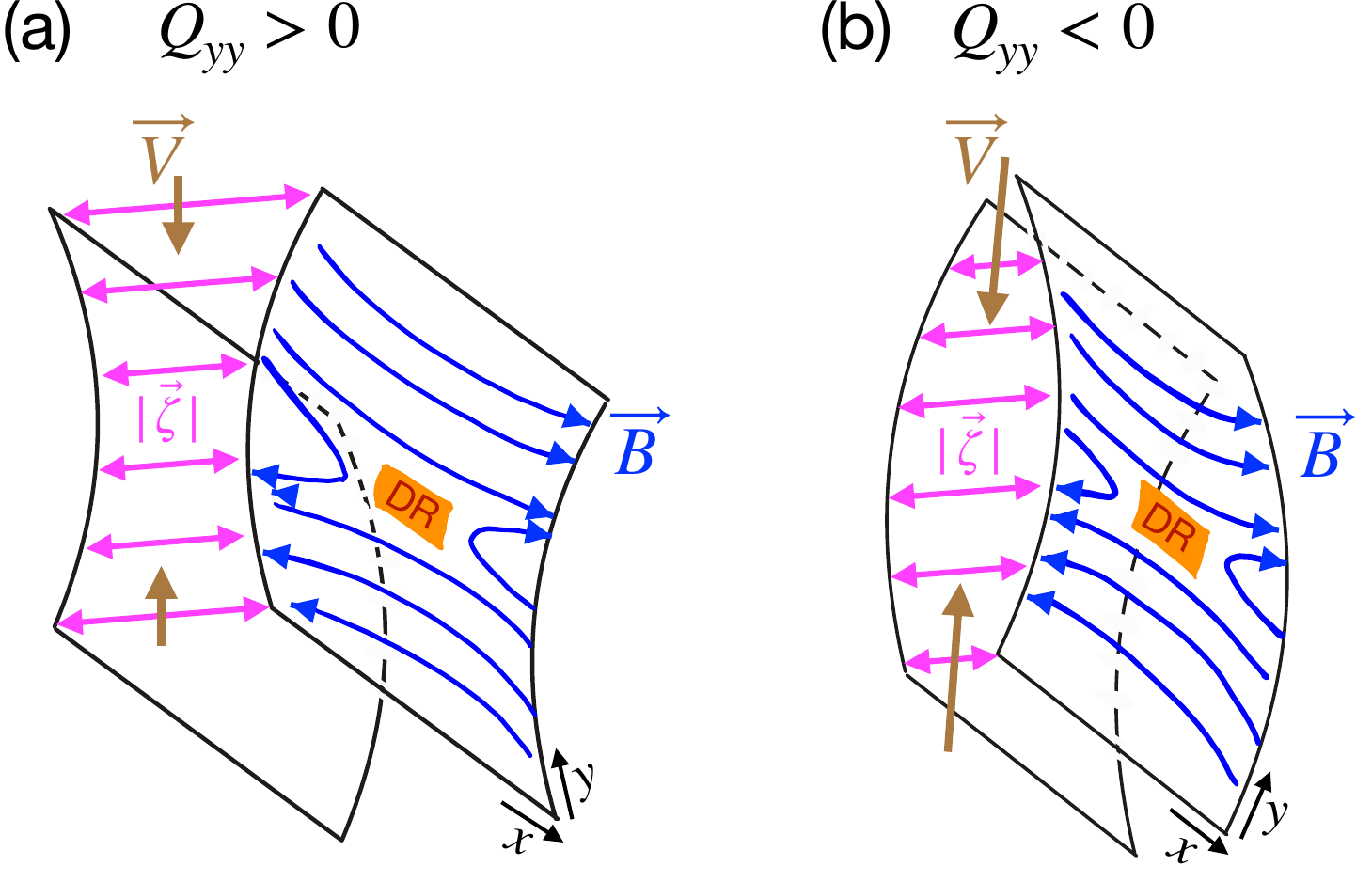}
    \caption{Depiction of the gravitational ``3D" effect. Here, it is assumed that the system is ``slab symmetric," so that $\bvec{h}=0$ and the distance between neighboring planes of reconnection would be uniform if not for the effect of gravity. For $Q_{yy}>0$ in panel (a), the distance between neighboring planes increases in the upstream direction, and it decreases for $Q_{yy}<0$ in panel (b). As a result of the same factors at play in Fig \ref{fig::3DEffect}, the drift speed is then decreased in the former case, and increased in the latter.}
    \label{fig::QEffect}
\end{figure}

In general, this exact same effect exists for any situation in which the separation between neighboring planes of reconnection is not uniform. Within our formulation, this consequence of Faraday's Law follows automatically from the principle that $\partial_t \psi = \znorm \tnorm E_r$ must be a uniform function in a steady state. We then have the same inverse dependence on $\znorm$ within the drift speed $\bs{V}_d = \partial_t \psi/(\znorm |\bvec{H}_r|)$. It is always valid to conceptualize this effect in the same terms: as illustrated by Fig.\@ \ref{fig::QEffect}, if $\znorm$ is greater on one side of the current sheet than the other, that means that the planes of reconnection are spread further apart on that side and more compressed together on the other side, and the drift speed will tend to be non-uniform in order to balance out the EMF. 

Using Eq.\@ (\ref{Result::FermiMetricFormula}) and evaluating $\znorm = \sqrt{g_{zz}}$ along the inflow line, we find how $\bvec{h}$ and $\mathbb{Q}$ will each contribute to this effect,
\begin{equation}
\begin{split}
    \bs{V}_d =& \frac{\partial_t \psi}{|\bvec{H}_r|} \Big(1 + \bs{h}_y y - \frac{1}{2}\left(Q_{yy} + \bs{h}_y^2 \right)y^2 \Big) \\
    &+ \mathcal{O}(y^3).
\end{split}
\end{equation}
In the cylindrically symmetric case described above, the extrinsic curvature of an X-line located at $r=r_{\mathtt{x}}$ is $\bvec{h}=-\hat{r}/r_{\mathtt{x}}$. In the absence of gravity, the effect vanishes locally when the X-line is geodesic ($\bvec{h}=0$), which we will identify with ``slab symmetric" reconnection. $Q_{yy}>0$ implies that the curvature of spacetime is ``splaying apart" neighboring planes of reconnection, causing the drift speed upstream to decrease, as illustrated by Fig.\@ \ref{fig::QEffect}(a). Conversely, $Q_{yy}<0$ implies that neighboring planes of reconnection are being compressed together upstream, causing $\bs{V}_d$ to increase upstream, as illustrated by Fig.\@ \ref{fig::QEffect}(b).

\subsection{Gravitationally Induced Magnetic Tension} \label{Subsection::MagneticTension}

We will now derive the form of Ampere's Law which describes the in-plane magnetic field. This law derives from the ``$z$-component" of the Gauss-Ampere Law,
\begin{equation}
\begin{split}
    J^z &= \partial_i F^{zi} + \Gamma^{\mu}_{\mu \nu} F^{z \nu} + \Gamma^z_{\mu \nu} F^{\mu \nu}. \\
\end{split}
\end{equation}
Since $\Gamma^{z}_{\mu \nu} = \Gamma^z_{\nu \mu}$ while $F^{\mu \nu} = -F^{\nu \mu}$, the final term must be zero. Using Eq.\@ (\ref{Result::ChristoffelSymbols}), $J^z$ becomes,
\begin{equation}
\begin{split}
    J^z = \partial_i F^{zi} +& \Big(\bvec{a} - \mathbb{S} \cdot \bvec{x} - \bvec{h}\\
    &+ \mathbb{Q} \cdot \bvec{x} - \frac{1}{3}\mathcal{K} \bvec{x}\Big)_i F^{zi}.
\end{split}
\end{equation}
We will define the vector $\bvec{\Psi}$ by the relation $\oneform{\bs{\Psi}} \equiv \text{Rot}(\bvec{H}_r)$, where $\bvec{H}_r\equiv \tnorm \bvec{B}_r$ is the ``auxiliary" field. $\bvec{H}_r$ is analogous to the auxiliary field of ``macroscopic" electromagnetism, as discussed in Appendix \ref{Appendix::MaxwellsEquations}. Using Eq.\@ (\ref{Result::CoordinatedPsi}), $\bvec{\Psi}$ has the following relationship with $F^{zi}$,
\begin{equation}
\begin{split}
    \tnorm \znorm F^{zi} &=-\tnorm \znorm g^{zz}g^{ij} \left( \oneform{d}\psi \right)_j \\
    =& -\frac{\tnorm}{\znorm} g^{ij} \left(\znorm \text{Rot}(\bvec{B}_r) \right)_j \\
    =& -\bs{\Psi}^i
\end{split}
\end{equation}
The ``spatial current density" associated with $J^z$, corresponding to the vector $\vec{j}$ defined in Appendix \ref{Appendix::MaxwellsEquations}, is $j^M \equiv \tnorm \hat{M} \cdot \vec{J} = \tnorm \znorm J^z$. In terms of $j^M$, Ampere's Law becomes,
\begin{equation}
\begin{split}
    j^M =& -\tnorm \znorm \partial_i \left( \frac{1}{\tnorm \znorm} \bs{\Psi}^i\right) \\
    &- \left(\bvec{a} - \mathbb{S} \cdot \bvec{x} - \bvec{h} + \mathbb{Q} \cdot \bvec{x} - \frac{1}{3}\mathcal{K} \bvec{x}\right) \cdot \bvec{\Psi} \\
    =& -\partial_i \bs{\Psi}^i + \bs{\Psi}^i \partial_i \ln(\tnorm \znorm) \\
    &- \left(\bvec{a} - \mathbb{S} \cdot \bvec{x} - \bvec{h} + \mathbb{Q} \cdot \bvec{x} - \frac{1}{3}\mathcal{K} \bvec{x}\right) \cdot \bvec{\Psi} \\
    =& -\partial_i \bs{\Psi}^i  + \frac{1}{3} \mathcal{K} \bvec{x} \cdot \bvec{\Psi}.
\end{split}
\end{equation}
It is most desirable to formulate this equation explicitly in terms of the ``covariant" components $\bs{\Psi}_i$, which are directly proportional to $\partial_i \psi$, and are therefore easier to relate to the field-line geometry. Truncated to order $\mathcal{O}(|\bvec{x}|)$, $\partial_i \bs{\Psi}^i = \bs{g}^{ij}\partial_i \bs{\Psi}_j + \bs{\Psi}_j \partial_i \bs{g}^{ij} = \delta^{ij} \partial_i \bs{\Psi}_j - \frac{1}{3} \mathcal{K} \bvec{x} \cdot \bvec{\Psi}$, and we obtain our final expression for Ampere's Law,
\begin{equation} \label{Result::AmperesLaw}
    \delta^{ij} \partial_i \bs{\Psi}_j - \frac{2}{3} \mathcal{K} \bvec{x} \cdot \bvec{\Psi} = -j^M.
\end{equation}

\begin{figure}
    \centering
    \includegraphics[width=0.8\linewidth]{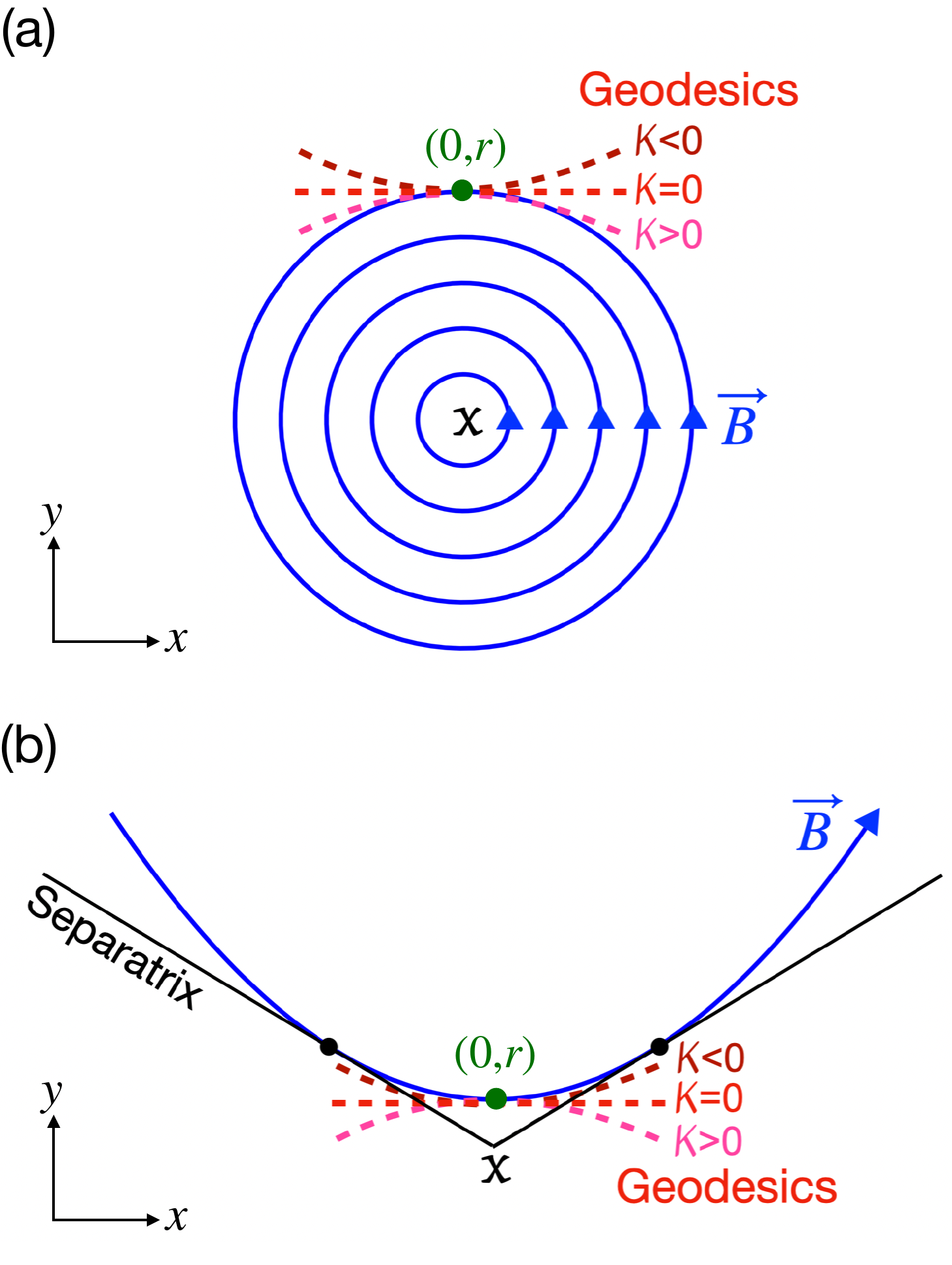}
    \caption{Illustrations of the gravitationally induced magnetic tension effect. In both panels, the extrinsic curvature of a hypothetical field-line geometry is visualized by direct comparison with geodesic curves, which are defined as curves having zero extrinsic curvature, depicted by the red dashed curves. The red geodesics curve inwards for $\mathcal{K}>0$, outward for $\mathcal{K}<0$, and are simply Euclidean straight lines for $\mathcal{K}=0$. Note that the direction in which the geodesics bend can easily verified using Eq.\@ (\ref{Result::MinimumCurvature}): the parabola which is tangent to $y_r(x)$ at the point $(0,r)$ and has zero extrinsic curvature at $(0,r)$ is $y=r(1 - \mathcal{K}x^2/3)$. Panel (a) visualizes the circular field-line case, while panel (b) visualizes the reconnecting field-lines of least curvature given by Eq.\@ (\ref{Statement::MinimalParabolas}). In general, the ``effective'' magnetic tension becomes larger when the magnetic field line deviates more from the geodesic.}
    \label{fig::TensionIllustration}
\end{figure}

Earlier it was implied that we should interpret the effect of $\mathcal{K}$ in this equation as producing a ``gravitationally induced magnetic tension." This may at first appear to be a problematic interpretation. Magnetic tension refers to the force which results from the \textit{extrinsic} curvature of field-lines, and throughout this paper, extrinsic and intrinsic curvature have been discussed as distinct considerations. However, it is possible for the global shape of the magnetic field-lines to be constrained such that the intrinsic Gaussian curvature of the plane of reconnection becomes a physically distinct factor in determining their extrinsic curvature. To illustrate this effect, we have provided in Appendix \ref{Appendix::TheWire} a solution to Eq.\@ (\ref{Result::AmperesLaw}) in the case of an infinitely thin wire along $\hat{M}$ carrying a current $I$, in the case where $\bvec{h}$, $\bvec{a}$, $\mathbb{S}$, and $\mathbb{Q}$ are each zero, but $\mathcal{K}$ is non-zero. The field-line geometry is found to be circular, that is, the field-lines are curves of constant $r \equiv |\bvec{x}|$, which we will note is sufficiently well approximated by $\sqrt{x^2 + y^2}$ since $\bvec{x} \cdot \bvec{x} =r^2= x^2 + y^2 + \mathcal{O}(|\bvec{x}|^5)$). The magnetic flux density is then purely a function of $r$, 
\begin{equation} \label{Result::WireDerivedSolution}
    |\bvec{B}| = \frac{I}{2\pi r}\left(1 + \frac{1}{3}\mathcal{K}r^2 + \mathcal{O}(r^3) \right).
\end{equation}
The correction involving $\mathcal{K}$ is the direct result of the term $-(2/3)\mathcal{K} \bvec{x} \cdot \bvec{\Psi}$ in Eq.\@ (\ref{Result::AmperesLaw}). We will now show that this same result could have been obtained had we known only that the field-line geometry must be circular, and postulated that the solution will satisfy the familiar magneto-static balance between magnetic pressure and tension,
\begin{equation} \label{Statement::TensionHypothesis}
    \vec{\nabla} \frac{|\bvec{B}|^2}{2} = |\bvec{B}|^2 \nabla_{\hat{\bs{b}}} \hat{\bs{b}}, \qquad \Rightarrow \vec{\nabla} \ln|\bvec{B}| = \nabla_{\hat{\bs{b}}} \hat{\bs{b}},
\end{equation}
where $\hat{\bs{b}} \equiv \bvec{B}/|\bvec{B}|$, and $\nabla_{\hat{\bs{b}}} \hat{\bs{b}}$ is the extrinsic field-line curvature. We will represent a given field-line by the column vector $(x,y)^T$, so that $\hat{\bs{b}} = (\dot{x}, \dot{y})^T$ where the dot denotes the derivative with respect to arc length. The extrinsic curvature of the field-lines is then given by the following equation,
\begin{equation} \label{Statement::InPlaneCurvatureRelation}
\begin{split}
        (\nabla_{\hat{\bs{b}}} \hat{\bs{b}})^i \dot{=}& \begin{pmatrix}
        \ddot{x} \\ \ddot{y}
    \end{pmatrix} + \Gamma^i_{jk} \bs{b}^j \bs{b}^k \\
    =& \begin{pmatrix}
        \ddot{x} \\ \ddot{y}
    \end{pmatrix} + \frac{2}{3}\mathcal{K}  \begin{pmatrix}
        \dot{y}^2 & -\dot{x} \dot{y} \\
        -\dot{x} \dot{y} & \dot{x}^2
    \end{pmatrix} \begin{pmatrix}
        x \\ y
    \end{pmatrix}.
\end{split}
\end{equation}
Since the field-lines are circular, $(\ddot{x}, \ddot{y})^T$ must yield the Euclidean result $-\hat{r}/r$ where $\hat{r}\equiv \bvec{x}/r$. If we evaluate the other term at a point $(0, r)$, at which we can also identify $\hat{r}\approx (0,1)^T$ and $\dot{y}=0$, the extrinsic curvature is,
\begin{equation} \label{Result::CircleCurvature}
    \nabla_{\hat{\bs{b}}} \hat{\bs{b}} = \left(-\frac{1}{r} + \frac{2}{3} \mathcal{K} r + \mathcal{O}(r^2)\right) \hat{r}.
\end{equation}
$\mathcal{K}>0$ therefore contributes an outward magnetic tension to the circular field-lines, whereas $\mathcal{K}<0$ contributes an inward tension. This is visualized in Fig \ref{fig::TensionIllustration}(a). Plugging back into Eq.\@ (\ref{Statement::TensionHypothesis}), we recover the original result Eq.\@ (\ref{Result::WireDerivedSolution}) derived in Appendix \ref{Appendix::TheWire},
\begin{equation}
\begin{split}
    &\partial_r \ln(|\bvec{B}|) = -\frac{1}{r} + \frac{2}{3}\mathcal{K}r, \\
    &\hookrightarrow \ln(|\bvec{B}|) = \ln\left(\frac{1}{r}\left(1 + \frac{1}{3}\mathcal{K}r^2\right)\right) + \text{constant}, \\
    &\hookrightarrow |\bvec{B}| \propto \frac{1}{r} + \frac{1}{3}\mathcal{K} r^.
\end{split}
\end{equation}

In general, there is no \textit{necessary} relationship between extrinsic and intrinsic curvature, and interpreting this term as an additional source of tension is a contingent fact which relies on there being some consistency in the field-line shape relative to the coordinates $(x,y)$. The situation of a current-carrying wire is one such case, where $(\ddot{x}, \ddot{y})^T$ is determined entirely by the constraint $x^2 + y^2= \text{constant}$, isolating the term $\Gamma^i_{jk} \bs{b}^j \bs{b}^k$ as a physically distinct contribution to the curvature. In the case of magnetic reconnection, it is reasonable to place a constraint on the field-line geometry which has the same effect. 

It is typical that the separatrix lines which characterize the field-line geometry of reconnection should tend towards ``straight lines" as a result of the overall tendency of the system to minimize magnetic tension wherever possible. This tendency is extended into curved space by the assumption that the separatrix curves approximate geodesics within the plane. One of the defining characteristics of the Riemann normal coordinate system, which can be directly verified by setting Eq.\@ (\ref{Statement::InPlaneCurvatureRelation}) to zero, is that all geodesics which pass through the origin approximately satisfy $y = \mu x$ for some constant $\mu$. We can derive a lower bound on the curvature of field-lines upstream of the diffusion region by noting that the field-line intersecting the point $(0,r)^T$ must have at least the curvature of the parabola $y=y_r(x)$, where $y_r(x)$ is defined as the unique parabola which intersects the point $(0,r)^T$ and has only two points of intersection with the separatrix curve $y=\mu |x|$. It is straightforward to derive the formula for these parabolas,
\begin{equation} \label{Statement::MinimalParabolas}
    y_r(x) = r + \frac{1}{4} \frac{\mu^2}{r}x^2.
\end{equation}
We once again have a family of curves which have a consistent form in our coordinate system independently of the intrinsic curvature of the space. We now evaluate their extrinsic curvature along $x=0$ using Eq.\@ (\ref{Statement::InPlaneCurvatureRelation}), which determines a lower bound for the field-line curvature along the inflow line,
\begin{equation} \label{Result::MinimumCurvature}
\begin{split}
    \min \left[(\nabla_{\hat{\bs{b}}} \hat{\bs{b}})^y\Big|_{(0,r)} \right] &= \ddot{y}\Big|_{(0,r)}  + \frac{2}{3}\mathcal{K}r \dot{x}^2\Big|_{(0,r)} \\
    &= \dot{x}^2 \left( y_r^{\prime \prime}(0) + \frac{2}{3}\mathcal{K}r \right) \\
    &= \left(g_{xx} \Big|_{(0,r)} \right)^{-1} \left( \frac{\mu^2}{2r} + \frac{2}{3}\mathcal{K} r \right) \\
    &= \frac{\mu^2}{2r} + \frac{\mu^2 + 4}{6}\mathcal{K}r ,
\end{split}
\end{equation}
where the prime denotes derivative with respect to $x$. The least possible field-line curvature necessitated by the separatrix slope $\mu$ depends explicitly on the intrinsic Gaussian curvature $\mathcal{K}$, where $\mathcal{K}>0$ again contributes an additional outward magnetic tension and $\mathcal{K}<0$ an inward tension. This is illustrated in Fig \ref{fig::TensionIllustration}b. In general, outward magnetic tension in this region is an inhibiting effect on reconnection which tends to decrease the rate, since under the conditions of force-free equilibrium, it must be balanced with a gradient in the magnetic field strength which lowers the density of magnetic flux that enters the diffusion region. We will find, therefore, that $\mathcal{K}>0$ will enhance this effect, whereas $\mathcal{K}<0$ will inhibit the effect, enabling faster reconnection.

\section{Leading Order Correction to the Reconnection Rate} \label{Section::GeneralizedFormulae}

\begin{figure}
    \centering
    \includegraphics[width=0.9\linewidth]{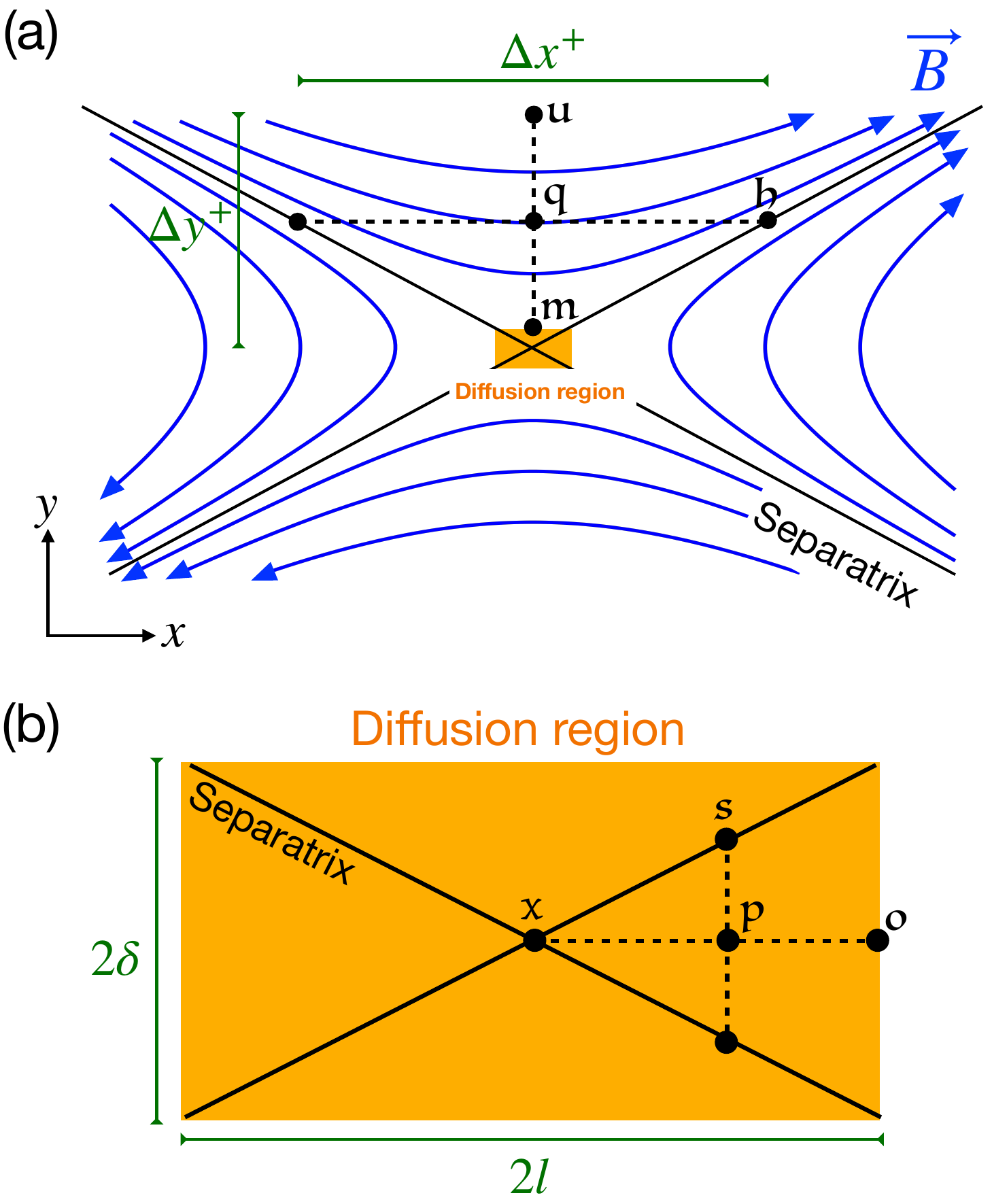}
    \caption{Illustration of the distinction between the diffusion region (b) and the larger mesoscale region (a). Points are labeled on the top half of the mesoscale region and the right half of the diffusion region which are used in Section \ref{Section::GeneralizedFormulae} for finite difference approximations. The point $\mathfrak{m}$ lies at $(0,\delta)$, the point $\mathfrak{u}$ lies at $(0,\Delta y^+)$, and the point $\mathfrak{h}$ lies at $(\Delta x^+/2,\Delta y^+/2)$. The origin point $(0,0)$, which is the X-point, is labeled $\mathtt{x}$, the point $\mathfrak{o}$ lies at $(l,0)$, and the point $\mathfrak{s}$ lies at $(l/2,\delta/2)$. In the middle of the top half of the mesoscale region, and the middle of the right half of the diffusion region, are the points $\mathfrak{q}$ and $\mathfrak{p}$ respectively, which lie at $(0, \Delta y^+/2)$ and $(l/2,0)$.}
    \label{fig::LocalRegion}
\end{figure}

Using the foundation presented in Section \ref{Section::Theory}, we will now derive formulae for the rate of reconnection following almost exactly the same approach taken in Ref.\@ \citep{Liu2017}, correcting for the electromagnetic NICS effects described in Section \ref{Section::Overview} to leading order. Firstly, this means drawing distinction between the microscopic (kinetic) scale diffusion region and the mesoscale region according to the simplified picture, as illustrated in Fig.\@ \ref{fig::LocalRegion}, in which the diffusion region is a rectangular region within the plane of reconnection having width $2\delta$ in the $y$-direction and length $2l$ in the $x$-direction, whereas the mesoscale region is a much larger rectangle having width $2\Delta y$ and width $2\Delta x$. In contrast with the original theory, we will not assume that this greater rectangle is perfectly centered about the X-point, following the discussion in Section \ref{Subsection::LightSpeed} on the meaning of ``symmetric" reconnection. Instead, there will be a difference between the size of the mesoscale region between the two sides of the current sheet, denoted by $\Delta y^{\pm}$ and $\Delta x^{\pm}$, such that $\Delta y = (\Delta y^+ + \Delta y^-)/2$ and $\Delta x = (\Delta x^+ + \Delta x^-)/2$. At the same time, following the discussion in Section \ref{Subsection::MagneticTension}, we will assume that the separatrix lines are geodesics through the origin which have the form $y \propto x$. As in Ref.\@ \citep{Liu2017}, we will assume a correspondence between the slope of these lines inside and outside of the diffusion region, so that $\Delta y^+/\Delta x^+ = \Delta y^-/\Delta x^- = \delta / l$. This correspondence, together with the assumption of symmetric reconnection where $\bar{y}(\Delta y^+) = -\bar{y}(-\Delta y^-)$, means that the geometry of the system is fully determined by the two parameters: $\mu \equiv \delta /l$, which we will call the ``localization parameter," and the distance scale $\Delta y$. 

In Section \ref{Section::Theory}, the scalar rate of reconnection $\mathcal{R}$ is defined in relation to a specific range of $\psi$, where the distance scales $w_y^{\pm}$ and $w_y \equiv (w_y^+ + w_y^-)/2$ are determined by the value $\Delta \psi$. One could use this foundation to calculate a variety of different rates for the same physical system, that is, the same overall field-sheet structure. For example, we could choose $\Delta \psi$ such that $w_y=\delta$, and then the rate would be $\delta/\Delta t$ where $\Delta t$ is the time scale over which the flux contained in the diffusion region is reconnected. However, this rate would be determined entirely by physics within the diffusion region and, as stated in the introduction, we will be taking it as a given that this would not require any general relativistic considerations, which is encapsulated by the approximations,
\begin{equation} \label{Statement::SpecialRelativisticDiffusionRegion}
    \vec{\boldsymbol{a}} l \approx 0, \qquad \mathbb{R} l^2 \approx 0.
\end{equation}
Instead, we will derive the rate which most closely corresponds to the ``local rate" in Ref.\@ \citep{Liu2017}, which here corresponds to the choice $w_y=\Delta y$. That is, we aim to derive the ratio $\Delta y/\Delta t$, where $\Delta t$ is now the time scale over which the flux contained in the mesoscale region reconnects. We then have a correspondence between the remaining distance scales $w_y^{\pm} = \Delta y^{\pm}$ as well as $\bar{w}_y = \Delta \bar{y} \equiv \bar{y}(\Delta y^{+})$. We seek only the leading order correction for NICS effects. For this purpose, it is convenient to define the dimensionless parameters which, for a leading order correction to be valid, we must assume to be small,
\begin{equation}
\begin{alignedat}{2}
    \bs{\bar{a}} &\equiv \Delta y \bvec{a} \cdot \hat{\bs{N}}, \quad \bar{S} &\equiv \Delta y^2 \hat{\bs{N}} \cdot \mathbb{S} \cdot \hat{\bs{N}}, \\
    \bar{\mathcal{K}} &\equiv \Delta y^2 \mathcal{K}, \qquad \bar{Q} &\equiv \Delta y^2 \hat{\bs{N}} \cdot \mathbb{Q} \cdot \hat{\bs{N}}.
\end{alignedat}
\end{equation}
The original theory assumed a ``slab symmetric" system such that the extrinsic X-line curvature $\bvec{h}$ is zero. Since we only aim to correct the original theory for NICS effects, we will also assume that $\bvec{h}=0$ in this section. 

Recall that $\partial_t \psi$ is a uniform function in a steady-state, and so its value may be determined by evaluating it at any point. We choose the point labeled $\mathfrak{o}$ in Fig.\@ \ref{fig::LocalRegion}(b), where the frozen-in condition $\nabla_{\vec{U}} \psi=0$ must be satisfied,
\begin{equation}
\begin{split}
    &\partial_t \psi + \bvec{V} \cdot \bdel \psi\Big|_{\mathfrak{o}}=0, \\
    &\hookrightarrow \partial_t \psi = \bs{V}^L_{\mathfrak{o}}  \bs{B}^N_{\mathfrak{o}} = \mu \bs{V}^L_{\mathfrak{o}} \bs{H}^L_{\mathfrak{m}}.
\end{split}
\end{equation}
Here, we have used the fact that $\bvec{H}_r\approx \bvec{B}_r$ in the diffusion region, and that $\bdel \cdot \bvec{B}_r=0$ implies that $\delta \bs{B}^L_{\mathfrak{m}} \approx l\bs{B}^N_{\mathfrak{o}}$, with $\mathfrak{m}$ labeled in Fig.\@ \ref{fig::LocalRegion}(a). Now we relate the value of $\langle \partial_{\bar{s}} \psi \rangle$ to the auxiliary field $\bvec{H}_r$,
\begin{equation}
\begin{split}
    \langle\partial_{\bar{s}}\psi \rangle  &= \frac{1}{\Delta \bar{y}} \int_0^{\Delta \bar{y}} \partial_{\bar{y}} \psi d\bar{y} \\
    &= \frac{1}{\Delta \bar{y}} \int_0^{\Delta \bar{y}} \tnorm \partial_y \psi d\bar{y} = \langle \znorm \bs{H^L} \rangle \\
    &\approx (\znorm_{\mathfrak{u}} \bs{H}^L_{\mathfrak{u}} + \bs{H}^L_{\mathfrak{m}})/2 \\
    &= \frac{1}{2}\bs{H}^L_{\mathfrak{u}} \left(1 + \bs{H}^L_{\mathfrak{m}} / \bs{H}^L_{\mathfrak{u}} + \bar{Q}/2 \right),
\end{split}
\end{equation}
where the point $\mathfrak{u}$ is also labeled in Fig \ref{fig::LocalRegion}a. We then have a formula for the harmonically averaged drift speed $\bar{\bs{V}}_d$,
\begin{equation} \label{Result::DriftSpeedFormula}
    \bar{\bs{V}}_d = \frac{\partial_t \psi}{\langle\partial_{\bar{s}}\psi \rangle} = \mu \frac{2\bs{H}^L_{\mathfrak{m}}/\bs{H}^L_{\mathfrak{u}}}{1 + \bs{H}^L_{\mathfrak{m}}/\bs{H}^L_{\mathfrak{u}} + \bar{Q}/2} \bs{V}^L_{\mathfrak{o}}.
\end{equation}

Following Ref.\@ \citep{Liu2017}, we will neglect entirely the tensor $\mathbb{F}_g$ within the mesoscale region. This amounts to two separate assumptions, the first of which being that the reconnection is purely ``anti-parallel" with $B_g \approx 0$. If the plasma flow velocity is also purely tangent to the plane of reconnection, then the frozen-in condition requires that $\oneform{\bs{E}_g} = \left(\bvec{V} \cdot \hat{M} \right) \text{Rot}(\bvec{B}_r)=0$. Therefore, under the assumption that $\bvec{V}$ and $\bvec{B}$ are purely within the plane, $\twoform{F_g}=0$ and $\twoform{F}=\twoform{F_r}=(1/\tnorm)\oneform{\bs{\Psi}} \wedge \oneform{M} +E_r \oneform{M} \wedge \oneform{T}$, where $\oneform{\bs{\Psi}} \equiv \text{Rot}(\bvec{H}_r)$, as defined in Section \ref{Subsection::MagneticTension}. The original theory is based on the assumption that the plasma in the mesoscale region is in a force-free equilibrium, satisfying $\twoform{F} \cdot \vec{J}=0$. Given $\twoform{F_g}=0$, the force-free condition requires that the current density along $\hat{M}$ is zero,
\begin{equation}
    \bvec{\Psi} \cdot \twoform{F} \cdot \vec{J} = \frac{1}{\tnorm}\left(|\bvec{\Psi}|^2 \oneform{M}\right) \cdot \vec{J} = \frac{|\bvec{\Psi}|^2}{\tnorm^2} j^M = 0.
\end{equation}
We will therefore impose the homogeneous form of the Ampere's Law Eq.\@ (\ref{Result::AmperesLaw}) to approximate the ratio $\bs{H}^L_{\mathfrak{m}}/\bs{H}^L_{\mathfrak{u}}$. We will apply the same method as in Ref.\@ \citep{Liu2017}, which is to approximate the derivatives of $\bvec{\Psi}$ at the point $\mathfrak{q}$ using its value at the points $\mathfrak{m}$, $\mathfrak{h}$, and $\mathfrak{u}$, shown in Fig.\@ \ref{fig::LocalRegion}(a),
\begin{equation} \label{Result::FDApproxAtQ}
\begin{split}
    \partial_y \bs{\Psi}_y\Big|_{\mathfrak{q}} &\approx \frac{\bs{\Psi}_y \Big|_{\mathfrak{u}}- \bs{\Psi}_y \Big|_{\mathfrak{m}}}{\Delta y^+} = \frac{\bs{H}^L_{\mathfrak{u}} - \bs{H}^L_{\mathfrak{m}}}{\Delta y^+}, \\
    -\partial_x \bs{\Psi}_x \Big|_{\mathfrak{q}} &\approx \frac{-2\bs{\Psi}_{x}}{\Delta x^+} \Big|_{\mathfrak{h}} \approx \frac{2\mu^2\bs{\Psi}_{y}}{\Delta y^+} \Big|_{\mathfrak{h}} \\
    &\approx \frac{2\mu^2}{\Delta y^+} \frac{\bs{\Psi}_y \Big|_{\mathfrak{u}} + \bs{\Psi}_y \Big|_{\mathfrak{m}}}{2} \\
    &= \frac{\mu^2}{\Delta y^+} (\bs{H}^L_{\mathfrak{u}} + \bs{H}_{\mathfrak{m}}^L),
\end{split}
\end{equation}
where we have used five separate approximations and identities: (1) we approximate $\bs{\Psi}_y \big|_{\mathfrak{h}}$ as the average value of $\bs{\Psi}_x$ between the points $\mathfrak{m}$ and $\mathfrak{u}$; (2) $\bs{\Psi}_x \big|_{y=\mu x}$$\approx- \bs{\Psi} \big|_{y=-\mu x}$; (3) $\Delta x^{+} \equiv \Delta y^+/\mu$, as defined in the first paragraph of this section; (4) $\bs{\Psi}_y$$= \bs{H}^L$ along the $y$-axis, as shown by Eq.\@ (\ref{Statement::FInflowDecomposition});  and (5) $\bs{\Psi}_x\big|_{\mathfrak{h}}$$= -\mu \bs{\Psi}_y\big|_{\mathfrak{h}}$. The last statement follows from the fact that, as mentioned in Section \ref{Subsection::MagneticTension}, $\bs{\Psi}_i \propto \partial_i \psi$, and since the separatrix must be a constant-$\psi$ curve which we assume to satisfy $y=\mu x$, we have $d\psi(x,\mu x)$$= 0 $$= (\partial_x \psi + \mu \partial_y \psi)\big|_{y=\mu x} dx$. Therefore, $(\bs{\Psi}_x/\bs{\Psi}_y) \big|_{\mathfrak{h}}$$=(\partial_x \psi/\partial_y \psi)\big|_{\mathfrak{h}}$$=-\mu$.

Substituting the approximations Eq.\@ (\ref{Result::FDApproxAtQ}) into Eq.\@ (\ref{Result::AmperesLaw}),
\begin{equation} \label{Result::FDForceFreeExpansion}
\begin{split}
    \frac{\bs{H}^L_{\mathfrak{u}} - \bs{H}^L_{\mathfrak{m}}}{\Delta y^+} =& \frac{\mu^2}{\Delta y^+} \left(\bs{H}^L_{\mathfrak{u}} + \bs{H}^L_{\mathfrak{m}} \right) \\
    &+ \frac{2}{3} \mathcal{K} \frac{\Delta y^+}{2} \frac{\bs{H}^L_{\mathfrak{u}} + \bs{H}^L_{\mathfrak{m}} }{2}.
\end{split}
\end{equation}
Referring back to the reasoning used in Section \ref{Subsection::LightSpeed}, we note that $(\Delta y^+)^2 = \Delta y^2 + \mathcal{O}(\Delta y^3)$, and therefore $\mathcal{K} (\Delta y^+)^2 = \bar{\mathcal{K}} + \mathcal{O}(\Delta y^3)$. We then solve Eq.\@ (\ref{Result::FDForceFreeExpansion}) for the desired ratio $\bs{H}^L_{\mathfrak{m}}/\bs{H}^L_{\mathfrak{u}}$,
\begin{equation}
    \frac{\bs{H}^L_{\mathfrak{m}}}{\bs{H}^L_{\mathfrak{u}}} \approx \frac{1 - (\mu^2 + \mathcal{\bar{K}}/6)}{1 + (\mu^2 + \mathcal{\bar{K}}/6)}.
\end{equation}
Plugging this into Eq.\@ (\ref{Result::DriftSpeedFormula}), and then substituting the result into the formula for the symmetric rate of reconnection Eq.\@ (\ref{Result::SymmetricRate}), we obtain our final expression for the rate,
\begin{equation}
\begin{split}
    \mathcal{R} =& \mu \left(1 - \mu^2 - \frac{1}{6}\mathcal{\bar{K}} - \frac{1}{4}\bar{Q} \left(1 - \mu^4 \right)\right) \\
    & \times \left(1 - \frac{1}{6}\bar{S} + \frac{2}{3}\bar{\bs{a}}^2 \right) \bs{V}_{\mathfrak{o}}^L .
\end{split}
\end{equation}
However, this formula is not complete on its own without a formula for the outflow speed $\bs{V}^L_{\mathfrak{o}}$. 

Since we are neglecting NICS effects within the diffusion region, the determination of $\bs{V}^L_{\mathfrak{o}}$ will be exactly the same as in Ref.\@ \citep{Liu2017}. It is found by imposing the special relativistic momentum equation at the point $\mathfrak{p}$, shown in Fig.\@ \ref{fig::LocalRegion}(b),
\begin{equation}
    \rho U^L \partial_x U^L - \bs{B}^N \partial_y \bs{B}^L + \partial_x \frac{(\bs{B}^N)^2}{2} = 0,
\end{equation}
where $\rho$ is the proper mass density of the plasma evaluated at the X-point, which we will assume to be approximately uniform throughout the diffusion region. We then make a finite difference approximation of the derivative $\partial_y \bs{B}^L$,
\begin{equation}
\begin{split}
    \partial_y \bs{B}^L \Big|_{\mathfrak{p}} &\approx \frac{2\bs{B}^L_{\mathfrak{s}}}{\delta} = \frac{2l}{\delta^2} \bs{B}^N_{\mathfrak{s}} \approx \frac{2l}{\delta^2} \frac{\bs{B}^N_{\mathtt{x}} + \bs{B}^N_{\mathfrak{o}}}{2} \\
    &= \frac{\mu \bs{B}^L_{\mathfrak{m}}}{\mu^2 l} = \bs{B}^L_{\mathfrak{m}}/(\mu l),
\end{split}
\end{equation}
where we have used the fact that $\bs{B}^N/\bs{B}^L=\mu$ along the separatrix. Mirroring the approximation of $\bs{\Psi}_y \Big|_{\mathfrak{h}}$ made previously, we have also approximated here that $\bs{B}^N_{\mathfrak{s}}$ is the average value of $\bs{B}^N$ between the points $\mathtt{x}$ and $\mathfrak{o}$. The finite difference approximation of the total momentum equation is then,
\begin{equation}
\begin{split}
    \rho \frac{(U^L_{\mathfrak{o}})^2 - (U^L_{\mathtt{x}})^2}{2l} =& \frac{\bs{B}^N_{\mathtt{x}} + \bs{B}^N_{\mathfrak{o}}}{2}\frac{\bs{B}^L_{\mathfrak{m}}}{\mu l} - \\
    &-\frac{(\bs{B}^N_{\mathfrak{o}})^2 - (\bs{B}^N_{\mathtt{x}})^2}{2l}, \\
    \hookrightarrow (U^L_{\mathfrak{o}})^2 =&(\bs{B}^L_{\mathfrak{m}})^2 - \mu^2 (\bs{B}^L_{\mathfrak{m}})^2,
\end{split}
\end{equation}
where we have assumed that the X-point is also the ``stagnation point" where $U^L=0$. We now have a formula for the outflow speed,
\begin{equation} \label{Result::PrimaryOutflowCondition}
    \frac{(\bs{V}^L_{\mathfrak{o}})^2}{1 - (\bs{V}^L_{\mathfrak{o}})^2} = \left(1 - \mu^2 \right) \frac{(\bs{B}^L_{\mathfrak{m}})^2}{\rho}.
\end{equation}
This is the exact same relationship found in Ref.\@ \citep{Liu2017}. Just as in the original theory, however, the magnetization of the plasma in the diffusion region $(\bs{B}^L_{\mathfrak{m}})^2/\rho$ is not an independent variable, but a dependent variable determined by physics in the mesoscale region. The relevant independent variables here are $\Delta \psi$, $\Delta y$, and $\rho$ and so this formula should instead be expressed in terms of the ``quasi-magnetization" $\sigma \equiv (\Delta \psi/\Delta y)^2/\rho$. Using the fact that $\Delta \psi = \Delta \bar{y} \langle \partial_{\bar{s}} \psi \rangle = \Delta y \langle \partial_{\bar{s}} \psi \rangle/\mathcal{C}$, where $\mathcal{C} \equiv (\mathcal{C}^+ + \mathcal{C}^{-})/2=\Delta y/\Delta \bar{y}$, the formula for the outflow speed is now,
\begin{equation} \label{Result::UltimateOutflowCondition}
\begin{split}
    \frac{(\bs{V}^L_{\mathfrak{o}})^2}{1 - (\bs{V}^L_{\mathfrak{o}})^2} &= \left(1 - \mu^2 \right) \left(\frac{2\bs{H}^L_{\mathfrak{m}}/\bs{H}^L_{\mathfrak{u}}}{\znorm_{\mathfrak{u}} + \bs{H}^L_{\mathfrak{m}}/\bs{H}^L_{\mathfrak{u}}} \right)^2 \mathcal{C}^2  \sigma  \\
    &= \left(1 - \mu^2 \right)A^2 \sigma,
\end{split}
\end{equation}
where,
\begin{equation}
\begin{split}
    A =& \Big(1 - \mu^2 - \frac{1}{6}\mathcal{\bar{K}} - \frac{1}{4}\bar{Q} \left(1 - \mu^4 \right)\Big)  \\
    &\times\Big(1 - \frac{1}{6}\bar{S} + \frac{2}{3}\bar{\bs{a}}^2 \Big)
\end{split}
\end{equation}
Because we are assuming that the diffusion region is small and therefore $\bar{\bs{V}}_d$ is approximately the harmonic average of the plasma flow speed itself, $\bar{\bs{V}}_d \leq 1$. Since $\mathcal{R}=\bs{\bar{V}}_d \mathcal{C}$ (see Eq.\@ (\ref{Result::SymmetricRate})), the average effective speed of light $\mathcal{C}$ should be interpreted as the maximum possible rate of reconnection. 

We call $\sigma$ the ``quasi-magnetization" since it does not have any immediate relation to the actual magnetization of the plasma as measured by any particular observer in their local reference frame. However, we will find that this parameter will serve to generalize certain expectations for how the rate of reconnection should compare with the Alfv\'en speed. In particular, we will find that $\mathcal{R} \sim 0.1V_A$ where $V_A \equiv \sqrt{\sigma/(1+\sigma)}$ is the ``quasi-Alfv\'en speed." 

\begin{figure}
    \centering
    \includegraphics[width=1.0\linewidth]{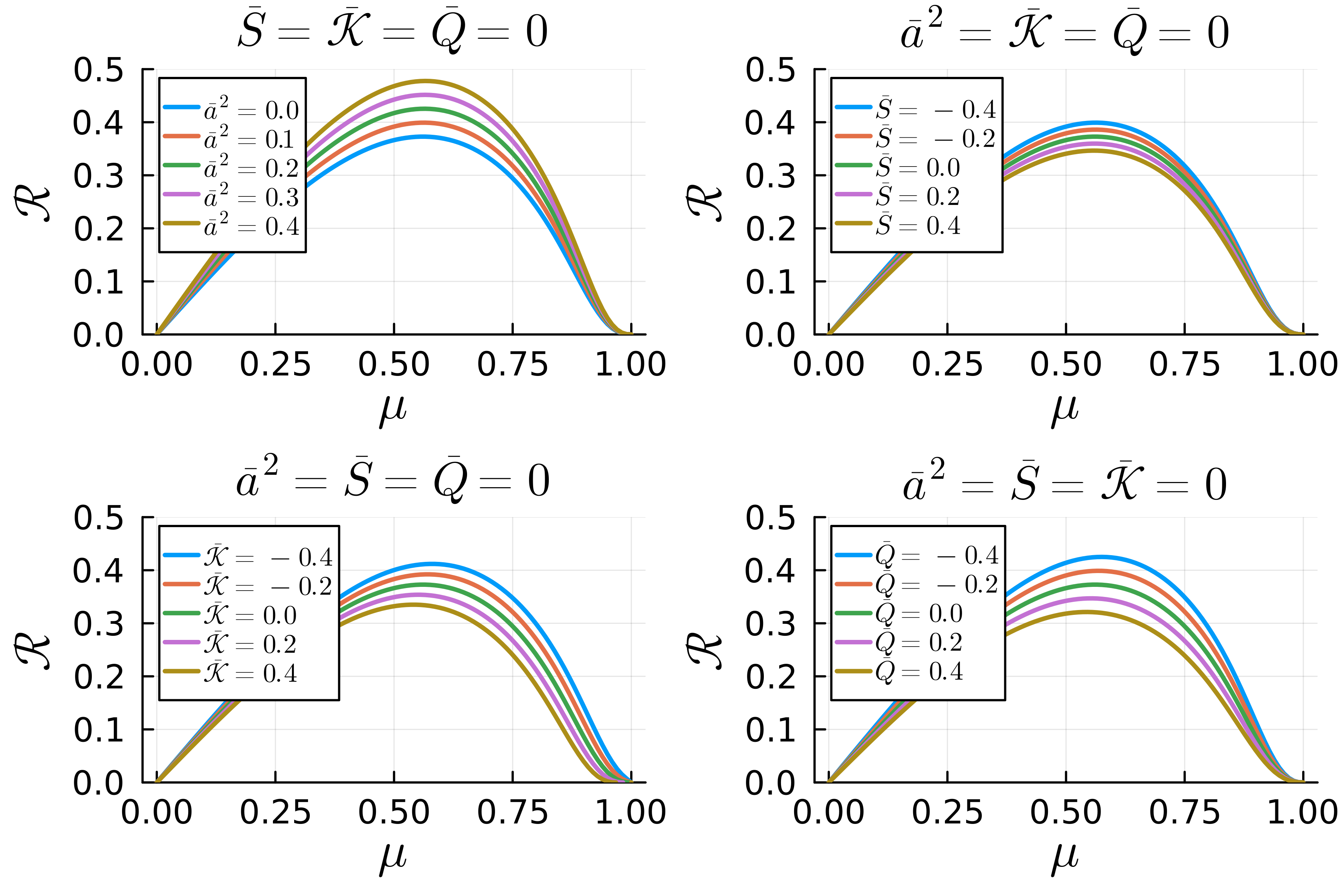}
    \caption{Top panels: the parameters $\bs{\bar{a}}^2$ and $\bar{S}$ are isolated respectively, showing the time dilation and contraction effect on the rate as discussed in Section \ref{Subsection::LightSpeed}. Bottom left: $\bar{\mathcal{K}}$ is isolated, showing the effect of gravitational induced magnetic tension as discussed in \ref{Subsection::MagneticTension}. Bottom right: $\bar{Q}$ is isolated, showing the effect of the gravitational distortion of the system's ``3D" shape as discussed in Section \ref{Subsection::3DEffect}. In all panels, the rate is calculated for a ``high magnetization" case where $V_A = 0.99$.}
    \label{fig::RateCurves}
\end{figure}

We now have a complete set of equations which determine the rate of reconnection:
\begin{equation} \label{Result::MostGeneralRate}
\begin{split}
    \mathcal{R} =& \mu A \bs{V}^L_{\mathfrak{o}} \\
    \frac{(\bs{V}^L_{\mathfrak{o}})^2}{1 - (\bs{V}^L_{\mathfrak{o}})^2} =& \left(1 - \mu^2 \right)A^2 \sigma \\
    A \equiv& \Big(1 - \mu^2 - \frac{1}{6}\mathcal{\bar{K}} - \frac{1}{4}\bar{Q} \left(1 - \mu^4 \right)\Big) \\
    &\times\Big(1 - \frac{1}{6}\bar{S} + \frac{2}{3}\bar{\bs{a}}^2 \Big)
\end{split}
\end{equation}
Fig.\@ \ref{fig::RateCurves} shows quantitative calculations which aim to isolate the effects of each NICS-related parameter, which are fully consistent with the discussions on these effects given in Section \ref{Section::Overview}: (1) $\bs{\bar{a}}^2$ and $\bar{S}$ affect the rate due to their effect on the effective speed of light $\mathcal{C}\equiv (\mathcal{C}^+ + \mathcal{C}^-)/2$, which appears firstly in the formula for symmetric reconnection $\mathcal{R}=\bs{\bar{V}}_d \mathcal{C}$ (Eq.\@ (\ref{Result::SymmetricRate})) and also in the outflow condition Eq.\@ (\ref{Result::UltimateOutflowCondition}); (2) $\bar{Q}$ affects the rate by creating a non-uniform distance between planes of reconnection ($\znorm dz$), which is directly related to $\bs{\bar{V}}_d$ via Eq.\@ (\ref{Result::DriftSpeedFormula}), and also determines the value of $(\bs{B}_{\mathfrak{m}}^L)$ given fixed $\sigma$, as shown in Eq.\@ (\ref{Result::UltimateOutflowCondition}); (3) $\mathcal{K}$ contributes a ``gravitationally induced magnetic tension" which increases or decreases the gradient in $\bs{H}^L$ necessitated by force-free equilibrium, increasing or decreasing the ratio $\bs{H}^L_{\mathfrak{m}}/\bs{H}^L_{\mathfrak{u}}$, which appears in both equations in Eq.\@ (\ref{Result::MostGeneralRate}). 

\begin{figure}
    \centering
    \includegraphics[width=0.9\linewidth]{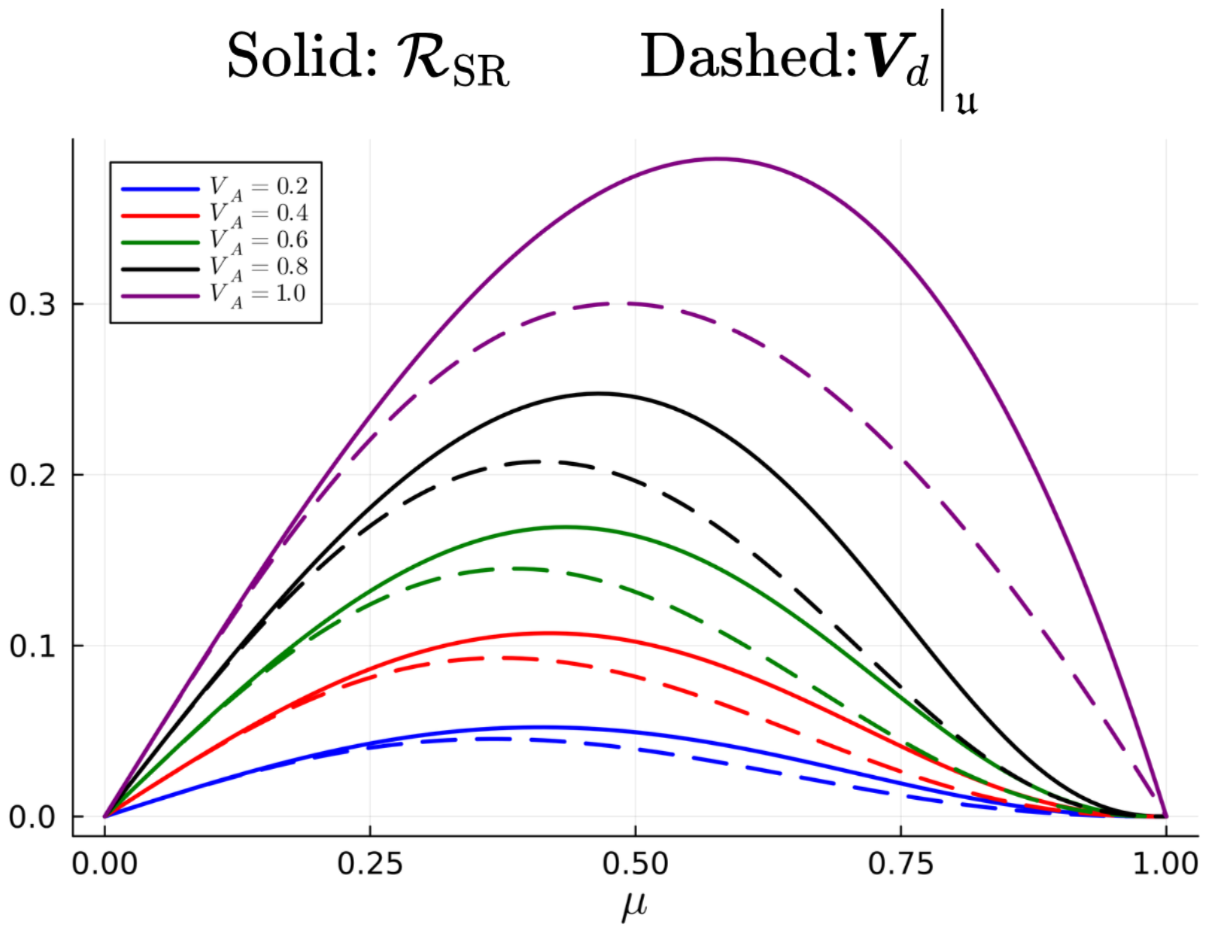}
    \caption{The solid curves represent the value of $\mathcal{R}_{\text{SR}}$ for different values of $V_A$, while the dashed curves represent the value of $\bs{V}_d \Big|_{\mathfrak{u}}$ in the ``special relativistic" limit," which is the same thing as $\mathcal{R}_{\text{SR}}/(1 + \mu^2)$. The difference between the solid and dashed curves of the same color are the irreducible difference between our theory and Ref.\@ \citep{Liu2017} in this limit, that is, it is the difference which is necessitated by the differing ways that the rate of reconnection is defined between the two theories and cannot be eliminated by a different choice of ``normalization" or parameterization.}
    \label{fig::CompareTheories}
\end{figure}

In the case where these parameters are zero, the rate becomes $\mathcal{R}_{\text{SR}}$ which we call the ``special relativistic rate." This has a very simple formula,
\begin{equation} \label{Result::SRRate}
    \mathcal{R}_{\text{SR}} = \mu \left( 1 - \mu^2\right) \sqrt{\frac{\left(1 - \mu^2 \right)^3 \sigma}{1 + \left(1 - \mu^2 \right)^3 \sigma}}. 
\end{equation}
There are some superficial differences between this formula and the original result derived in Ref.\@ \citep{Liu2017}. The original theory defined the reconnection rate as a speed normalized to the mesoscale Alfv\'en speed, whereas our scalar definition is purely a quantity of speed (normalized to the speed of light in our units). Our theory also makes a different choice of representing the degree of freedom represented by the plasma magnetization, where we have defined the quasi-magnetization $\sigma$ to be based on the total amount of flux contained within the system, whereas the magnetization parameter used in Ref.\@ \citep{Liu2017} depends on the density of magnetic flux within the plasma at the point that it enters the system. The more substantial difference is that $\mathcal{R}_{\text{SR}} = \bar{\bs{V}}_d$, whereas in the original theory, the rate is directly related to $\bs{V}_d \Big|_{\mathfrak{u}}$. In this ``special relativistic limit," 
\begin{equation}
\begin{split}
    \lim_{\bs{\bar{a}},\bar{S}, \bar{\mathcal{K}}, \bar{Q} \rightarrow 0} \bs{V}_d \Big|_{\mathfrak{u}} &= \frac{\mu \bs{H}^{L}_{\mathfrak{m}}}{\bs{H}^L_{\mathfrak{u}}} \bs{V}^L_{\mathfrak{o}} = \mu \frac{1 - \mu^2}{1 + \mu^2} \bs{V}^{L}_{\mathfrak{u}} \\
    &= \frac{\mathcal{R}_{\text{SR}}}{1 + \mu^2}.
\end{split}
\end{equation}
In Fig \ref{fig::CompareTheories}, we directly compare $\mathcal{R}_{\text{SR}}$ and $\bs{V}_d \Big|_{\mathfrak{u}}$ to show that this difference ultimately does not result in a very large quantitative difference.

\section{Magnetic Reconnection Around A Static and Spherically Symmetric Gravitational Source} \label{Section::CentralBodyProblem}

We will now explore the implications of the formulae derived in the previous section for magnetic reconnection around gravitational sources. It is reasonable to assume that the gravitational field of the plasma undergoing reconnection can be neglected, and so the gravitational source in question simply becomes a separate entity from the plasma itself. Under this assumption, the spacetime in the direct vicinity of the system in question must be a ``vacuum solution" of the Einstein Field Equations. Solutions of this kind must be ``Ricci flat," i.e.\@, they satisfy $\text{Ric}_{\mu \nu} \equiv g^{\alpha \beta} R_{\mu \alpha \nu \beta} = 0$ \citep{Hobson2011}. By imposing this condition at the X-sheet, we obtain the relation,
\begin{equation}
    \text{Ric}_{yy} \Big|_{\mathtt{x}} = S_{yy} + \mathcal{K} - Q_{yy} = 0.
\end{equation}
In our formulae, we can then make the substitution $\bar{Q}=\bar{S}+\bar{\mathcal{K}}$. This reduces the NICS effects in our theory down to the three parameters $\{ \bs{\bar{a}}^2, \bar{S}, \bar{\mathcal{K}}\}$. For simplicity, we will find expressions for these parameters which assume that plasma is accreting with zero angular momentum around a gravitational source is static and spherically symmetric such that the metric is given by the Schwarzschild metric. 

\subsection{NICS Effects on Accretion Plasma With Zero Angular Momentum} \label{Subsection::ZAMOEffects}

\begin{figure}
    \centering
    \includegraphics[width=0.9\linewidth]{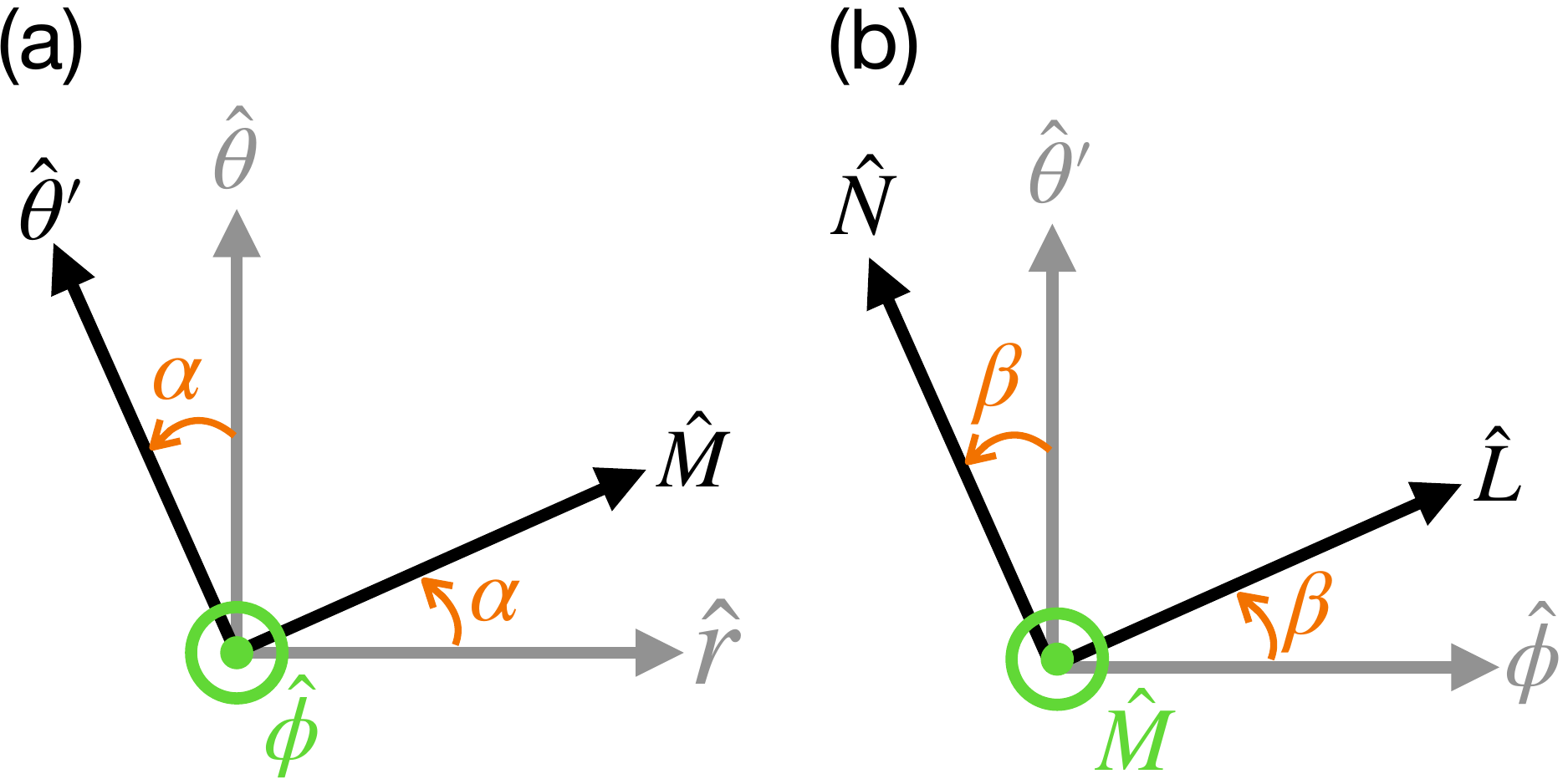}
    \caption{ As shown in (a), $\hat{M}$ is assumed to be a linear combination of $\{ \hat{r}, \hat{\theta}\}$, where $\alpha$ is the angle between $\hat{M}$ and $\hat{r}$. The vector $\hat{\theta}^{\prime}$ is also defined to be a linear combination of $\{ \hat{r}, \hat{\theta}\}$ and is orthogonal to $\hat{M}$, with $\alpha$ also specifying the angle between $\hat{\theta}^{\prime}$ and $\hat{\theta}$. From these definitions, the plane of reconnection must be spanned by $\hat{\phi}$ and $\hat{\theta}^{\prime}$ (shown in (b)), where $\beta$ determines the orientation of $\hat{\bs{L}}$ and $\hat{\bs{N}}$ relative to these vectors in the same way that $\alpha$ determines the orientation of $\{ \hat{M}, \hat{\theta}^{\prime}\}$ relative to $\{\hat{r}, \hat{\theta} \}$.}
    \label{fig::AlphaBeta}
\end{figure}

We show in Appendix \ref{Appendix::SchRiemann} that within any zero angular momentum frame, the Riemann curvature tensor of the Schwarzschild metric has the following expression,
\begin{equation}
\begin{split}
    \mathbb{R}/\kappa = &-(\oneform{T} \wedge \oneform{r}) \otimes (\oneform{T} \wedge \oneform{r}) + \frac{1}{2} (\oneform{T} \wedge \oneform{\theta}) \otimes (\oneform{T} \wedge \oneform{\theta}) \\
    &+ \frac{1}{2} (\oneform{T} \wedge \oneform{\phi}) \otimes (\oneform{T} \wedge \oneform{\phi}) - \frac{1}{2} (\oneform{r} \wedge \oneform{\theta}) \otimes (\oneform{r} \wedge \oneform{\theta}) \\
    &- \frac{1}{2} (\oneform{r} \wedge \oneform{\phi}) \otimes (\oneform{r} \wedge \oneform{\phi}) + (\oneform{\theta} \wedge \oneform{\phi}) \otimes (\oneform{\theta} \wedge \oneform{\phi}),
\end{split}
\end{equation}
where $\kappa \equiv r_s/r^3$, $r\equiv 2GM$ is the Schwarzschild radius of the gravitational source, and $r$ the Schwarzschild radial coordinate. The 1-forms $\{\oneform{r},\oneform{\theta},\oneform{\phi} \}$ correspond to a set of orthonormal basis vectors $\{\hat{r}, \hat{\theta}, \hat{\phi} \}$ spanning the subspace orthogonal to $\hat{T}$, which conform to their expected meaning: $\hat{r}$ points ``radially" away from the gravitational source, and $\{ \hat{\theta}, \hat{\phi} \}$ span the ``angular directions." We then find simple relations for $\mathcal{\bar{K}}$ and $\bar{S}$ in terms of the orientation of vectors $\{\hat{\bs{L}}, \hat{\bs{N}}, \hat{M} \}$ relative to the basis $\{\hat{r}, \hat{\theta}, \hat{\phi} \}$,
\begin{equation} \label{Result::SandKMostGeneral}
\begin{alignedat}{2}
    &\bar{S}/\bar{\kappa} &&= N_r^2 - \frac{1}{2}\left( N_{\theta}^2 + N_{\phi}^2\right) = \frac{1}{2}\left(3N_r^2-1 \right),\\
    &\bar{\mathcal{K}}/\bar{\kappa} &&= -\frac{1}{2}(L_{[r}N_{\theta]})^2 - \frac{1}{2}(L_{[r}N_{\phi]})^2  +(L_{[\theta}N_{\phi]})^2  \\
    & &&= -\frac{1}{2}M_{\phi}^2 - \frac{1}{2}M_{\theta}^2 + M_r^2 = \frac{1}{2}\left(3M_r^2-1 \right),
\end{alignedat}
\end{equation}
where $\bar{\kappa} \equiv \Delta y^2 \kappa$, and the brackets denote ``anti-symmetrization," i.e.\@ $A_{[\mu}B_{\nu]} \equiv A_{\mu}B_{\nu}-A_{\nu}B_{\mu}$. We will specify the orientation of $\hat{\bs{N}}$ and $\hat{M}$ using angles $\alpha$ and $\beta$ defined by the following,
\begin{equation}
\begin{split}
    \begin{pmatrix}
        \hat{M} \\ \hat{\theta}^{\prime}
    \end{pmatrix} &\equiv
    \begin{pmatrix}
        \cos(\alpha) & \sin(\alpha) \\
        -\sin(\alpha) & \cos (\alpha)
    \end{pmatrix} 
    \begin{pmatrix}
        \hat{r} \\ \hat{\theta}
    \end{pmatrix}, \\ 
    \begin{pmatrix}
        \hat{\bs{L}} \\ \hat{\bs{N}}
    \end{pmatrix} &\equiv
    \begin{pmatrix}
        \cos(\beta) & \sin(\beta) \\
        -\sin(\beta) & \cos (\beta)
    \end{pmatrix} 
    \begin{pmatrix}
         \hat{\phi} \\ \hat{\theta}^{\prime}
    \end{pmatrix}.
\end{split}
\end{equation}
$\alpha$ is then the angle formed by $\hat{M}$ and $\vec{r}$, and $\beta$ is the angle formed by $\hat{\bs{N}}$ and $\hat{\theta}^{\prime}$, where $\{ \hat{\theta}^{\prime}, \hat{\phi} \}$ span the plane of reconnection (see Fig.\@ \ref{fig::AlphaBeta}). 

\begin{figure*}
    \centering
    \includegraphics[width=0.9\linewidth]{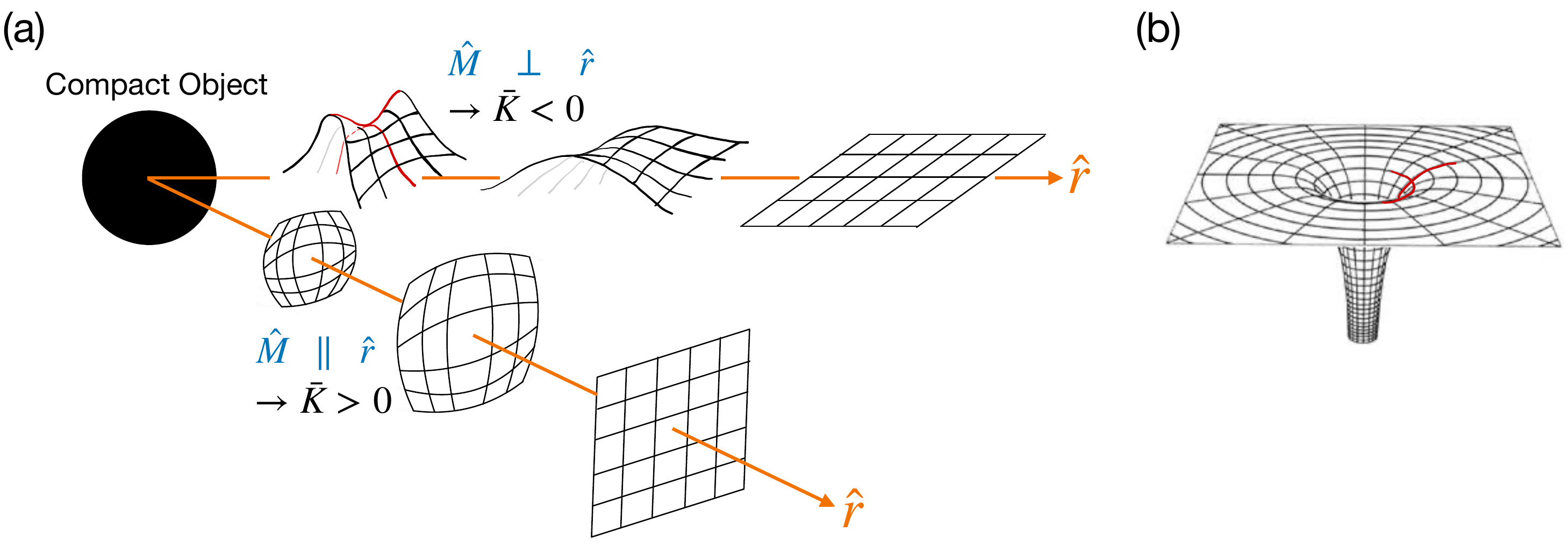}
    \caption{In (a), the intrinsic Gaussian curvature of the plane of reconnection is illustrated with ``embedding diagrams," in which it is shown how the plane would need to be \textit{extrinsically} curved within a Euclidean space in order to produce the same effect of intrinsic curvature. Recall that Gaussian curvature is the product of a surface's ``principal curvatures," and its sign is therefore determined by whether the principal curvatures point in the same direction or in opposite directions \citep{Sternberg2013}. For $\hat{M} \parallel \hat{r}$ (i.e.\@, $\alpha=0$), $\bar{\mathcal{K}} >0$ and the plane can be then represented by a ``dome shaped" surface. For $\hat{M} \perp \hat{r}$ (i.e.\@, $\alpha=\pi/2$), $\mathcal{\bar{K}} < 0$ and the plane can instead be represented by a ``saddle shaped" surface. In (b), we draw a connection to the famous embedding diagram for the ``equatorial plane" of the Schwarzschild metric (i.e.\@, the result of suppressing the Schwarzschild time coordinate $t_s$ and one of the angular coordinates $\{\theta, \phi \}$), which is known as ``Flamm's Paraboloid" \citep{Flamm2015}. The ``saddle shaped" surfaces shown in panel (a) are merely a ``patch" of the embedding diagram (i.e.\@, the saddle structure depicted by the crossing red curves) shown in (b).}
    \label{fig::Embeddings}
\end{figure*}

From these definitions, we find that $M_r = \cos(\alpha)$ and $N_r=\sin(\alpha)\cos(\beta)$, and we obtain simple expressions for $\bar{S}$ and $\mathcal{\bar{K}}$,
\begin{equation} \label{Result::SAndKFormulae}
\begin{split}
    \bar{S} &= \frac{1}{2}\left(3\sin^2(\alpha)\cos^2(\beta) - 1 \right) \bar{\kappa}, \\
    \mathcal{\bar{K}} &= \frac{1}{2}\left(3\cos^2(\alpha) - 1 \right) \bar{\kappa}.
\end{split}
\end{equation}
When $\hat{\bs{N}}$ points radially ($\alpha=\pi/2$ and $\beta=0$), $\bar{S}$ is positive, corresponding to tidal stretching along the inflow line. When $\hat{\bs{N}}$ is purely angular ($\alpha=0$ or $\beta=\pi/2)$), there will instead be tidal compression along the inflow line. The geometry of the plane of reconnection for different orientations is illustrated in Fig.\@ \ref{fig::Embeddings}, where it is shown that when $\hat{M}$ points radially ($\alpha=0$) it has positive intrinsic Gaussian curvature, and will be negatively curved when $\hat{M}$ points in an angular direction. 

The assumptions we have made about the spacetime do not in any way determine the value of $\bs{\bar{a}}^2$. However, in the case of an accretion plasma, we may place a plausible upper bound on this parameter. We will assume that the ``natural" motion of the plasma is a freely-falling plummet directly into the gravitating body. If there is a net linear acceleration, then, it is because the accretion encounters some resistance to this natural motion, resulting in a net acceleration outwards in the direction $\hat{r}$. If under ordinary conditions we can expect that the plasma is at most halted by this resistance, i.e.\@, $dr/dt_p\leq0$ where $t_p$ is the proper time of the frame co-moving with the plasma, then the acceleration of the system is at most the acceleration needed for a ``static" system where $dr/dt_p=0$. We calculate this acceleration as $\vec{a}_s$ in Appendix \ref{Appendix::SchAcceleration}, from which we obtain the desired upper bound,
\begin{equation}
    \bs{\bar{a}}^2 \leq \Delta y^2 (\vec{a}_s \cdot \hat{\bs{N}})^2 = \frac{1}{4}\frac{N_r^2}{r/r_s-1} \bar{\kappa}.
\end{equation}
Comparing this value with the expressions for $\bar{S}$ and $\mathcal{\bar{K}}$, we note that $\bar{\bs{a}}^2$ will be much smaller than the other NICS parameters except when $r_s<r<1.25r_s$. Outside of this very narrow range immediately outside the event horizon of a black hole, we conclude that it is reasonable to approximate the system as inertial for the purposes of predicting the rate of reconnection. 

\subsection{Overall Gravitational Effect on the Average Rate}

\begin{figure}
    \centering
    \includegraphics[width=1.0\linewidth]{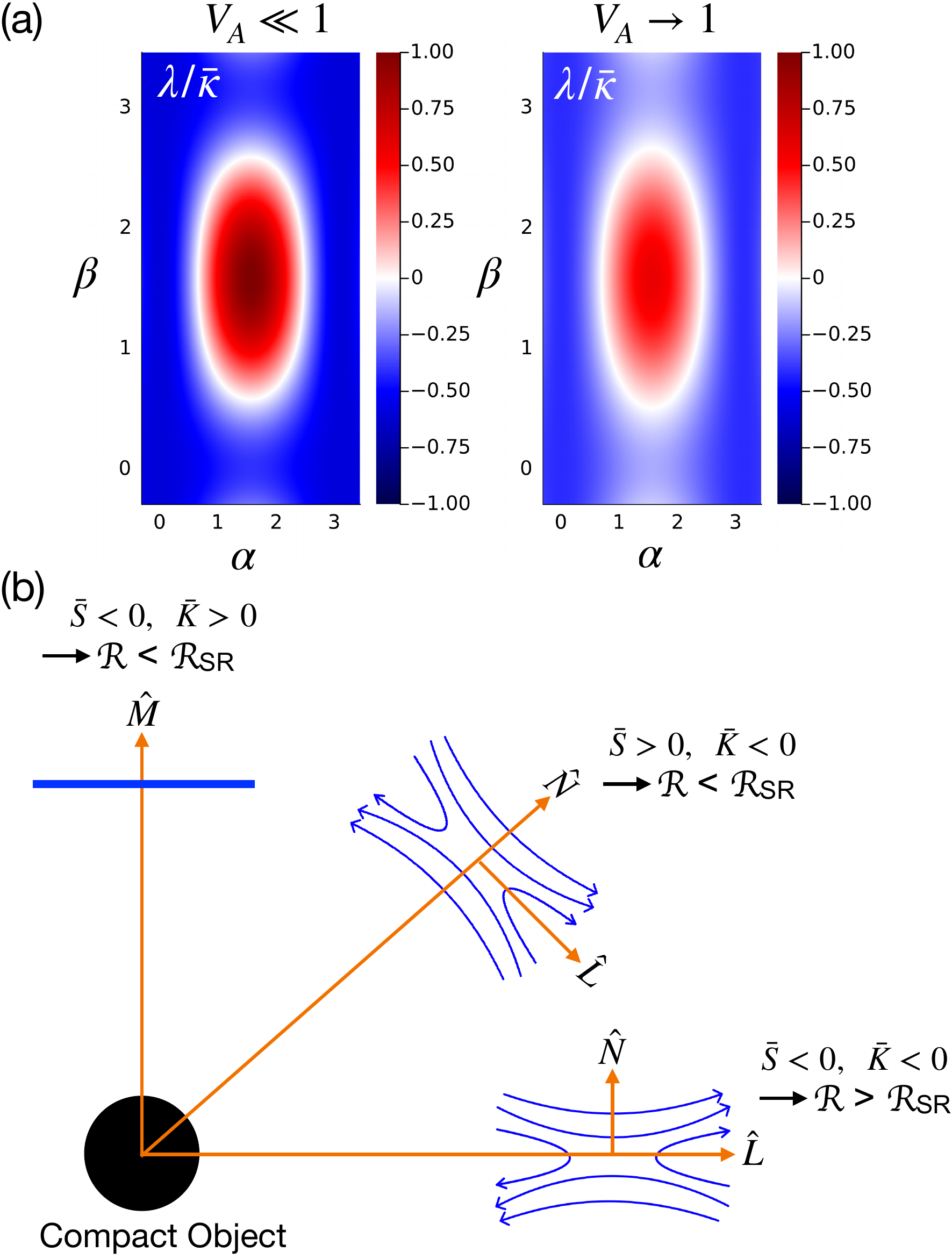}
    \caption{In (a), a heatmap displaying the value of $\lambda/\bar{\kappa}$ for the low and high magnetization limits. In (b), we illustrate three example configurations and their associated effect on the reconnection rate.}
    \label{fig::LambdaIllustration}
\end{figure}

The question of the overall effect on the reconnection rate can be simplified if we average $\mathcal{R}$ over $\mu$. This can easily be done analytically in two limits: the low magnetization limit ($V_A \ll 1$) and the high magnetization limit ($V_A \rightarrow 1)$. In the low magnetization limit, $\mathcal{R}$ has the Taylor expansion around $V_A=0$,
\begin{equation}
\begin{split}
    \mathcal{R} =& \mu (1 - \mu^2)^{3/2} \left( 1 - \mu^2 - \frac{1}{3}\bar{\mathcal{K}} - \frac{1}{2}\bar{Q}(1 - \mu^4) \right) \\
    &\times \left(1 - \frac{1}{3} \bar{S} + \frac{4}{3} \bs{\bar{a}}^2 \right) V_{A} \\
    &+ \mathcal{O}(\Delta y^3) + \mathcal{O}(V_A^2).
\end{split}
\end{equation}
$\mathcal{R}$ in the high magnetization limit is found simply from $\bs{V}_{\mathfrak{o}}^L \rightarrow 1$,
\begin{equation}
\begin{split}
    \lim_{V_A \rightarrow 1} \mathcal{R} =& \mu \left(1 - \mu^2 - \frac{1}{6}\mathcal{\bar{K}} - \frac{1}{4}\bar{Q}(1 - \mu^4) \right) \\
    & \times\left( 1 - \frac{1}{6}\bar{S} + \frac{2}{3} \bs{\bar{a}}^2\right) + \mathcal{O}(\Delta y^3).
\end{split}
\end{equation}
Now we simply calculate the average rate $\langle \mathcal{R} \rangle \equiv \int_0^1 \mathcal{R} d\mu$ in both limits,
\begin{equation}
\begin{split}
    \langle \mathcal{R} \rangle \Big|_{V_A \ll 1} &= \frac{V_A}{7}\left(1 - 0.4\bar{6} \bar{\mathcal{K}} - 0.6\bar{1}\bar{Q} - 0.\bar{3}\bar{S} + 1.\bar{3}\bs{\bar{a}}^2 \right), \\
    \langle \mathcal{R} \rangle \Big|_{V_A \rightarrow 1} &= \frac{1}{4}\left(1 - 0.\bar{3}\mathcal{\bar{K}} - 0.\bar{3}\bar{Q}  - 0.1\bar{6}\bar{S} + 0.\bar{6} \bs{\bar{a}}^2  \right),
\end{split}
\end{equation}
where the bar over digits denotes repeated decimals, i.e.\@ $1/3=0.\bar{3}$. Both expressions are of the form $\langle \mathcal{R} \rangle = \langle \mathcal{R}_{\text{SR}} \rangle (1 + \lambda)$, where $\lambda$ encapsulates the gravitational effect, and where $\langle \mathcal{R}_{\text{SR}} \rangle \Big|_{V_A \ll 1} = V_A/7$ and $\langle \mathcal{R}_{\text{SR}} \rangle \Big|_{V_A \rightarrow 1} = 1/4$. Applying the substitution $\bar{Q} = \bar{S} + \mathcal{\bar{K}}$, the approximation $\bs{\bar{a}}^2 \approx 0$, and the expressions for $\bar{S}$ and $\mathcal{\bar{K}}$ given in Eq.\@ (\ref{Result::SAndKFormulae}), we determine $\lambda$ in both limits,
\begin{equation} \label{Result::Lambdas}
\begin{split}
    \lambda \Big|_{V_A \ll 1} =& \big( 1.0\bar{1} - 1.61\bar{6} \cos^2(\alpha) \\
    &- 1.41\bar{6} \sin^2(\alpha)\cos^2(\beta) \big) \bar{\kappa}, \\
   \lambda \Big|_{V_A \rightarrow 1} =& \big(0.58\bar{3} - \cos^2(\alpha) \\
   &- 0.75 \sin^2(\alpha)\cos^2(\beta) \big) \bar{\kappa}.
\end{split}
\end{equation}
The value of $\lambda / \bar{\kappa}$ is shown in Fig.\@ \ref{fig::LambdaIllustration} for both limits. In both cases, it is seen that $\lambda$ reaches its greatest \textit{positive} value at $\alpha=\pi/2$ and $\beta=\pi/2$, which is the case where $\bar{S}$ and $\bar{\mathcal{K}}$ obtain their greatest \textit{negative} value. It is only in configurations that approximate this one that all of the gravitational effects work in tandem to enhance the rate of reconnection. As either $\alpha$ or $\beta$ deviate from this configuration, the gravitational effects will begin working against each other. $\lambda=0$ along a closed curve within the $(\alpha,\beta)$ plane encircling the point $(\pi/2,\pi/2)$, and outside of the curve, the overall gravitational effect will be to decrease the rate. 

\subsection{Summary}

In this paper, we have provided a fully gauge-independent framework for the theory of magnetic reconnection. Our definition of the process itself and its rate (Sections \ref{Subsection::InvariantReconnectionDefinition} and \ref{Subsection::RateDefinition}) rely only on the assumption that the system is ``2D," so that there is a defined ``flux function" given by $\oneform{d}\psi \equiv \twoform{F} \cdot \vec{\zeta}$. Beyond this, the bulk of the theoretical foundation we provide applies specifically to reconnection in a steady-state, which is a case having a natural choice of coordinate representation (Eq.\@ (\ref{Statement::GaugeConstraint})), in which the rate has a familiar relationship to the drift speed $\bs{V}_d \equiv E_r/|\bvec{B}_r|$ (Eq.\@ (\ref{Result::StationaryRate})), and in which the metric is found to have a local approximation in which its deviation from the simple Minskowski metric $\eta \equiv \text{diag}(-1,1,1,1)$ depends only on parameters with an immediate physical interpretation: the net acceleration of the system $\bvec{a}$, the curvature of the X-line in its own rest frame, and on the Riemann curvature tensor evaluated along the X-sheet $\mathbb{R}$ (Eq.\@ (\ref{Result::FermiMetricFormula})). This framework constitutes a major step towards eliminating all ambiguity in the physical interpretation of our results. Its only weakness in this regard is that it is not yet fully general: for the gauge condition Eq.\@ (\ref{Statement::GaugeConstraint}) to be valid, it is assumed that the Killing vectors $\vec{\tau}$ and $\vec{\zeta}$ are orthogonal to each other and also hypersurface orthogonal. We proceeded as though these assumptions are physically reasonable. It will be the subject of our future work to go beyond these assumptions, which we expect to better clarify the actual limitations of this initial version of our theory. 

In Section \ref{Section::Overview}, we presented an overview of the different electromagnetic NICS effects on magnetic reconnection which are associated with the ``general relativistic" parameters seen in the metric formula (Eq.\@ (\ref{Result::FermiMetricFormula})). The first effect derives from the linear acceleration $\bvec{a}$ and tidal tensor $\mathbb{S}$, which are found to contract or dilate light crossing times within the system, which will increase or decrease the maximum possible rate $(\mathcal{C}^+ + \mathcal{C}^-)/2$. In the case of ``symmetric reconnection" (which we covariantly define in Section \ref{Subsection::LightSpeed}), the rate of reconnection is found to be directly proportional to this maximum rate (as given by Eq.\@ (\ref{Result::SymmetricRate})), where $\bvec{a}$ is seen to always increase the rate while $S_{yy}$ increases(decreases) the rate when it is negative(positive). The second effect is caused by the gravitational distortion of the system's ``3D" shape, such that the equivalent of a ``slab symmetric" situation (i.e.\@, that of a geodesic X-line where $\bvec{h}=0$) exhibits a non-uniform electric field in a steady-state, increasing(decreasing) the rate when $Q_{yy}$ is negative(positive). Finally, we derived the form of Ampere's Law which governs the in-plane magnetic field generated by the out-of-plane current density (Eq.\@ (\ref{Result::AmperesLaw})), and showed how the way that the field-line geometry of reconnection is constrained necessitates greater or lesser amounts of magnetic tension opposing the inflow plasma depending on the intrinsic Gaussian curvature of the plane of reconnection $\mathcal{K}$, which we call ``gravitationally induced magnetic tension." Just as with $S_{yy}$ and $Q_{yy}$, $\mathcal{K}$ also increases(decreases) the rate when it is negative(positive). 

Section \ref{Section::Overview} provides the conceptual basis for understanding the result of Section \ref{Section::GeneralizedFormulae}, in which we simply generalize the theory of \citet{Liu2017} to account for the electromagnetic NICS effects (Eq.\@ (\ref{Result::MostGeneralRate})). We only correct the theory to account for the \textit{electromagnetic} effects because we assume, as was assumed in the original theory, that the plasma in the mesoscale region reaches a force-free equilibrium. It is important to note that this assumption excludes at the outset the most ``obvious" effect of gravity, which is the ``acceleration due to gravity" felt by the plasma itself. Accounting for this would require us to include the ``inertial term" in the momentum equation, which plays no role in force-free equilibrium. Therefore, one should not mistake our inclusion of the tidal tensor $\mathbb{S}$ as the inclusion of ``tidal force," which our theory neglects entirely. It is possible that a general relativistic treatment of the non-electromagnetic forces acting on the plasma is necessary to accurately describe plasma in general relativistic situations, which is a direction that we intend to explore in future works. 

Finally, in Section \ref{Section::CentralBodyProblem} we applied Eq.\@ (\ref{Result::MostGeneralRate}) to a very simple context: a plasma which accretes with zero angular momentum onto a static and spherically symmetric gravitational source. The result is encapsulated by Eq.\@ (\ref{Result::Lambdas}), in which the significance of the effect of gravity is encapsulated by the parameter $\lambda$ and calculated within both the high and low magnetization limits. In both cases, the magnitude of lambda is typically on the order of the dimensionless parameter $\bar{\kappa} \equiv r_s \Delta y^2/r^3$, whereas the \textit{sign} of $\lambda$ (i.e.\@, whether or not the rate is increased or decreased by gravity) is purely a question of the orientation of the system relative to the gravitational source (as shown in Fig.\@ \ref{fig::LambdaIllustration}). This analysis enables us to form very straightforward expectations for what kinds of plasma systems should exhibit faster or slower reconnection than in non-gravitational contexts. However, it is important to be aware of the lack of realism within the premises of this analysis: accretion plasmas generally have significant angular momentum, and the compact objects around which we may expect these gravitational effects to be significant are often not static. In future works, we aim to generalize this analysis to account for a system with arbitrary motion relative to a gravitational source with non-zero spin.

\section{References}

\bibliography{cite}
\bibliographystyle{apalike}

\appendix

\section{3+1 Electrodynamics} \label{Appendix::MaxwellsEquations}

Here we give a brief overview of the ``3+1" representation of electrodynamics in general relativity, as formulated by \citet{Komissarov2004, Komissarov2011}. We begin with an arbitrary gauge constraint of the type usually desired, in which a global time coordinate is defined such that ``absolute space" means the 3D submanifolds of spacetime defined by constant $t$. The spacetime metric is then decomposed such that the line element takes the form,
\begin{equation} \label{Statement::General3Plus1Coordinates}
    ds^2 = (\beta^2-\alpha^2) dt^2 + 2\beta_idx^i dt + \gamma_{ij}dx^idx^j.
\end{equation}
$\alpha$ is called the ``lapse function," $\beta^i \equiv \gamma^{ij}\beta_j$ the ``shift vector," and $\gamma_{ij}$ is the metric induced on `` absolute space," i.e.\@, on the 3D spacelike hypersurfaces having constant $t$. In this section, we will adopt the convention that Latin indices $\{ i,j,k,l,\dots\}$ index exclusively the spatial coordinates which parameterize absolute space, while Greek indices $\{\mu,\nu,\sigma,\lambda,\dots\}$ index all spacetime coordinates. Let $g \equiv \text{det}g_{\mu \nu}$, then we define the electromagnetic vector fields,
\begin{equation} \label{Statement::EBDHDefinitions}
\begin{split}
    E_{i} &= F_{it}, \\
    B^i &= \frac{\sqrt{-g}}{2}\varepsilon^{ijk} F_{jk}, \\
    D^i & = \frac{\sqrt{-g}}{2}\varepsilon^{ijk} (\star \mathbb{F})_{jk}, \\
    H_i &= (\star \mathbb{F})_{ti},
\end{split}
\end{equation}
where $\varepsilon_{ijk}$ is the Levi-Civita symbol, and $\star$ is the Hodge star operator. These fields are entirely tangent to ``absolute space." Therefore, within this formulation, it is best to conceptualize them as ``3-vectors" living within the tangent space of absolute space described by the ``spatial metric" $\gamma_{ij}$. Strictly within this section, we will adopt the convention that their indices are raised and lowered by $\gamma_{ij}$, and so $\vec{E} \equiv \gamma^{ij} E_i \vec{\partial_j}$ and $\vec{H} \equiv \gamma^{ij} H_i\vec{\partial}_j$. If $\vec{\nabla}^{(3)}$ is the ``spatial" covariant derivative defined in relation to the spatial metric $\gamma_{ij}$, then electrodynamics is described by the 3D time dependent partial differential equations,
\begin{equation} \label{Statement::3Plus1Maxwell}
\begin{split}
    \vec{\nabla}^{(3)} \cdot \vec{B} &= 0, \\
    \vec{\nabla}^{(3)} \cdot \vec{D}&= \rho,\\
    \frac{1}{\sqrt{\gamma}} \partial_{t}(\sqrt{\gamma} \vec{B}) + \vec{\nabla}^{(3)} \times \vec{E} &= 0, \\ 
    \frac{1}{\sqrt{\gamma}} \partial_{t}(\sqrt{\gamma} \vec{D}) + \vec{\nabla}^{(3)} \times \vec{H} &= -\vec{j},
\end{split}
\end{equation}
where $\gamma \equiv \text{det}\gamma_{ij}$, and the electric 4-current $\vec{J}$ has the decomposition $\alpha \vec{J} \equiv \rho\vec{\partial_t} + \vec{j}$. This system of equations is completed with the ``constitutive relations,"
\begin{equation} \label{Statement::ConstitutiveRelations}
\begin{split}
    \vec{E} &=  \alpha \vec{D} + \vec{\beta} \times \vec{B}, \\
    \vec{H} &= \alpha \vec{B} - \vec{\beta} \times \vec{D}.
\end{split}
\end{equation}
This description clearly mirrors the ``macroscopic" formulation of electromagnetism within in a polarized and/or magnetized medium, in which $\vec{D}$ and $\vec{H}$ are called ``auxiliary fields." 

At the beginning of Section \ref{Section::Theory}, we motivated our gauge-independent approach to defining the process of reconnection with the statement that ``magnetic behavior" is, in part, something ``the physicist chooses." We will now clarify what is meant by this statement. Note that Eq.\@ (\ref{Statement::3Plus1Maxwell}), taken together with Eq.\@ (\ref{Statement::ConstitutiveRelations}), couples the electromagnetic 3-vectors $\{\vec{E},\vec{D},\vec{B},\vec{H} \}$ to only the zeroth and first order derivatives of $\{\alpha, \beta_i, \gamma_{ij} \}$, which is to say, to only the zeroth and first order derivatives of the metric tensor $g_{\mu \nu}$. On the other hand, the Riemann curvature tensor depends on the \textit{second order} derivatives of the metric. Eq.\@ (\ref{Statement::3Plus1Maxwell}) therefore does not feature an explicit dependence on the intrinsic geometry of spacetime, and it is problematic to interpret the ``effects" of $\{\alpha, \beta_i, \gamma_{ij} \}$ as merely ``gravitational effects." This statement may at first be surprising if it assumed that the absence of gravity \textit{necessitates} the usual coordinate representation of spacetime where $ds^2 = \eta_{\mu \nu}dx^{\mu} dx^{\nu}\equiv -dt^2 + \delta_{ij}dx^i dx^j$. Here it is important to appreciate that the choice of coordinates is always nonphysical, even in a flat spacetime. 

It may be instructive to show this fact directly. First, we will assume that spacetime is flat and $x^{\mu}$ is a ``Minkowskian" coordinate system satisfying $ds^2=\eta_{\mu \nu}dx^{\mu}dx^{\nu}$. Under this assumption, Eq.\@ (\ref{Statement::ConstitutiveRelations}) becomes the trivial constitutive relations in a vacuum $\vec{D}=\vec{E}$ and $\vec{H}=\vec{B}$, and Eq.\@ (\ref{Statement::3Plus1Maxwell}) becomes the standard 3+1 form of Maxwell's Equations in special relativity. Lets now consider the effect of an \textit{arbitrary} coordinate transformation $x^{\mu} \rightarrow x^{\mu^{\prime}}$. We will define $J^{\mu}_{\nu} \equiv \partial x^{\mu}/\partial x^{\nu^{\prime}}$, which is the Jacobian of the inverse map $x^{\mu^{\prime}} \rightarrow x^{\mu}$. The ``covariant" components of the metric transform as $g_{\mu^{\prime} \nu^{\prime}} = J^{\lambda}_{\mu^{\prime}} J^{\sigma}_{\nu^{\prime}} \eta_{\lambda \sigma}$. Matching components with Eq.\@ (\ref{Statement::General3Plus1Coordinates}), we obtain a system of equations from which one could determine the variables $\{\alpha, \beta_{i^{\prime}}, \gamma_{i^{\prime} j^{\prime}} \}$ within the new coordinate system,
\begin{equation} \label{Result::GaugeVariablesFlatSpacetime}
\begin{split}
    g_{t^{\prime} t^{\prime}} = \gamma^{i^{\prime} j^{\prime}} \beta_{i^{\prime}} \beta_{j^{\prime}} - \alpha^2 &= -(J^t_{t^{\prime}})^2 + \delta_{ij} J^i_{t^{\prime}} J^j_{t^{\prime}}, \\
    g_{t^{\prime} i^{\prime}} = \beta_{i^{\prime}} &= -J^t_{t^{\prime}} J^t_{i^{\prime}} + \delta_{jk} J^j_{t^{\prime}} J^k_{i^{\prime}}, \\
    g_{i^{\prime} j^{\prime}} = \gamma_{i^{\prime} j^{\prime}} &= -J^t_{i^{\prime}} J^t_{j^{\prime}} + \delta_{kl}J^k_{i^{\prime}}J^l_{j^{\prime}}.
\end{split}
\end{equation}
Since the intrinsic geometry of spacetime is independent of our choice of coordinates, Eq.\@ (\ref{Statement::3Plus1Maxwell}) and Eq.\@(\ref{Statement::ConstitutiveRelations}), taken with Eq.\@ (\ref{Result::GaugeVariablesFlatSpacetime}), must describe the \textit{exact same physics} as the standard formulation using the coordinates $x^{\mu}$. This is sufficient to prove that the inclusion of $\{\alpha, \beta_{i}, \gamma_{ij} \}$ in Eq.\@ (\ref{Statement::3Plus1Maxwell}) cannot have a strictly physical interpretation. In the flat spacetime case, the difference in how the electromagnetic 3-vectors seen in Eq.\@ (\ref{Statement::3Plus1Maxwell}) and Eq.\@ (\ref{Statement::ConstitutiveRelations}) behave is entirely due to a difference in how spacetime is ``sliced" by $t$ or by $t^{\prime}$, which is a very broad and entirely non-physical degree of freedom. While the coupling between the electromagnetic 3-vectors and $g_{\mu \nu}$ must \textit{in part} owe something to the real physical effects of gravity on electromagnetism in the case where spacetime is curved, it is not clear within this framework how these can be distinguished from the ``effects" which are just the inevitable result of being forced to choose a gauge other than $g_{\mu \nu}=\eta_{\mu \nu}$. It is precisely this difficulty in distinguishing between ``magnetic behavior" that has a clear physical interpretation from behavior that derives merely from our choices in representation which motivates the gauge-independent foundation developed in Sections \ref{Subsection::InvariantReconnectionDefinition}-\ref{Subsection::RateDefinition}.

\section{Field-Sheets and Magnetic Field-Lines} \label{Appendix::FieldSheets}

In Section \ref{Subsection::InvariantReconnectionDefinition}, we refer to the surfaces of constant $\psi$ and $z$ as the ``field-sheets" of $\twoform{F_r}=\oneform{d}\psi \wedge \oneform{d}z$. This is the term used by \citet{Gralla2014} to refer, in general, to surfaces of constant $\phi_1$ and $\phi_2$ given a Faraday 2-form expressed as $\oneform{d}\phi_1 \wedge \oneform{d}\phi_2$. It was stated that the ``spatial" density of the field-sheets corresponds to the density of magnetic field-lines, whereas their ``temporal" density corresponds with the electric field. However, in Section \ref{Subsection::InvariantReconnectionDefinition}, no explicit justification of this interpretation was made except within an inertial reference frame in a flat spacetime. Here, we will show that this is the correct way to interpret field-sheets in general. 

Field-sheets can only represent Faraday 2-forms, or parts of Faraday 2-forms, which are closed and simple forms, so that they can be expressed as $\oneform{d}\phi_1 \wedge \oneform{d}\phi_2$ for some pair of ``Euler potentials" $\phi_1$ and $\phi_2$. This is the same thing as assuming the electromagnetic fields are ``degenerate," meaning that the scalar $\vec{E} \cdot \vec{B}=0$. Proving this is simple: using Eq.\@ (\ref{Statement::FOrthonormal}), it is clear that $\twoform{F} \wedge \twoform{F}=(\vec{E}\cdot \vec{B}) \kform{4}{dV}$, where $\kform{4}{dV}=\oneform{e}^t \wedge \oneform{e}^x \wedge \oneform{e}^y \wedge \oneform{e}^z$ is the spacetime volume element. It is clear that $\oneform{d}\phi_1 \wedge \oneform{d}\phi_2\wedge \oneform{d}\phi_1 \wedge \oneform{d}\phi_2=0$, and therefore, $\vec{E}\cdot \vec{B}=0$. From Eq.\@ (\ref{Statement::FOrthonormal}), note that $\oneform{E}=\twoform{F} \cdot \vec{e}_t$, and therefore $\vec{E}\cdot \vec{B}=\oneform{E}(\vec{B})=\vec{B} \cdot \twoform{F} \cdot \vec{e}_t=0$. For this relation to hold for any choice of $\vec{e}_t$, we must have $\vec{B} \cdot \twoform{F}=\oneform{d}\phi_1(\vec{B})\oneform{d}\phi_2 - \oneform{d}\phi_2(\vec{B})\oneform{d}\phi_1=0$. It follows, then, that $\vec{B} \perp \vec{\nabla}\phi_1$ and $\vec{B} \perp \vec{\nabla}\phi_2$.

\begin{figure}
    \centering
    \includegraphics[width=0.8\linewidth]{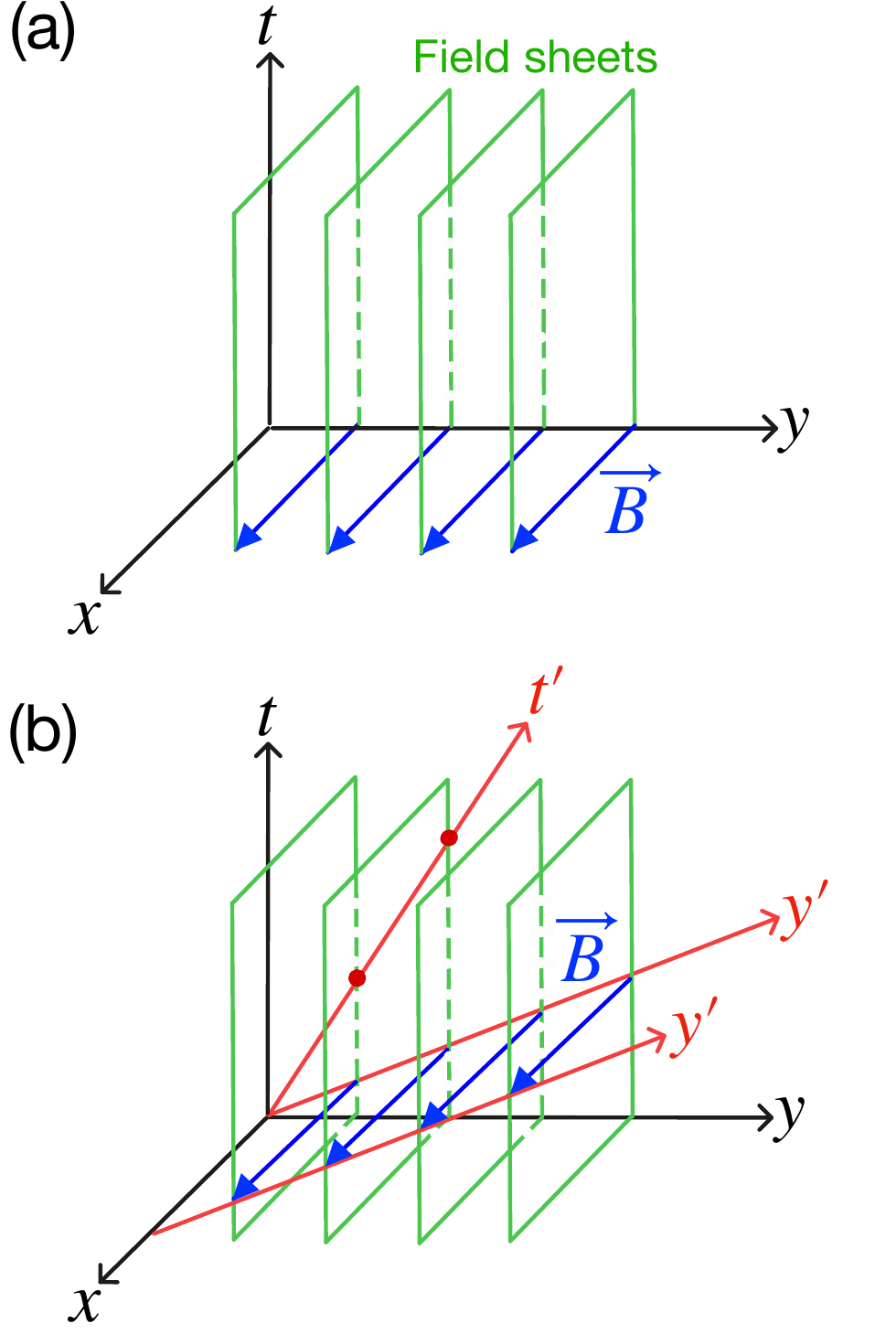}
    \caption{Illustration of the relationship between field-sheets and field-lines of a constant magnetic field: here $\twoform F=B^x\oneform{d}y\wedge\oneform{d}z$, thus $\phi_1=B^xy$ and $\phi_2=z$. In these 3D illustrations we suppress the $z$ dimension. (a) The field-sheets corresponding to a uniform magnetic field with zero electric field within the coordinate system $\{x, y, t\}$. The magnetic field-lines are the cross section of the field-sheet structure within a surface of constant $t$. (b) The same field-sheets, now illustrating the observation in a Lorentz boosted (in the $y$-direction) frame with coordinate $\{ x, y^{\prime}, t^{\prime}\}$. The magnetic field-lines in the new frame corresponds to the cross section of the field-sheet structure within a surface of constant $t^{\prime}$. The density of (blue) field lines is proportional to the magnetic field strength. The electric field (perpendicular to both $\vec{B}$ and the relative velocity) in the new frame is proportional to the density of red dots, where $t^{\prime}$-axis pierces field sheets. The ``motion'' of magnetic field lines in each frame can be illustrated if one takes different $t$-slices from past to future in panel (a), and different $t^{\prime}$-slices in panel (b).}
    \label{fig::FieldSheets}
\end{figure}

In general, therefore, any magnetic field $\vec{B}$ deriving from a Faraday 2-form $\twoform{F}=\oneform{d}\phi_1 \wedge \oneform{d}\phi_2$ will lie tangent to the surfaces of constant $\phi_1$ and $\phi_2$, that is, to the field-sheets. The magnetic field-lines are therefore merely the result of taking a cross section of the field-sheets within a ``time slice," i.e.\@, within a spacelike surface orthogonal to $\vec{e}_t$, as explained in Section \ref{Subsection::InvariantReconnectionDefinition}. When the field-sheets are timelike, they can then be interpreted as the ``world-sheets" of field-line motion. Conversely, the electric field is $\oneform{E}=\twoform{F}\cdot \vec{e}_t =
\oneform{d}\phi_2(\vec{e}_t) \oneform{d}\phi_1 -\oneform{d}\phi_1(\vec{e}_t) \oneform{d}\phi_2$. 
Therefore, $\vec{E}$ is zero if $\vec{e}_t$ lies within the field-sheets, and is only non-zero if there is a non-zero field-sheet density along $\vec{e}_t$. A simple example demonstrating these facts is shown in Fig.\@ \ref{fig::FieldSheets}.

\section{Calculations in Fermi Coordinates}

\subsection{Derivation of $g_{tt}$ and $g_{zz}$} \label{Appendix::FermiGttGzz}

We seek Taylor expansions of the functions $|\vec{\tau}|^2$ and $|\vec{\zeta}|^2$ about the X-sheet,
\begin{equation}
\begin{split}
    |\vec{\tau}|^2 =& 1 + \boldsymbol{x}^i \left(\nabla_i |\vec{\tau}|^2 \right)\Big|_{\mathtt{x}} + \frac{1}{2} \boldsymbol{x}^i \boldsymbol{x}^j \left(\nabla_i \nabla_j |\vec{\tau}|^2 \right)\Big|_{\mathtt{x}} \\
     &+ \mathcal{O}(|\vec{\boldsymbol{x}}|^3), \\
    |\vec{\zeta}|^2 =& 1 + \boldsymbol{x}^i \left( \nabla_i |\vec{\zeta}|^2 \right)\Big|_{\mathtt{x}} + \frac{1}{2} \boldsymbol{x}^i \boldsymbol{x}^j \left( \nabla_i \nabla_j |\vec{\zeta}|^2 \right)\Big|_{\mathtt{x}} \\
    &+ \mathcal{O}(|\vec{\boldsymbol{x}}|^3).
\end{split}
\end{equation}
Using Eq.\@ (\ref{Statement::SecondaryGaugeConstraint}), the first order terms are given by $\bvec{a}$ and $\bvec{h}$,
\begin{equation}
\begin{split}
    &\vec{\nabla} |\vec{\tau}|^2 \Big|_{\mathtt{x}} = \tnorm^2_{\mathtt{x}} \vec{\nabla} \ln(\tnorm^2) \Big|_{\mathtt{x}}= 2\vec{\nabla} \ln(|\vec{\tau}|) \Big|_{\mathtt{x}} = 2\vec{\boldsymbol{a}}, \\
    &\vec{\nabla} |\vec{\zeta}|^2 \Big|_{\mathtt{x}} = \znorm^2_{\mathtt{x}} \vec{\nabla} \ln(\znorm^2) \Big|_{\mathtt{x}}= 2\vec{\nabla} \ln(|\vec{\zeta}|) \Big|_{\mathtt{x}} = -2\vec{\boldsymbol{h}}.
\end{split}
\end{equation}
As for the second order terms, consider a pair of identical geodesics within the plane of reconnection having the same tangent vector $\vec{v}$, each passing through the X-point, but one at the coordinate time $t$ and the other at $t + \epsilon$. We consider the displacement vector between these geodesics as a function of the arc length parameter $s$, which must satisfy the equation of geodesic deviation \citep{Hobson2011}:
\begin{equation} \label{Result::JacobiFieldCondition}
\begin{split}
    \frac{d^2}{ds^2} \left(\epsilon \tau^{\mu} \right) &\equiv v^{\alpha} v^{\beta} \nabla_{\alpha} \nabla_{\beta} \left(\epsilon \tau^{\mu} \right) = -R^{\mu}_{\alpha \nu \beta} v^{\alpha} v^{\beta} \left(\epsilon \tau^{\nu} \right), \\
    &\hookrightarrow \nabla_{\alpha} \nabla_{\beta} \tau^{\mu} = - R^{\mu}_{\alpha \nu \beta} \tau^{\nu}.
\end{split}
\end{equation}
We are relying on the assumption that there exist geodesics which lie entirely within the plane of reconnection. This follows from the fact that the coordinates $t$ and $z$ are orthogonal to the coordinates $x$ and $y$, which means that $\Gamma^z_{ij}$ and $\Gamma^t_{ij}$, where $i$ and $j$ index only the $x$ and $y$ components, are necessarily zero. This is shown explicitly by Eq.\@ (\ref{Result::GammaUpperTAndZ}) in the following section. If we consider a hypothetical geodesic which is tangent to the plane of reconnection at some point $p$, then $\dot{z}\big|_{p}=\dot{t}\big|_{p}=0$ (where the dot is the derivative with respect to arc length), and we have from the geodesic equation $\ddot{z}\big|_p=-\Gamma^z_{ij} \dot{\bs{x}}^i \dot{\bs{x}}^j\big|_p=0$ and $\ddot{t}\big|_p=-\Gamma^t_{ij}\dot{\bs{x}}^i \dot{\bs{x}}^j\big|_p=0$. Therefore, a geodesic that is tangent to the plane of reconnection at one point remains tangent at all points. 

Now we evaluate the second derivative of $|\vec{\tau}|^2$ at $\mathtt{x}$,
\begin{equation}
\begin{split}
    \frac{1}{2} \left(\nabla_{i} \nabla_{j} |\vec{\tau}|^2\right) \Big|_{\mathtt{x}} =& - (\nabla_i \tau_{\mu})(\nabla_j \tau^{\mu})\Big|_{\mathtt{x}} - \tau_{\mu} \left(\nabla_{i} \nabla_{j} \tau^{\mu}\right)\Big|_{\mathtt{x}} \\ 
    =& -\eta_{\mu \nu} \left(\Gamma^{\mu}_{it} \Gamma^{\nu}_{jt} \right) \Big|_{\mathtt{x}} + R_{t i t j}\Big|_{\mathtt{x}} \\
    =& -\frac{\eta_{\mu \nu} \eta^{\mu \lambda} \eta^{\nu \sigma}}{4} \left(\partial_{i} g_{t\lambda} + \partial_t g_{i \lambda} - \partial_{\lambda} g_{it} \right)\Big|_{\mathtt{x}} \\
    & \times\left(\partial_{j} g_{t\sigma} + \partial_t g_{j \sigma} - \partial_{\sigma} g_{jt} \right)\Big|_{\mathtt{x}} + R_{titj}\Big|_{\mathtt{x}} \\
    =& -\frac{\eta^{\lambda \sigma}}{4} \left(\partial_i g_{t\lambda} - \partial_{\lambda} g_{ti} \right) \left(\partial_jg_{t \sigma} - \partial_{\sigma}g_{tj}\right)\Big|_{\mathtt{x}} \\
    &+ R_{titj}\Big|_{\mathtt{x}}\\ 
    =& \boldsymbol{a}_i \boldsymbol{a}_j + R_{titj}\Big|_{\mathtt{x}}.
\end{split}
\end{equation}
Again, following the same steps but for $|\vec{\zeta}|^2$, we also have $\frac{1}{2} \left(\nabla_{i} \nabla_{j} |\vec{\zeta}|^2\right) \Big|_{\mathtt{x}} = \boldsymbol{h}_i \boldsymbol{h}_j - R_{zizj}\Big|_{\mathtt{x}}$. In substance, this method for calculating the second order terms in the Taylor expansion of the metric is the same method used in Ref.\@ \citep{Sternberg2013}, which uses ``Jacobi fields" (the general term for the solutions to Eq.\@ (\ref{Result::JacobiFieldCondition})) to calculate the formula for the Riemann coordinate system Eq.\@ (\ref{Statement::RiemannMetricFormula}). Putting it all together,
\begin{equation}
\begin{split}
    &g_{tt} = -(1 + \vec{\boldsymbol{x}} \cdot \vec{\boldsymbol{a}})^2 + S_{ij} \boldsymbol{x}^i \boldsymbol{x}^j + \mathcal{O}(|\vec{\boldsymbol{x}}|^3), \\
    &g_{zz} = (1 - \vec{\boldsymbol{x}} \cdot \vec{\boldsymbol{h}})^2 + Q_{ij} \boldsymbol{x}^i \boldsymbol{x}^j + \mathcal{O}(|\vec{\boldsymbol{x}}|^3),
\end{split}
\end{equation}
where,
\begin{equation}
\begin{split}
    &S_{\mu \nu} \equiv -R_{\alpha \mu \beta \nu} T^{\alpha} T^{\beta} \Big|_{\mathtt{x}},\\
    &Q_{\mu \nu} \equiv -R_{\alpha \mu \beta \nu} M^{\alpha} M^{\beta} \Big|_{\mathtt{x}}. \\
\end{split}
\end{equation}

\subsection{Derivation of Christoffel Symbols} \label{Appendix::FermiChristoffel}

We use the formula $\Gamma^{\mu}_{\alpha \beta}$ $\equiv$ $\frac{1}{2}g^{\mu \lambda}(\partial_{\alpha}g_{\lambda \beta}$ $+$ $\partial_{\beta} g_{\lambda \alpha}$ $-$ $\partial_{\lambda}g_{\alpha \beta})$. 
Since the metric formula is itself truncated to order $\mathcal{O}(|\bvec{x}|^2)$, all formulae derived here will be truncated to order $\mathcal{O}(|\bvec{x}|)$. Throughout, the indices $(i,j)$ will strictly reference $x$ and $y$ components. 

First, we seek the components of the form $\Gamma^t_{\mu \nu}$ and $\Gamma^{z}_{\mu \nu}$,
\begin{equation} \label{Result::GammaUpperTAndZ}
\begin{split}
    \Gamma^{t}_{\mu \nu} &= \frac{1}{2}g^{tt} \left(\partial_{\mu} g_{t\nu} + \partial_\nu g_{t \mu}\right), \\
    \Gamma^{z}_{\mu \nu} &= \frac{1}{2}g^{zz} \left(\partial_{\mu} g_{z\nu} + \partial_\nu g_{z \mu}\right).
\end{split}
\end{equation}
Because the $t$ and $z$ coordinates are orthogonal to each other and all other coordinates, only the components of the form $\Gamma^{t}_{ti} = \partial_i \ln(\tnorm)$ and $\Gamma^{z}_{zi}=\partial_i \ln(\znorm)$ will be nonzero. Using Eq.\@ (\ref{Result::TAndZFormulae}),
\begin{equation}
\begin{split}
    \Gamma^{t}_{ti} &= \left(\bvec{a} - \mathbb{S} \cdot \bvec{x} \right)_i, \\
    \Gamma^{z}_{zi} &= \left(-\bvec{h} + \mathbb{Q} \cdot \bvec{x} \right)_i.
\end{split}
\end{equation}
Now we find the components $\Gamma^i_{t\mu}$,
\begin{equation}
\begin{split}
    \Gamma^i_{t\mu} &= \frac{1}{2}\bs{g}^{ij} \left(\partial_{t} g_{j\mu} + \partial_{\mu} g_{tj} - \partial_j g_{t\mu}\right) \\
    &= -\frac{1}{2}\partial^i g_{t \mu} = \frac{1}{2} \delta^t_{\mu} \partial^i \tnorm^2 .
\end{split}
\end{equation}
Similarly, $\Gamma^{i}_{z\mu} = \frac{1}{2}\delta_{\mu}^z \partial^i \znorm^2$. Again using Eq.\@ (\ref{Result::TAndZFormulae}),
\begin{equation}
\begin{split}
    \Gamma^i_{tt} &= \bs{a}^i + \left(\bs{a}^i\bs{a}_j -S^i_j \right) \bs{x}^j,\\
    \Gamma^i_{zz} &= \bs{h}^i - \left(Q^i_j + \bs{h}^i\bs{h}_j \right) \bs{x}^j.
\end{split}
\end{equation}
The only remaining non-zero Christoffel symbols are those purely involving the indices $(i,j)$. First, note that $\Gamma^x_{xx} = \frac{1}{2}\partial_x g_{xx} = 0$ and $\Gamma^y_{yy} = \frac{1}{2}\partial_y g_{yy}=0$, and so the non-zero components must be of the form $\Gamma^x_{yi}$ or $\Gamma^y_{xi}$. We also note that, truncated to order $\mathcal{O}(|\bvec{x}|)$, these are the same as $\Gamma_{xyi}$ and $\Gamma_{yxi}$,
\begin{equation}
\begin{split}
    \Gamma^{x}_{yi} &= \frac{1}{2}\left(\partial_y g_{xi} + \partial_i g_{xy} - \partial_x g_{iy} \right) \\
    &= \frac{1}{3} \mathcal{K} \left(2x \delta_{yi} - y\delta_{xi} \right),\\
    \Gamma^y_{xi} &= \frac{1}{2}\left(\partial_x g_{yi} + \partial_i g_{xy} - \partial_y g_{ix} \right) \\
    &= \frac{1}{3}\mathcal{K} \left(2y\delta_{xi} - x\delta_{yi} \right).
\end{split}
\end{equation}
Putting it all together, the non-zero Christoffel symbols are:
\begin{equation}
\begin{split}
    \Gamma^t_{ti} &= \left(\bvec{a} - \mathbb{S} \cdot \bvec{x} \right)_i, \\
    \Gamma^z_{zi} &= \left(-\bvec{h} + \mathbb{Q} \cdot \bvec{x} \right)_i, \\
    \Gamma^i_{tt} &= \bs{a}^i + \left(\bs{a}^i\bs{a}_j -S^i_j  \right) \bs{x}^j,\\
    \Gamma^i_{zz} &= \bs{h}^i - \left(Q^i_j + \bs{h}^i\bs{h}_j \right) \bs{x}^j,\\
    \Gamma^x_{yi} &= \frac{1}{3} \mathcal{K} \left(2x \delta_{yi} - y\delta_{xi} \right),\\
    \quad \Gamma^y_{xi} &= \frac{1}{3}\mathcal{K} \left(2y\delta_{xi} - x\delta_{yi} \right).
\end{split}
\end{equation}

\section{Magnetic Field Due to Current-Carrying Wire for $\mathcal{K} \neq 0$} \label{Appendix::TheWire}

Suppose in the place of an X-line, we instead have an infinitely thin wire passing through $\mathtt{x}$ along $\hat{M}$ carrying a current $I$. We will assume at first a ``slab symmetric" case $\bvec{h}=0$ in the absence of NICS effects, so that $\tnorm$ and $\znorm$ are uniformly $1$ and $\bs{\Psi}_i = \partial_i \psi$. In this case, $\delta^{ij} \partial_i \partial_j \psi = -j^M$ where $j^M$ is merely a point source at $\mathtt{x}$ corresponding to the current through the wire. Because this situation has rotational symmetry within the plane, $\psi$ must be of the form $\psi(r)$ where $r\equiv \sqrt{x^2 + y^2}$, and it is determined by the ODE,
\begin{equation}
    \psi^{\prime \prime} + \frac{1}{r} \psi^{\prime} = -j^M.
\end{equation}
This equation has the well known solution $\psi = (I/2\pi) \ln(r)$, which gives the expected magnetic flux density $|\bs{B}| = I/(2\pi r)$. Still neglecting all other NICS effects, we will now vary the parameter $\mathcal{K}$ slightly from zero, and seek solutions of the form,
\begin{equation}
    \psi = \frac{I}{2\pi} \ln(r) + \tilde{\psi},
\end{equation}
where the perturbation $\tilde{\psi}$ is on the order $\mathcal{O}(\mathcal{K})$. Eq.\@ (\ref{Result::AmperesLaw}) is now,
\begin{equation}
    \delta^{ij} \partial_i \partial_j \tilde{\psi} - \frac{I}{3\pi}\mathcal{K} \bs{x}^i \partial_i \ln(r) = \mathcal{O}(\mathcal{K}^2).
\end{equation}
Note that this equation can still be treated as a partial differential equation in 2D Euclidean space, and that the rotational symmetry is unbroken. Therefore, $\tilde{\psi} = \tilde{\psi}(r)$, and we again formulate the problem as an ODE,
\begin{equation}
    \tilde{\psi}^{\prime \prime} + \frac{1}{r}\tilde{\psi}^{\prime} = \frac{I}{3\pi} \mathcal{K}, \quad \Rightarrow \tilde{\psi}(r) = \frac{I}{12\pi} \mathcal{K} r^2.
\end{equation}
Adding this perturbation to $(I/2\pi)\ln(r)$, we obtain the final result,
\begin{equation}
\begin{split}
    \psi(r) &= \frac{I}{2\pi}\left(\ln(r) + \frac{1}{6} \mathcal{K}r^2 \right) + \mathcal{O}(\mathcal{K}^2), \\
    &\hookrightarrow |\bvec{B}| = \frac{I}{2\pi}\left(\frac{1}{r} + \frac{1}{3}\mathcal{K} r \right).
\end{split}
\end{equation}

\section{Calculations Involving the Schwarzschild Metric} \label{Appendix::Schwarzschild}

Here we detail important calculations for the usage of the Schwarzschild solution for the final derivations provided in this paper. These calculations will draw from formulae provided by Ref.\@ \citep{Mueller2009}, using the Painleve-Gullstrand coordinates $\{t_{pg},r,\theta,\phi\}$ and the Schwarzschild coordinates $\{ t_s,r,\theta,\phi\}$. We will let $r_s \equiv 2GM$ be the Schwarzschild radius.

\subsection{Components of the Riemann Curvature Tensor Within A Zero Angular Momentum Frame} \label{Appendix::SchRiemann}

In Painleve-Gullstrand coordinates, the Schwartzchild metric is expressed with the following line element,
\begin{equation}
\begin{split}
    ds^2 =& -dt_{pg}^2 + \left(dr + \sqrt{r_s/r} dt \right)^2 \\
    &+ r^2 \left(d\theta^2 + \sin^2(\theta) d\phi^2 \right).
\end{split}
\end{equation}
Freely-falling objects which have zero angular momentum and have energy equal to the stationary observer at infinity have world-lines tangent to the vector field $\vec{\nabla}t_{pg}$. We may define a natural orthonormal frame corresponding to the local frame of these objects \citep{Mueller2009},
\begin{equation}
\begin{alignedat}{2}
    &\hat{T}_0 \equiv \vec{\partial}_{t_{pg}} - \sqrt{\frac{r_s}{r}}\vec{\partial_r}, && \qquad \hat{r}_0 \equiv \vec{\partial_r}, \\ 
    &\hat{\theta} \equiv \frac{1}{r} \vec{\partial_{\theta}}, && \qquad \hat{\phi} \equiv \frac{1}{r\sin(\theta)} \vec{\partial_{\phi}}.
\end{alignedat}
\end{equation}
Only within this section, we will adopt a more convenient notation convention in which $\oneform{a}\oneform{b} \equiv \oneform{a} \wedge \oneform{b}$, and $(\mathbb{X},\mathbb{Y}) \equiv \mathbb{X} \otimes \mathbb{Y}$. We can then express the Riemann curvature tensor in the following way \citep{Mueller2009},
\begin{equation} \label{Statement::RiemannTensorInFreeFall}
\begin{split}
    \mathbb{R}/\kappa =& (\oneform{T_0} \oneform{r_0},\oneform{T_0} \oneform{r_0}) + \frac{1}{2}(\oneform{T_0} \oneform{\theta},\oneform{T_0} \oneform{\theta}) + \frac{1}{2}(\oneform{T_0} \oneform{\phi},\oneform{T_0} \oneform{\phi}) \\
    &- \frac{1}{2}(\oneform{r_0} \oneform{\theta},\oneform{r_0}\oneform{\theta}) - \frac{1}{2}(\oneform{r_0} \oneform{\phi},\oneform{r_0}\oneform{\phi}) + (\oneform{\theta} \oneform{\phi},\oneform{\theta}\oneform{\phi}),
\end{split}
\end{equation}
where $\kappa \equiv r_s/r^3$. Every zero angular momentum frame $\{\hat{T}, \hat{r}, \hat{\theta}, \hat{\phi} \}$ can be expressed as a simple Lorentz transformation,
\begin{equation}
\begin{split}
    \hat{T}_0 &\equiv \cosh(\xi) \hat{T} + \sinh(\xi) \hat{r}, \\
    \hat{r}_0 &\equiv \sinh(\xi) \hat{T} + \cosh(\xi) \hat{r},
\end{split}
\end{equation}
where $\xi$ is the relative rapidity between the two frames. We now show that $\mathbb{R}$ has the same form within an arbitrary zero angular momentum frame, that is, that Eq.\@ (\ref{Statement::RiemannTensorInFreeFall}) would remain valid with $\hat{T}_0 \rightarrow \hat{T}$ and $\hat{r}_0 \rightarrow \hat{r}$. We will prove this via direct substitution. First, consider the 2-form $\oneform{T_0}\oneform{r_0}$,
\begin{equation}
\begin{split}
    \oneform{T_0} \oneform{r_0} =& \left(\cosh(\xi) \oneform{T} + \sinh(\xi) \oneform{r} \right) \\
    &\wedge \left(\sinh(\xi) \oneform{T} + \cosh(\xi) \oneform{r} \right) \\
    =& \cosh^2(\xi) \oneform{T} \oneform{r} + \sinh^2(\xi) \oneform{r} \oneform{T} = \oneform{T} \oneform{r}.
\end{split}
\end{equation}
Clearly, $\oneform{T_0} \oneform{\theta} = \cosh(\xi) \oneform{T} \oneform{\theta} + \sinh(\xi) \oneform{r} \oneform{\theta}$ and $\oneform{r_0} \oneform{\theta} = \sinh(\xi) \oneform{T} \oneform{\theta} + \cosh(\xi) \oneform{r} \oneform{\theta}$, and therefore,
\begin{equation}
\begin{split}
    (\oneform{T_0}\oneform{\theta},\oneform{T_0}\oneform{\theta}) =& \cosh^2(\xi)(\oneform{T} \oneform{\theta},\oneform{T}\oneform{\theta})  \\
    &+ \cosh(\xi) \sinh(\xi) \left((\oneform{T}\oneform{\theta},\oneform{r}\oneform{\theta} ) + (\oneform{r} \oneform{\theta},\oneform{T}\oneform{\theta})\right) \\
    &+ \sinh^2(\oneform{r}\oneform{\theta},\oneform{r}\oneform{\theta}),  \\
    (\oneform{r_0} \oneform{\theta},\oneform{r_0} \oneform{\theta}) =& \sinh^2(\xi)(\oneform{T} \oneform{\theta},\oneform{T}\oneform{\theta}) \\
    &+ \cosh(\xi) \sinh(\xi) \left((\oneform{T}\oneform{\theta},\oneform{r}\oneform{\theta} ) + (\oneform{r} \oneform{\theta},\oneform{T}\oneform{\theta})\right) \\
    &+ \cosh^2(\xi)(\oneform{r}\oneform{\theta},\oneform{r}\oneform{\theta}), \\ \\
    \hookrightarrow (\oneform{T_0}\oneform{\theta},\oneform{T_0}\oneform{\theta}) -&  (\oneform{r_0} \oneform{\theta},\oneform{r_0} \oneform{\theta}) = (\oneform{T} \oneform{\theta},\oneform{T}\oneform{\theta}) - (\oneform{r} \oneform{\theta},\oneform{r}\oneform{\theta}).
\end{split}
\end{equation}
Following the same steps, $(\oneform{T_0}\oneform{\phi},\oneform{T_0}\oneform{\phi}) - (\oneform{r_0} \oneform{\phi},\oneform{r_0} \oneform{\phi}) = (\oneform{T}\oneform{\phi},\oneform{T}\oneform{\phi}) - (\oneform{r} \oneform{\phi},\oneform{r} \oneform{\phi})$. We have therefore proved that the tensors $(\oneform{T_0}\oneform{r_0},\oneform{T_0}\oneform{r_0})$, $(\oneform{T_0}\oneform{\theta},\oneform{T_0}\oneform{\theta}) - (\oneform{r_0} \oneform{\theta},\oneform{r_0} \oneform{\theta})$, and $(\oneform{T_0}\oneform{\phi},\oneform{T_0}\oneform{\phi}) - (\oneform{r_0} \oneform{\phi},\oneform{r_0} \oneform{\phi})$ are exactly preserved by the substitution $\oneform{T_0} \rightarrow \oneform{T}$ and $\oneform{r_0} \rightarrow \oneform{r}$. By inspection, it follows directly that the formula Eq.\@ (\ref{Statement::RiemannTensorInFreeFall}) is also preserved by these substitutions, and so we have (now using the standard notation) the formula used in Section \ref{Section::CentralBodyProblem},
\begin{equation}
\begin{split}
    \mathbb{R}/\kappa = &-(\oneform{T} \wedge \oneform{r}) \otimes (\oneform{T} \wedge \oneform{r}) + \frac{1}{2} (\oneform{T} \wedge \oneform{\theta}) \otimes (\oneform{T} \wedge \oneform{\theta}) \\
    &+ \frac{1}{2} (\oneform{T} \wedge \oneform{\phi}) \otimes (\oneform{T} \wedge \oneform{\phi}) - \frac{1}{2} (\oneform{r} \wedge \oneform{\theta}) \otimes (\oneform{r} \wedge \oneform{\theta}) \\
    &- \frac{1}{2} (\oneform{r} \wedge \oneform{\phi}) \otimes (\oneform{r} \wedge \oneform{\phi}) + (\oneform{\theta} \wedge \oneform{\phi}) \otimes (\oneform{\theta} \wedge \oneform{\phi}).
\end{split}
\end{equation}

\subsection{Acceleration of the Static Frame} \label{Appendix::SchAcceleration}

Now we will use the Schwarzschild coordinate system, in which the line element is,
\begin{equation}
\begin{split}
    ds^2 =& -(1 - r_s/r)dt_{s}^2 + \frac{1}{1 - r_s/r}dr^2 \\
    &+ r^2(d\theta^2 + \sin^2(\theta)d\phi^2).
\end{split}
\end{equation}
``Static" observers which have zero angular momentum and for which $r$ is a constant of motion follow world-lines tangent with $\vec{\nabla} t_s$. We may define the orthonormal basis corresponding to this frame \citep{Mueller2009}:
\begin{equation}
\begin{alignedat}{2}
    &\hat{T}_s \equiv \frac{1}{\sqrt{1 - r_s/r}} \vec{\partial_{t_s}}, && \qquad \hat{r} \equiv \sqrt{1 - r_s /  r} \vec{\partial_r}, \\ 
    &\hat{\theta} \equiv \frac{1}{r} \vec{\partial_{\theta}}, && \qquad \hat{\phi} \equiv \frac{1}{r\sin(\theta)} \vec{\partial_{\phi}}.
\end{alignedat}
\end{equation}
Since this line element has the same form as Eq.\@ (\ref{Statement::GaugeConstraint}), we may calculate the acceleration of the static frame $\vec{a}_s \equiv \nabla_{\hat{T}_s} \hat{T}_s$ using Eq.\@ (\ref{Statement::SecondaryGaugeConstraint}) where $\tnorm \rightarrow \sqrt{1 - r_s/r}$,
\begin{equation}
    \vec{a}_s = \vec{\nabla} \ln \left(\sqrt{1 - r_s/r} \right) = \frac{r_s}{2r^2} \frac{1}{\sqrt{1 - r_s/r}} \hat{r}.
\end{equation}
Note that this acceleration diverges at the event horizon $r=r_s$, since this is the threshold where ``static" motion in the sense meant here becomes physically impossible. At $r=r_s$, curves of constant $r$ are null curves (light trajectories), and for $r<r_s$ they are spacelike curves (super-luminal trajectories).

\end{document}